\documentclass[12pt]{article}
\pdfoutput=1

\usepackage{booktabs}
\usepackage[a4paper,text={16.8cm,22.4cm}]{geometry}
\usepackage{amsmath,amsfonts,braket,slashed,amssymb,tikz,bm,psfrag,graphicx,color,dsfont,euscript}
\usepackage{multicol}
\usepackage[small,labelfont=bf]{caption}

\usepackage{tikz}
\usepackage{adjustbox}
\usetikzlibrary{decorations.pathmorphing}
\usetikzlibrary{decorations.markings}

\RequirePackage[sort&compress,square,comma,numbers]{natbib}
\allowdisplaybreaks
\usepackage{epsfig,amssymb,amsmath,psfrag,epstopdf,color}
\pdfoutput=1
\usepackage{graphicx}
\usepackage[symbol]{footmisc}

\usepackage{paralist}
\usepackage{hyperref}
\allowdisplaybreaks
\usepackage{mathrsfs}

\usepackage[abs]{overpic}
\usepackage{mathtools}
\usepackage{subfigure}

\newcommand{\beq}{\begin{equation}}
\newcommand{\eeq}{\end{equation}}
\def\bea#1\eea{\begin{align}#1\end{align}}

\def \be  {\begin{equation}}
\def \ee  {\end{equation}}
\def \ba  {\begin{eqnarray}}
\def \ea  {\end{eqnarray}}

\usepackage{graphicx}
\graphicspath{ {./figure/} }

\allowdisplaybreaks

\begin{document}
\begin{titlepage}
\begin{center}
\Large\bf
Factorization of photon induced processes in ultra-peripheral heavy ion collisions
\end{center}

\vspace{0.3cm}
\begin{center}
Yu-Cheng Hui$^{a}$~\footnote[2]{kad86556@gmail.com}
\\
\vspace{0.7cm} 
{\sl ${}^a$Department of Physics and Center for Field Theory and Particle Physics, Fudan University, Shanghai, China}
\end{center}

\vspace{0.8cm}
\begin{abstract}
In this study, we investigate photon-photon scattering in ultra-peripheral heavy ion collisions (UPCs). We start by deriving an effective Lagrangian from first principles and then apply factorization techniques from Soft-Collinear effective theory (SCET). This approach allows us to decompose the photon-photon scattering cross-section into two primary factors: the generalized transverse momentum distributions (GTMDs) and the hard scattering amplitude. We further analyze the emission of soft photons by final state leptons, incorporating a soft function into the cross section through an evolution method. Our analysis yields detailed predictions for observable angular correlations among the final state leptons. Specifically, we calculate the angular correlations characterized by the azimuthal parameters $\langle\cos 2\phi\rangle$ and $\langle\cos 4\phi\rangle$, highlighting the influence of initial photons polarization and recoil effects from final state soft photons.
%In ultra-peripheral heavy ion collisions (UPCs), photon-photon scattering has been observed in various experiments. Calculating observables for photon-photon scattering and comparing them with experimental results through numerical simulations would be highly meaningful. This article begins with first principles, constructing an effective Lagrangian. Subsequently, employing the factorization methods within Soft-collinear effective theory (SCET), the photon-photon scattering cross-section is factorized into two components: the transverse phase space PDF and the hard function. Following this, we consider the process wherein final state leptons emit soft photons. Therefore, by employing the method of evolution, the soft function can be convolved into the cross-section. Leveraging the obtained results, we provide a comprehensive calculation of the observable angular correlation. Finally, utilizing the results obtained through theoretical calculations, we conduct numerical simulations regarding the angular correlation of final state leptons. The numerical results elucidate the $\langle\cos 2\phi\rangle$ and $\langle\cos 4\phi\rangle$ azimuthal anomalies induced by the polarization of the initial photon and final state soft photons radiation. Our work establishes a seamless connection with the Glauber model, which is extensively employed in heavy-ion collision studies. Additionally, it provides insights for subsequent calculations concerning the quark-gluon plasma (QGP) effect on final leptons in non-central and central collisions.
\end{abstract}

\end{titlepage}

\tableofcontents
\vspace{8mm}

\section{Introduction}
\label{sec:intro}

In relativistic heavy-ion collisions, the interaction between two charged heavy ions generates exceptionally intense electromagnetic field, with magnitudes reaching up to $10^5$T \cite{skokov2009estimate,bzdak2012event,deng2012event,roy2015event,pu2016bjorken}. These fields can be effectively modeled using the equivalent photon approximation (EPA) \cite{Williams:1934ad,wang2021lepton}, where the strong electromagnetic field is treated as a polarized photon field. The resulting photons, often referred to as Breit-Wheeler photons, can be approximated as on-shell. When the photon wavelength is much longer than the nuclear radius, these photons can couple with the charged particles inside the nucleus, leading to collective motion \cite{wang2021lepton}. The photon number density is proportional to the square of the nuclear charge, significantly enhancing the cross section for two-photon scattering in ultra-peripheral collisions (UPC). In UPCs, where the nuclei remain intact and only quasi-real photon interactions occurs, the clean experimental environment provides a unique opportunity to probe physics beyond the Standard Model. For example, precise measurements of lepton pairs produced in UPCs allow stringent tests of the Standard Model, particularly for anomalous magnetic and electric dipole moments, thereby offering constraints on new physics \cite{
czarnecki2001muon,giudice2012testing,kurz2014hadronic,kurz2015light,kurz2016electron,liu2019light,liu2020light,aebischer2021effective,xiao2022multi,cirigliano2021determining,particle2022review,wen2023single,Cao:2023juc,Shao:2023bga,Cao:2021lmj}. Additionally, UPCs provide a valuable electromagnetic probe for investigating the properties of the quark-gluon plasma (QGP) formed in heavy-ion collisions \cite{pu2023coherent,klein2019acoplanarity}.

Significant progress has been made in UPC studies. For instance, the STARlight software package \cite{Klein:2016yzr}, based on the EPA, has been developed to explore photon-induced dilepton production. However, theoretical predictions using the EPA often show significant deviations from experimental data due to the absence of critical information regarding photon transverse momentum and polarization. To address these discrepancies, various approaches \cite{wang2021lepton,Williams:1934ad,Shao:2023bga,Shao:2022stc,li2020impact,Wang:2022gkd,Krauss:1997vr,Harland-Lang:2020veo} have been developed to compute polarization-dependent scattering cross sections as functions of the transverse momemtum of lepton pairs and the impact parameter. These refined cross sections enable detailed studies of azimuthal modulation in lepton pairs produced in both ultra-peripheral and peripheral heavy-ion collisions.

Despite these advancements, several avenues remain open for exploration in UPC studies. One promising direction is to investigate the influence of the QGP on the behavior of final-state lepton pairs. In UPCs, the total transverse momentum of the lepton pair, $p_{l_{1\perp}}+p_{l_{2\perp}}$, is remarkably small, typically around 10 MeV, and significantly smaller than the transverse momentum of each individual lepton, $p_{l_{1\perp}}$, $p_{l_{2\perp}}$. However, measurements by the ATLAS \cite{aaboud2018observation} and STAR \cite{adam2018low} collaborations have shown that, in both peripheral and central heavy-ion collisions, the total transverse momentum of lepton pairs is notably larger than in UPCs, a phenomenon known as the broadening effect. This effect is particularly observed in the small transverse momentum region, where photon scattering dominates lepton pair production in peripheral and central collisions \cite{aaboud2018observation,adam2018low,alice2015measurement}. The broadening likely results from the interaction of the lepton pair with electromagnetic fields and charged particles within the QGP, leading to multiple electromagnetic scatterings as the pair traverses the medium.

Preliminary theoretical research on the QGP’s contribution \cite{klein2019acoplanarity} to transverse momentum broadening has drawn analogies with QCD jet theory \cite{Baier:1996sk,Baier:1996vi}. In QCD, particles undergo scattering with quarks and gluons, whereas in QED, lepton pairs primarily interact with charged quarks. A QED-type Wilson line can describe the exchange of multiple photons between leptons and the medium, allowing the treatment of lepton pairs as dipoles. Analogous to the computation of the color dipole in QCD \cite{iancu2001nonlinear,mueller1999parton,Pasechnik:2023mdd}, the QED saturation scale $\langle\hat{q}L\rangle$, where $L$ is the medium length and $\hat{q}$ can be extracted from experimental data \cite{burke2014extracting}, can be determined. The multiple scattering amplitude between dipoles and the medium is then calculated as:
\begin{align*}
    \langle\mathcal{U}(b_\perp+\frac{1}{2}r_\perp)\mathcal{U}^{\dagger}(b_\perp-\frac{1}{2}r_\perp)\rangle
    =e^{-\frac{\langle\hat{q}L\rangle r^2_\perp}{4}},
\end{align*}
where $\mathcal{U}$ represents the QED-type Wilson line, $b_\perp$ is the impact parameter, and $r_\perp$ denotes the dipole size. Incorporating this result into the scattering cross section yields a modified cross section that accounts for the QGP’s influence. Subsequent analysis has examined the effect of the QGP’s magnetic field \cite{adam2018low} on transverse momentum broadening. However, further refinement and detailed investigations into the QGP's properties are necessary.

Building upon earlier work, this paper aims to establish a quantum framework for impact parameter-dependent scattering cross sections in UPCs by integrating QED factorization \cite{wu2021factorization} with the classical Glauber model \cite{miller2007glauber}. This framework is intended to facilitate the use of UPC final-state products as electromagnetic probes for future studies of the QGP.

Using Effective Field Theory (EFT), we derive the average angular correlation of lepton pairs from the Standard Model Lagrangian and validate our calculations with simulation data. Specifically, we analyze the angular correlation of $\mu^+\mu^-$ pairs from ultra-peripheral Au-Au collisions at 200 GeV and the angular correlation of $\tau^+\tau^-$ pairs produced in ultra-peripheral Pb-Pb collisions at 5020 GeV under LHC experimental conditions.

This paper is organized as follows: In Section 2, we construct the effective Lagrangian for the process $\gamma+\gamma\rightarrow l+\bar{l}+\gamma_s$. Section 3 describes the factorization of the UPC cross-section into high-energy and non-perturbative components using a wave packet description of nuclei. In Section 4, we establish the equivalence between the classical photon distribution and generalized transverse momentum distributions (GTMDs). Section 5 discusses the decoupling transformation of the soft function. Section 6 focuses on the angular correlation of final-state lepton pairs. Section 7 explores modifications to the angular correlation due to soft radiation. Numerical results are presented in Section 8, followed by conclusions in Section 9.

\section{The effective lagrangian}
\label{sec:lag}
In the theory of quantum electrodynamics (QED), the process $\gamma\gamma\rightarrow l\bar{l}\gamma_s$ involves the emission of soft photons in the final state, denoted by $\gamma_s$. In this process, there exist three degrees of freedom, corresponding to the photon field $A^{\mu}$ and the fermion fields which correspond to the leptons and anti-leptons, denoted as $\psi$ and $\bar{\psi}$, respectively. However, in effective field theory, we can project the complete QED field $\psi$ with projection operator $P_v=\frac{1+\slashed{v}}{2}$ to fields corresponding to different velocity. Then velocity becomes one of the characteristics of the field, with different electrons equipped with different velocities corresponding to different fermion fields after scattering.

Meanwhile, through a decoupling transformation, soft photons will decouple from the fermion field. However, collinear photons scattered in the initial state still have interaction vertices with the fermion field, therefore, soft photons and collinear photons correspond to different degrees of freedom. As a result, corresponding to process $\gamma \gamma \rightarrow l \bar{l} \gamma_s$, we have a total of six degrees of freedom. These include the gauge invariant collinear photon field $\mathcal{A}^{\mu}_c$ originating from initial state scattering, the gauge invariant soft photon field $\mathcal{A}^{\mu}_s$ arising from final state radiation, and the lepton fields $h_{v_1}$, $\bar{h}_{v_1}$, $h_{v_2}$, and $\bar{h}_{v_2}$ corresponding to different ejection velocities $v_1$ and $v_2$, respectively.

Firstly, we can express the kinematic term of $\mathcal{A}^{\mu}_s$ and $\mathcal{A}^\mu_c$ as $-\frac{1}{4}(\mathcal{F}^s_{\mu\nu})^2$ and $-\frac{1}{4}(\mathcal{F}^c_{\mu\nu})^2$, where $\mathcal{F}^s_{\mu\nu}=\partial_{\mu}\mathcal{A}^s_{\nu}-\partial_{\nu}\mathcal{A}^s_{\mu}$, $\mathcal{F}^c_{\mu\nu}=\partial_{\mu}\mathcal{A}^c_{\nu}-\partial_{\nu}\mathcal{A}^c_{\mu}$. When it comes to the general gauge field, we need to construct a gauge-invariant block, the derivative can be represented as covariant derivative
\begin{align}
iD_{\mu}=i\partial_{\mu}-eA_\mu,
\end{align}
where $A_{\mu}$ represents the photon field in the QED theory. In QED theory, the gauge field that appears in the covariant derivative also includes both soft field and collinear field. However, in the effective theory, as we will see below, with the decoupling transformation, $\mathcal{A}^s$ is decoupled from the covariant derivative. Next, we construct the kinematic term of the lepton field. In fact, this term should be written in a gauge-invariant form as 
\begin{align}
\bar{h}_{v_i}(x)i(v_i\cdot D)h_{v_i}(x),
\end{align}
where
\begin{align}
i(v_i\cdot D)=iv_i\cdot \partial-ev_i\cdot A_s(x)-ev_i\cdot A_c(x).
\end{align}
However, the term $\bar{h}_{v_i}(x)(v_i\cdot A_c)h_{v_i}(x)$ is prohibited due to momentum conservation. Furthermore, as a result of the aforementioned decoupling transformation, the soft field is decoupled from the covariant derivative, thus resulting in 
\begin{align*}
i(v_i\cdot D)\rightarrow iv_i\cdot\partial.
\end{align*}
Finally, we construct the interaction term. The term with dimension 4 has the form $\mathcal{A}^{\mu}_c\bar{h}_{v_i}\Gamma_{\mu}h_{v_j}$, $\Gamma_{\mu}$ is a tensor in spinor space. This term is prohibited for the reason that in UPC scattering photons are treated on shell, so we don't consider the situation of photon decay.\\
The term with dimension 5 can be constructed as $\mathcal{C}\mathcal{A}^{\mu}_n\mathcal{A}^{\nu}_{\bar{n}}\bar{h}_{v_1}\Gamma_{\mu\nu}h_{v_2}$, where $\Gamma_{\mu\nu}$ is a tensor in spinor space. For clarity, we replace index $c$ to $n$ and $\bar{n}$, which corresponding to the direction of light cone and anti-light cone. $\mathcal{C}=\mathcal{C}_0Z^{-1}$ is a renormalized Wilson coefficient. And $\mathcal{C}_0$ equipped by subscript 0 represents the bare quantity. $Z^{-1}$ renormalizes both the Wilson coefficient and the operator simultaneously, given by
\begin{align}
\mathcal{C}_0\langle\mathcal{A}^{\mu}_n\mathcal{A}^{\nu}_{\bar{n}}\bar{h}_{v_1}\Gamma_{\mu\nu}h_{v_2}\rangle_0&=\mathcal{C}_0Z^{-1}\langle\mathcal{A}^{\mu}_n\mathcal{A}^{\nu}_{\bar{n}}\bar{h}_{v_1}\Gamma_{\mu\nu}h_{v_2}\rangle=\mathcal{C}\langle\mathcal{A}^{\mu}_n\mathcal{A}^{\nu}_{\bar{n}}\bar{h}_{v_1}\Gamma_{\mu\nu}h_{v_2}\rangle.
\end{align}
The details about renormalized Wilson coefficient can be found in \cite{Becher:2014oda}.

 Based on the above analysis, we could write down the complete effective Lagrangian corresponding to the double photon scattering process in UPC, up to soft photon radiation correction.
 \begin{align}
    \mathcal{L}_{eff}=-\frac{1}{4}(\mathcal{F}^s_{\mu\nu})^2-\frac{1}{4}(\mathcal{F}^c_{\mu\nu})^2+\sum\limits^2\limits_{i=1}\bar{h}_{v_i}(x)i(v_i\cdot D)h_{v_i}(x)+\mathcal{C}\mathcal{A}^{\mu}_n\mathcal{A}^{\nu}_{\bar{n}}\bar{h}_{v_1}\Gamma_{\mu\nu}h_{v_2}.
\end{align}

\section{Factorization}
\label{sec:fac}
 First of all, let us briefly review wave packet expansion in field theory. The expansion of a single particle state can be expressed as the composition of its corresponding momentum eigenstates
\begin{align}
\ket{\phi}=\int \frac{d^3 \vec{p}}{(2\pi)^3\sqrt{2E_p}}\phi(\vec{p})\ket{\vec{p}}.
\end{align}
Consider a scattering process $A+B \rightarrow n~~particles$. The incoming wave function can be expanded as
\begin{align}
&\ket{\phi_A}_{in}=\int \frac{d^3\vec{p}_A}{(2\pi)^3\sqrt{2E_A}}\phi(\vec{p}_A)\ket{\vec{p}_A}_{in},\\ \notag
&\ket{\phi_B}_{in}=\int\frac{d^3\vec{p}_B}{(2\pi)^3\sqrt{2E_B}}\phi(\vec{p}_B)e^{-ip_{B\perp}\cdot b_\perp}\ket{\vec{p}_B}_{in},\\ \notag
\end{align}
where $b_\perp$ represents the impact parameter. The probability of a scattering event between a particle A and a particle B resulting in the final state of n particles occupying a specific point in phase space is defined as 
\begin{align}
P=\prod\limits_{f}\frac{d^3p_f}{(2\pi)^3}\frac{1}{2E_f}\left |_{out}\braket{\vec{p}_1\vec{p}_2\dots|\phi_A\phi_B}_{in}\right|^2.
\end{align}
Through the analogy of the quantum field theory, which expands the incoming particle state and the outgoing particle state, we can also expand the incoming heavy ion states into their corresponding momentum eigenstates.

Let's briefly introduce the notation we used in our passage. $p_A$ and $p_B$ are the momentum eigenvalues of collided heavy ions A and B, respectively, while $p^{\prime}_A$ and $p^{\prime}_B$ are their corresponding momenta in the complex wave packets. With momentum in the wave packets and the corresponding momentum in the complex wave packets, a pair of new independent variables, $\mathcal{P}$ and $q$, can be reconstructed. The exact definition is as follows: 
\begin{align}
\label{notation}
&\mathcal{P}_{A\perp}=\frac{1}{2}(p_{A\perp}+p_{A\perp}^{\prime}),~~~~~~~~~
\mathcal{P}_{B\perp}=\frac{1}{2}(p_{B\perp}+p^{\prime}_{B\perp}),\\ \notag
&q_\perp=p_{A\perp}-p_{A\perp}^\prime=q_{A\perp},~~~~~~~~
q_\perp=p^{\prime}_{B\perp}-p_{B\perp}=q_{B\perp},\\ \notag
&\mathcal{P}_{Az}=\frac{1}{2}(p_{Az}+p^{\prime}_{Az})=p_{Az},~~~
\mathcal{P}_{Bz}=\frac{1}{2}(p_{Bz}+p^{\prime}_{Bz})=p_{Bz}. \notag
\end{align}
We use different subscript to denote different partical: $p_A$ and $p_B$ are momentum of wave packet, $p_{l_1}$ and $p_{l_2}$ denote momentum of final states leptons and we use $k_1$, $k_2$ to express scattering photons momentum. Here, we use the relation that $ p_{Az}=p_{Az}^{\prime}$ and $ p_{Bz}=p^{\prime}_{Bz}$. This is because we select the heavy ion beam direction as the z-direction, and the plane perpendicular to the beam as the impact plane which is denoted by $\perp$. Given that $p_z\gg p_\perp$, the momentum component in the z-direction is approximately equal to the corresponding z-direction momentum component in the complex wave packet. For a more detailed explanation, please refer to \cite{wu2021factorization}.

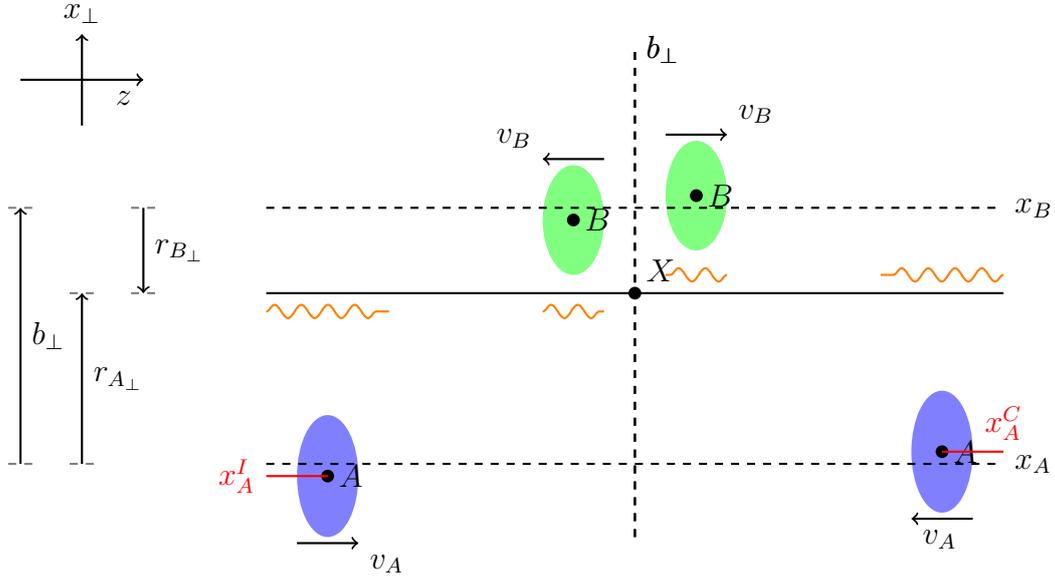
\begin{figure}[ht]
\begin{center}
\tikzset{every picture/.style={line width=0.8pt}}
\begin{tikzpicture} [x=23pt,y=23pt,yscale=1,xscale=1]
\draw[black, thick] (-6,0) -- (6,0);
\draw[black, dashed] (0,-4) -- (0,4) node[right]{$b_\perp$};
\fill[blue!50] (-5,-3) ellipse (0.5 and 1.0);
\filldraw[black] (-5,-3) circle (2pt) node[anchor=west]{$A$};
\draw[red, thick] (-5,-3) -- (-6,-3) node[anchor=east]{$x^I_A$};
\draw[->] (-10,3.5) -- (-8,3.5) node[below left] {$z$};
\draw[->] (-9,2.75) -- (-9,4.25) node[above] {$x_\perp$};
\draw[black, dashed] (0,-4) -- (0,4) node[right]{$b_\perp$};
\fill[blue!50] (5,-2.6) ellipse (0.5 and 1.0);
\filldraw[black] (5,-2.6) circle (2pt) node[anchor=west]{$A$};
\draw[red, thick] (5,-2.6) -- (6,-2.6) node[above]{$x^C_A$};
\draw[black, dashed] (-6,-2.8) -- (6,-2.8) node[right]{$x_A$};
\filldraw[black] (0,0) circle (2pt) node[above right]{$X$};
\fill[green!50] (-1,1.2) ellipse (0.5 and 0.9);
\fill[green!50] (1,1.6) ellipse (0.5 and 0.9);
\filldraw[black] (-1,1.2) circle (2pt) node[anchor=west]{$B$};
\filldraw[black] (1,1.6) circle (2pt) node[anchor=west]{$B$};
\draw[black, dashed] (-6,1.4) -- (6,1.4) node[right]{$x_B$};
\draw[->] (-5.5,-4.1) -- (-4.5,-4.1) node[below right] {$v_A$};
\draw[->] (5.5,-3.7) -- (4.5,-3.7) node[below right] {$v_A$};
\draw[->] (-0.5,2.2) -- (-1.5,2.2) node[above left] {$v_B$};
\draw[->] (0.5,2.6) -- (1.5,2.6) node[above right] {$v_B$};
\draw[orange, decorate, decoration=snake] (-6.0,-0.3) -- (-4.0,-0.3);
\draw[orange, decorate, decoration=snake] (-1.5,-0.3) -- (-0.5,-0.3);
\draw[orange, decorate, decoration=snake] (1.5,0.3) -- (0.5,0.3);
\draw[orange, decorate, decoration=snake] (6.0,0.3) -- (4.0,0.3);
\draw[->] (-8,1.4) -- (-8,0) node[midway,anchor=west]{$r_{B_\perp}$};
\draw[->] (-9,-2.8) -- (-9,0) node[midway,anchor=west]{$r_{A_\perp}$};
\draw[->] (-10,-2.8) -- (-10,1.4) node [midway,anchor=west]{$b_\perp$};
\draw[gray, dashed] (-8.2,1.4) -- (-7.8,1.4);
\draw[gray, dashed] (-8.2,0) -- (-7.8,0);
\draw[gray, dashed] (-9.2,0) -- (-8.8,0);
\draw[gray, dashed] (-9.2,-2.8) -- (-8.8,-2.8);
\draw[gray, dashed] (-10.2,1.4) -- (-9.8,1.4);
\draw[gray, dashed] (-10.2,-2.8) -- (-9.8,-2.8);
\end{tikzpicture}
\end{center}
\captionsetup{justification=raggedright,singlelinecheck=false}
\caption{A diagrammatic sketch corresponding to the process of heavy ion ultra-peripheral collision. $X$ marks the impact position of double photons. The direction of A, B ions velocity follows the direction of z-axion. $x_A$ and $x_B$ denote the center positions corresponding to ions A and B, respectively.}
\label{pbpb-fig1}
\end{figure}

In relativistic heavy ion collisions, the velocity of the heavy ions is approximately equal to the speed of light. Consequently, the velocity of heavy ions can be decomposed along the light cone direction and perpendicular to the direction of light cone. The light cone directions corresponding to ions A and B are denoted as $n_A$ and $n_B$, respectively. These directions can be expressed within a general coordinate system as 
\begin{align}
    n^\mu_A=\bar{n}^\mu_B=(1,~0,~0,~1),~~~n^\mu_B=\bar{n}^{\mu}_A=(1,~0,~0,~-1).
\end{align}
Fig.\ref{pbpb-fig1} is a diagrammatic sketch corresponding to the process of heavy ion ultra-peripheral collision,
where $X$ denotes the impact position of the photons. The direction of A, B ions velocity follows the direction of z-axion. Meanwhile $x_A$ and $x_B$ denote the center positions corresponding to ions A and B, respectively. The relation can be expressed as:
$r_{A_\perp}=X_\perp-x_{A_\perp}$, $r_{B_\perp}=X_\perp-x_{B_\perp}$ and $b_\perp=r_{A_\perp}-r_{B_\perp}$.

The research presented herein is firmly grounded in two fundamental prerequisites: 
\begin{align}
&1) \left|\mathcal{P}_{iz}\right|\gg\left|\mathcal{P}_{i\perp}\right|,\Delta p_\perp,\Delta p_z ;\\ \notag
&2)\left|b_\perp\right| \gg \Delta x_\perp,\notag
\end{align}
where $i=A,B$; $\Delta p_i=p_i-p^{\prime}_i$ and $\Delta x_\perp=x^C_\perp-x^{I}_\perp$. In UPC, it is possible to completely fulfill two of these fundamental conditions.

We will now commence with the effective Lagrangian formalism, employing the wave packet expansion method, to establish the proof of the factorization of the UPC process. In the process, two approximately on-shell photons scatter into final-state leptons, followed by the emission of final-state soft photons from these leptons. The probability of photons within heavy ions A and B, undergoing a scattering process resulting in final-state leptons and other products, is denoted as 
\begin{align}
    P(b_\perp)=&\int d^4X^{\prime}\int d^4X\int d\Gamma_{l_1}d\Gamma_{l_2}\\ \notag
    &\times\prod_x[d\Gamma_x]\sum\limits_{\{p_x\}}
    \bra{\phi_A\phi_B}\hat{O}^{\dagger}(X^{\prime})\ket{\{p_x\};p_{l_1},p_{l_2}}
    \bra{p_{l_1},p_{l_2};\{p_x\}}\hat{O}(X)\ket{\phi_A\phi_B}.\notag
\end{align}
The ions are not shattered, but in the final state, their momentum changes. This change is reflected in the conservation of four momentum, and we will rewrite the momentum of the ions and other products corresponding to nonperturbative states in the final state as $\{p_x\}$. $p_{l_1}$ and $p_{l_2}$ correspond to the momenta of leptons $l_1$ and $l_2$, while $\Gamma_{l_1}$, $\Gamma_{l_2}$ and $\Gamma_{x}$ signify the phase spaces of $l_1$, $l_2$, and other products $x$. In the initial step, it is crucial to factorize the generalized transverse momentum distributions (GTMD) corresponding to scattering photons and the high-energy scattering cross-section. For the sake of simplicity, we refrain from explicitly specifying the momentum and phase space of final-state soft radiation, which will be addressed in subsequent steps. With the cross section formula 
\begin{align}
\sigma_b=\int d^2 b_\perp P(b_\perp),
\end{align}
we derive 
\begin{align}
\frac{d\sigma_b}{d^2b_\perp}=&\int d^4X^{\prime}\int d^4 X\int d\Gamma_{l_1} d\Gamma_{l_2} \prod\limits_x[d \Gamma_x]\\ \notag
&\times\sum\limits_{\{p_x\}}\bra{\phi_A\phi_B}\hat{O}^{\dagger}(X^{\prime})\ket{\{p_x\};p_{l1}\,p_{l2}}\bra{p_{l1}\,p_{l2};\{p_x\}}\hat{O}(X)\ket{\phi_A\phi_B}.\notag
\end{align}
The differential cross section pertains to the scattering event of particle A with particle B, both situated within the reaction plane, with a separation denoted as $b_\perp$. The dependency of the scattering cross-section on the impact parameter $b_\perp$ arises from the expansion of the wave packet of $\ket{\phi_A\phi_B}$. Photons surrounding ion $A$ and ion $B$ undergo scattering at the four-dimensional spacetime point $X$. Guided by the principle of uncertainty, it is imperative to integrate over the entire four dimensional space-time coordinates.

By expanding the initial state using wave packets corresponding to momentum eigenstates, we obtain 
\begin{align}
&\ket{\phi_A\phi_B}=\int\frac{d^3\vec{p}_A}{(2\pi)^3\sqrt{2E_{pA}}}e^{-ip_{\perp A}\cdot x_{\perp A}}\phi_A(\vec{p}_A)\int\frac{d^3\vec{p}_B}{(2\pi)^3\sqrt{2E_{pB}}}e^{-ip_{\perp B}\cdot x_{\perp B}}\phi_B(\vec{p}_B)\ket{\vec{p}_A\vec{p}_B}.
\end{align} 
Substituting $\ket{\phi_A\phi_B}$ for $\ket{\vec{p}_A\vec{p}_B}$ in the aforementioned differential cross section yields the result expressed as 
\begin{align}
    \frac{d\sigma_b}{d^2b_\perp}
    =&\int d^4X^{\prime}\int d^4X\int d\Gamma_{l_1}d\Gamma_{l_2}
\int\frac{d^3\vec{p}_A}{(2\pi)^3\sqrt{2E_{pA}}}\frac{d^3\vec{p}^{\,\prime}_A}{(2\pi)^3\sqrt{2E^{\prime}_{pA}}}\frac{d^3\vec{p}_B}{(2\pi)^3\sqrt{2E_{pB}}}\\ \notag
&\times\frac{d^3\vec{p}^{\,\prime}_B}{(2\pi)^3\sqrt{2E^{\prime}_{pB}}}
\phi_A(\vec{p}_A)\phi_B(\vec{p}_B)\phi^*_A(\vec{p}^{\,\prime}_A)\phi^*_B(\vec{p}_B^{\,\prime})
e^{-i(p_{\perp A}-p^{\prime}_{\perp A})\cdot x_{\perp A}}e^{-i(p_{\perp B}-p^{\prime}_{\perp B})\cdot x_{\perp B}}\\ \notag
&\times \mathop{\ooalign{$\sum$\cr$\displaystyle{\int}$\cr}}\limits_{\{p_x\}}
[\bra{p^{\prime}_Ap^{\prime}_B}\hat{O}^{\dagger}(X^{\prime})\ket{\{p_x\};p_{l_1}p_{l_2}}
\bra{p_{l_1}p_{l_2};\{p_x\}}\hat{O}(X)\ket{p_Ap_B}] \notag
\end{align}
By substituting the effective Lagrangian interaction term $\mathcal{A}^{\mu}_{\bar{n}}\mathcal{A}^{\nu}_n\bar{h}_{v_1}\Gamma_{\mu\nu}h_{v_2}(X)$ for $\hat{O}$, the S-matrix can be derived. Meanwhile, employing the method of factorization, the S-matrix can be decomposed into perturbative high-energy scattering part and non-perturbative matrix elements corresponding to collinear photon field,
\begin{align}
        \frac{d\sigma_b}{d^2b_\perp}=&\int d^4X^{\prime}\int d^4X\int d\Gamma_{l_1}d\Gamma_{l_2}
        \int\frac{d^3\vec{p}_A}{(2\pi)^3\sqrt{2E_{p_A}}}\frac{d^3\vec{p}^{\,\prime}_A}{(2\pi)^3\sqrt{2E^{\prime}_{p_A}}}\frac{d^3\vec{p}_B}{(2\pi)^3\sqrt{2E_{p_B}}}\\ \notag
        &\times \frac{d^3\vec{p}^{\,\prime}_B}{(2\pi)^3\sqrt{2E^{\prime}_{p_B}}}
        \phi_A(\vec{p}_A)\phi_B(\vec{p}_B)\phi^{*}_A(\vec{p}^{\,\prime}_A)\phi^{*}_B(\vec{p}^{\,\prime}_B)e^{-i(p_{A_\perp}-p^{\prime}_{A_\perp})\cdot x_{A_\perp}}\\ \notag
        &\times e^{-i(p_{B_\perp}-p^{\prime}_{B_\perp})\cdot x_{B_\perp}}\\ \notag
        &\times\bra{p_{l_1}p_{l_2}}T\big\{\Gamma^{\alpha\beta}_{\mu\nu}\bar{h}^{v_1}_{\alpha}h^{v_2}_{\beta}(X)\big\}
        \bar{T}\big\{\Gamma^{\beta^{\prime}\alpha^{\prime}}_{\mu^{\prime}\nu^{\prime}}\bar{h}^{v_2}_{\beta^{\prime}}h^{v_1}_{\alpha^{\prime}}(X^{\prime})\big\}
        \ket{p_{l_1}p_{l_2}}\\ \notag
        &\times \mathop{\ooalign{$\sum$\cr$\displaystyle{\int}$\cr}}\limits_{\{p_x\}}
        \bra{p^{\prime}_Ap^{\prime}_B}
        \mathcal{A}^{\dagger\nu^{\prime}}_B\mathcal{A}^{\dagger\mu^{\prime}}_A
        (X^\prime)\ket{\{p_x\}}
        \bra{\{p_x\}}\mathcal{A}^{\mu}_A\mathcal{A}^{\nu}_B(X)\ket{p_Ap_B}. \notag
\end{align}
In the subsequent stage, starting with the S-matrix, it is imperative to derive $\mathcal{M}$ and concurrently derive the delta function for four-dimensional momentum conservation. For the perturbative matrix element, within the interaction picture, the field operator can still be approximately expanded as a creation operator and an annihilation operator, followed by contraction with the external state,
    \begin{align}
        &\bra{p_{l_1}p_{l_2}}T\big\{\Gamma^{\alpha\beta}_{\mu\nu}
        \bar{h}^{v_1}_{\alpha}h^{v_2}_{\beta}(X)\big\}
        \bar{T}\big\{\Gamma^{\beta^{\prime}\alpha^{\prime}}_{\mu^{\prime}\nu^{\prime}}
        \bar{h}^{v_2}_{\beta^{\prime}}h^{v_1}_{\alpha^{\prime}}(X^{\prime})\big\}
        \ket{p_{l_1}p_{l_2}}\\ \notag
        &=\sum\limits_{spin}\big[\bar{u}(v_1)\Gamma_{\mu\nu}\nu(v_2)e^{i(p_{l_1}+p_{l_2})\cdot X}\big]\times
        \big[\bar{\nu}(v_2)\Gamma^{\dagger}_{\nu^{\prime}\mu^{\prime}}u(v_1)e^{-i(p_{l_1}+p_{l_2})\cdot X^{\prime}}\big]. \notag
    \end{align}

In the context of perturbative field theory, when integrating over the four-dimensional spacetime coordinates $X$ and $X^{\prime}$, the Fourier factor induces the delta function which implies conservation of initial and final state momenta, transforming S-matrix into the product of $\mathcal{M}$ and the delta function. However, in this scenario, non-perturbative matrix elements also participate, and their dependence on $X$ and $X^{\prime}$ must be taken into account. To enable integrate over $X$ and $X^{\prime}$ and derive the conditions for conservation of initial and final state momenta, a translation operator should be employed to eliminate the dependence of $\mathcal{A}^{\mu}_A\mathcal{A}^{\nu}_B$ and $\mathcal{A}^{\dagger\nu^{\prime}}_B\mathcal{A}^{\dagger\mu^{\prime}}_A$ on $X$ and $X^{\prime}$. By translating the spacetime coordinates of the non-perturbative matrix elements and simultaneously introducing the lepton tensor into the differential scattering cross section, we obtain
    \begin{align}
        \frac{d\sigma_b}{d^2b_\perp}=&
        \int d^4X^{\prime}d^4X\int d\Gamma_{l_1}d\Gamma_{l_2}\int\frac{d^3\vec{p}_A}{(2\pi)^3\sqrt{2E_{p_A}}}\frac{d^3\vec{p}^{\,\prime}_A}{(2\pi)^3\sqrt{2E^{\prime}_{p_A}}}\frac{d^3\vec{p}_B}{(2\pi)^3\sqrt{2E_{p_B}}}\frac{d^3\vec{p}^{\,\prime}_B}{(2\pi)^3\sqrt{2E^{\prime}_{p_B}}}\\ \notag
        &\times \phi_A(\vec{p}_A)\phi_B(\vec{p}_B)\phi^{*}_A(\vec{p}^{\,\prime}_A)\phi^{*}_B(\vec{p}^{\,\prime}_B)e^{-i(p_{A_\perp}-p^{\prime}_{A_\perp})\cdot x_{A_\perp}}
        e^{-i(p_{B_\perp}-p^{\prime}_{B_\perp})\cdot x_{B_\perp}}\\ \notag
        &\times\sum\limits_{spin}
        \bar{u}(v_1)\Gamma_{\mu\nu}\nu(v_2)\bar{\nu}(v_2)\Gamma^{\dagger}_{\nu^{\prime}\mu^{\prime}}u(v_1)\\ \notag
&\times\mathop{\ooalign{$\sum$\cr$\displaystyle{\int}$\cr}}\limits_{\{p_x\}}
\bra{p^{\prime}_Ap^{\prime}_B}
\mathcal{A}^{\dagger\nu^{\prime}}_B\mathcal{A}^{\dagger\mu^{\prime}}_A(0)
\ket{\{p_x\}}\bra{\{p_x\}}\mathcal{A}^{\mu}_A\mathcal{A}^{\nu}_B(0)\ket{p_Ap_B}\\ \notag
&\times e^{i(p_{l_1}+p_{l_2})\cdot X}e^{i\big(\{p_x\}-p_A-p_B\big)\cdot X}
\times e^{-i(p_{l_1}+p_{l_2})\cdot X^{\prime}}
e^{i\big((p^{\prime}_A+p^{\prime}_B)-\{p_x\}\big)\cdot X^{\prime}}\\ \notag
=&\int d\Gamma_{l_1}d\Gamma_{l_2}\int\frac{d^3\vec{p}_A}{(2\pi)^3\sqrt{2E_{p_A}}}
\frac{d^3\vec{p}^{\,\prime}_A}{(2\pi)^3\sqrt{2E^{\prime}_{p_A}}}
\frac{d^3\vec{p}_B}{(2\pi)^3\sqrt{2E_{p_B}}}\frac{d^3\vec{p}^{\,\prime}_B}{(2\pi)^3\sqrt{2E^{\prime}_{p_B}}}\\ \notag
&\times \phi_A(\vec{p}_A)\phi_B(\vec{p}_B)\phi^{*}_A(\vec{p}^{\,\prime}_A)\phi^{*}_B(\vec{p}^{\,\prime}_B)
e^{-i(p_{A_\perp}-p^{\prime}_{A_\perp})\cdot x_{A_\perp}}
e^{-i(p_{B_\perp}-p^{\prime}_{B_\perp})\cdot x_{B_\perp}}\\ \notag
&\times \sum\limits_{spin}\bar{u}(v_1)
\Gamma_{\mu\nu}\nu(v_2)\bar{\nu}(v_2)\Gamma^{\dagger}_{\nu^{\prime}\mu^{\prime}}
u(v_1)\\ \notag
&\times\mathop{\ooalign{$\sum$\cr$\displaystyle{\int}$\cr}}\limits_{\{p_x\}}
\bra{p^{\prime}_Ap^{\prime}_B}
\mathcal{A}^{\dagger\nu^{\prime}}_B\mathcal{A}^{\dagger\mu^{\prime}}_A(0)\ket{\{p_x\}}
\bra{\{p_x\}}\mathcal{A}^{\mu}_A\mathcal{A}^{\nu}_B(0)\ket{p_Ap_B}\\ \notag
&\times (2\pi)^4\delta^{(4)}\big[p_{l_1}+p_{l_2}+\{p_x\}-p_A-p_B\big]
(2\pi)^4\delta^{(4)}\big[(p^{\prime}_A+p^{\prime}_B)-(p_{l_1}+p_{l_2}+\{p_x\})\big]. \notag
    \end{align}
To facilitate our exposition, we employ concise shorthand notation
\begin{align}
&\hat{M}^{\mu\nu}(0)=\mathcal{A}_A^{\mu}\mathcal{A}_B^{\nu}(0),~~~~~~~~~~~~~~~~~~~~\hat{M}^{\dagger\nu^{\prime}\mu^{\prime}}(0)=\mathcal{A}_B^{\dagger\nu^{\prime}}\mathcal{A}_A^{\dagger\mu^{\prime}}(0),\\ \notag
&L_{\mu\nu}L^{\dagger}_{\mu^{\prime}\nu^{\prime}}\equiv\sum_{spin}
\bar{\nu}(v_2)\Gamma^{\dagger}_{\nu^{\prime}\mu^{\prime}}u(v_1)\bar{u}(v_1)\Gamma_{\mu\nu}\nu(v_2). \notag
\end{align}
In the next step, we apply a Fourier transformation to one of the Delta functions. Next, we shift the spacetime coordinates of the $\hat{M}$ operator from $0$ to $-\frac{x}{2}$ and also reposition the coordinates of the conjugate operator from $0$ to $+\frac{x}{2}$ using a translation operator. With the operation, the field have regained the dependence on space-time coordinates.   
\begin{align}
&\bra{\{p_x\}}\hat{M}(0)\ket{p_Ap_B}(2\pi)^4\delta^{(4)}(p_A+p_B-p_{l_1}-p_{l_2}-\{p_x\})\\ \notag
&=\int d^4xe^{i(p_A+p_B-p_{l_1}-p_{l_2}-\{p_x\})\cdot x}
\bra{\{p_x\}}e^{i\hat{P}\cdot\frac{x}{2}}\hat{M}(-\frac{x}{2})e^{-i\hat{P}\cdot\frac{x}{2}}\ket{p_Ap_B}\\ \notag
&=\int d^4xe^{-i(p_{l_1}+p_{l_2})\cdot x}\times e^{i(p_A+p_B-\{p_x\})\cdot \frac{x}{2}}\bra{\{p_x\}}\hat{M}(-\frac{x}{2})\ket{p_Ap_B}. \notag
\end{align}
Using the same operation, I shift the conjugate operator by a distance of $+\frac{x}{2}$:
\begin{align}
\bra{p^{\prime}_Ap^{\prime}_B}\hat{M}^{\dagger}(0)\ket{\{p_x\}}
=e^{-i(p^{\prime}_A+p^{\prime}_B)\cdot \frac{x}{2}+i\{p_x\}\cdot \frac{x}{2}}\bra{p^{\prime}_Ap^{\prime}_B}\hat{M}^{\dagger}(\frac{x}{2})\ket{\{p_x\}}.
\end{align}
With the obtained results, the cross section can be represented as 
\begin{align}
\frac{d\sigma_b}{d^2b_\perp}=&
\int d\Gamma_{l_1}d\Gamma_{l_2}L^{\dagger}_{\mu^{\prime}\nu^{\prime}}
L_{\mu\nu}
\int d^4xe^{-i(p_{l_1}+p_{l_2})\cdot x}\\ \notag
&\times\int\frac{d^3\vec{p}_A}{(2\pi)^3\sqrt{2E_{pA}}}\frac{d^3\vec{p}^{\,\prime}_A}{(2\pi)^3\sqrt{2E^{\prime}_{pA}}}\frac{d^3\vec{p}_B}{(2\pi)^3\sqrt{2E_{pB}}}
\frac{d^3\vec{p}^{\,\prime}_B}{(2\pi)^3\sqrt{2E^{\prime}_{pB}}}
\phi_A(\vec{p}_A)\phi_B(\vec{p}_B)\\ \notag
&\times\phi^*_A(\vec{p}^{\,\prime}_A)\phi^*_B(\vec{p}_B^{\,\prime})
e^{-i(p_{\perp A}-p{\prime}_{\perp A})\cdot x_{\perp A}}e^{-i(p_{\perp B}-p^{\prime}_{\perp B})\cdot x_{\perp B}}\\ \notag
&\times\mathop{\ooalign{$\sum$\cr$\displaystyle{\int}$\cr}}\limits_{\{p_x\}}\bra{p^{\prime}_Ap^{\prime}_B}\mathcal{A}_B^{\dagger\nu^{\prime}}\mathcal{A}^{\dagger\mu^{\prime}}_A(\frac{x}{2})\ket{\{p_x\}}
\bra{\{p_x\}}\mathcal{A}_A^{\mu}\mathcal{A}_B^{\nu}(-\frac{x}{2})\ket{p_Ap_B}\\ \notag
&\times(2\pi)^4\delta^{(4)}(p_{l_1}+p_{l_2}+\{p_x\}-p_A^{\prime}-p_B^{\prime}).
\end{align}
By representing the phase space integration of final-state leptons in terms of rapidity and momentum $p_\perp$
\begin{align}
\int d\Gamma_l&=\int\frac{d^4p}{(2\pi)^4}(2\pi)\delta(p^2-m^2)\Theta(p^0)=\int\frac{dyd^2p_\perp}{2(2\pi)^3},
\end{align} 
the differential scattering cross section can be expressed as 
\begin{align}
&\frac{d\sigma_b}{dy_{l_1}d^2p_{l_1\perp}dy_{l_2}d^2p_{l_2\perp}d^2b_\perp}\\
&=
\begin{aligned}[t]
&\frac{1}{[2(2\pi)^3]^2} L^{\dagger}_{\mu^{\prime}\nu^{\prime}}L_{\mu\nu}\int d^4x e^{-i(p_{l_1}+p_{l_2})\cdot x}
\int\frac{d^3\vec{p}_A}{(2\pi)^3\sqrt{2E_{pA}}}\frac{d^3\vec{p}^{\,\prime}_A}{(2\pi)^3\sqrt{2E^{\prime}_{pA}}}\frac{d^3\vec{p}_B}{(2\pi)^3\sqrt{2E_{pB}}}\\ \notag
&\times\frac{d^3\vec{p}^{\,\prime}_B}{(2\pi)^3\sqrt{2E^{\prime}_{pB}}}
\phi_A(\vec{p}_A)\phi_B(\vec{p}_B)\phi^*_A(\vec{p}^{\,\prime}_A)\phi^*_B(\vec{p}_B^{\,\prime})
e^{-i(p_{\perp A}-p{\prime}_{\perp A})\cdot x_{\perp A}}e^{-i(p_{\perp B}-p^{\prime}_{\perp B})\cdot x_{\perp B}}\\ \notag
&\times\mathop{\ooalign{$\sum$\cr$\displaystyle{\int}$\cr}}\limits_{\{p_x\}}(2\pi)^4\delta^{(4)}(p_{l_1}+p_{l_2}+\{p_x\}-p^{\prime}_A-p^{\prime}_B)
[\bra{p^{\prime}_Ap^{\prime}_B}\hat{M}^{\dagger}(\frac{x}{2})\ket{\{p_x\}}\bra{\{p_x\}}\hat{M}(-\frac{x}{2})\ket{p_Ap_B}]. \notag
\end{aligned}
\end{align}
The integration with respect to variable $\vec{p}^{\,\prime}$ can be achieved using another $\delta$ function. Integrating over variables $p^{\prime}_{A_z}$ and $p^{\prime}_{B_z}$ is accomplished by the action of the longitudinal component of the delta function, leading to the subsequent emergence of a factor $\frac{1}{|v^{\prime}_{A_z}-v^{\prime}_{B_z}|}$. Subsequently, we employ the transverse component of the delta function to carry out the integration $\int\frac{d^2p^{\prime}_{B_\perp}}{(2\pi)^2}$:
\begin{align}
&\int\frac{d^3\vec{p}_A}{(2\pi)^3}\frac{d^3\vec{p}^{\,\prime}_A}{(2\pi)^3}\frac{d^3\vec{p}_B}{(2\pi)^3}\frac{d^3\vec{p}^{\,\prime}_B}{(2\pi)^3}
(2\pi)^4\delta^{(4)}(\{p_x\}+p_{l_1}+p_{l_2}-p_A^{\prime}-p_B^{\prime})\\ \notag
&=\frac{1}{\left|v^{\prime}_{Az}-v^{\prime}_{Bz}\right|}\int\frac{dp_{Az}}{2\pi}\int\frac{dp_{Bz}}{2\pi}\int\frac{d^2p_{A\perp}}{(2\pi)^2}\frac{d^2p_{B\perp}}{(2\pi)^2}\frac{d^2p^{\prime}_{A\perp}}{(2\pi)^2}. \notag
\end{align}
Employing the Jacobi determinant, we convert the integration from variable $p_\perp$ and $p^{\prime}_\perp$ to variable $\mathcal{P}_\perp$ and $q_\perp$
\begin{align}
\int\frac{d^2p_{A\perp}}{(2\pi)^2}\frac{d^2p_{B\perp}}{(2\pi)^2}\frac{d^2p^{\prime}_{A\perp}}{(2\pi)^2}
\longrightarrow 
\int\frac{d^2\mathcal{P}_{A\perp}}{(2\pi)^2}\frac{d^2q_{A\perp}}{(2\pi)^2}\frac{d^2\mathcal{P}_{B\perp}}{(2\pi)^2}.
\end{align}
To determine the dependence on the impact parameter $b_\perp$ we must perform an integration with respect to $q_{B_\perp}$. For this reason, we impose the constraint $q_{A_\perp}=q_{B_\perp}=q_\perp$. By using a delta function
\begin{align}
\int\frac{d^2q_{B_\perp}}{(2\pi)^2}(2\pi)^2\delta^{(2)}(q_{A_\perp}-q_{B_\perp})=1,
\end{align}
the differential cross section can be transformed into 
\begin{align}
&\frac{d\sigma_b}{dy_{l_1}d^2p_{\perp l_1}dy_{l_2}d^2p_{\perp l_2}d^2b_\perp}\\ \notag
=&\frac{1}{[2(2\pi)^3]^2}\int d^4x e^{-i(p_{l_1}+p_{l_2})\cdot x}L^{\dagger}_{\mu^{\prime}\nu^{\prime}}L_{\mu\nu}
\frac{1}{\sqrt{2E_{pA}}\sqrt{2E^{\prime}_{pA}}\sqrt{2E_{pB}}\sqrt{2E^{\prime}_{pB}}}\frac{1}{\left|v^{\prime}_{Az}-v^{\prime}_{Bz}\right|}\\ \notag
&\times\int\frac{d^2\mathcal{P}_{A\perp}}{(2\pi)^2}\frac{d^2\mathcal{P}_{B\perp}}{(2\pi)^2}\frac{d^2q_{A\perp}}{(2\pi)^2}\frac{d^2q_{B\perp}}{(2\pi)^2}
e^{-iq_{A\perp}\cdot x_{\perp A}+iq_{B\perp}\cdot x_{B\perp}}
(2\pi)^2\delta^{(2)}(q_{A\perp}-q_{B\perp})\\ \notag
&\times\int\frac{dp_{Az}}{2\pi}\frac{dp_{Bz}}{2\pi}\phi_A(\vec{p}_A)\phi^*_A(\vec{p}^{\,\prime}_A)\phi_B(\vec{p}_B)\phi^*_B(\vec{p}^{\,\prime}_B)
\mathop{\ooalign{$\sum$\cr$\displaystyle{\int}$\cr}}\limits_{\{p_x\}}[\bra{p^{\prime}_Ap^{\prime}_B}\hat{M}^{\dagger}(\frac{x}{2})\ket{\{p_x\}}\bra{\{p_x\}}\hat{M}(-\frac{x}{2})\ket{p_Ap_B}]. \notag
\end{align}
In experiments, the particle wave packet can not be directly measured. Since the differential cross section is an experimentally observable, it should not directly depend on the wave packets. For this reason, we can define the Wigner function and integrate the wave packets out \cite{wu2021factorization}. 

The Wigner function depends not only on momentum but also on spatial coordinates. Therefore, to define the transverse Wigner function, one must integrate out the longitudinal momentum of the wave packets and subsequently perform a Fourier transformation to convert the transverse momentum of the wave packets into transverse spatial coordinates,
\begin{align}
&\int\frac{dp_{z}}{2\pi}\phi(\vec{p})\phi^*(\vec{p}^{\,\prime})\\ \notag
&=\int\frac{dp_{z}}{2\pi}[\phi(p_{\perp},p_{z})\phi^*(p^{\prime}_{\perp},p_{z})]\\ \notag
&=\int\frac{dp_{z}}{2\pi}[\int d^2t_{\perp} e^{ip_{\perp}\cdot t_{\perp}}\int dz e^{ip_{z}\cdot z} \widetilde{\phi}(t_{\perp},z)
\int d^2t^{\prime}_{\perp}e^{-ip^{\prime}_{\perp}\cdot t^{\prime}_{\perp}}\int dz^{\prime} e^{-ip_{z}\cdot z^{\prime}}\widetilde{\phi}^*(t^{\prime}_{\perp},z^{\prime})]. \notag
\end{align}
Let's redefine variables $\chi_\perp$ and $X_\perp$ such that 
\begin{align}
t_\perp=X_\perp-\frac{\chi_\perp}{2},~~~~~
t_\perp^{\prime}=X_\perp+\frac{\chi_\perp}{2}.
\end{align}
Through variable replacement, we obtain
\begin{align}
&\int\frac{dp_{z}}{2\pi}[\int d^2X_{\perp}d^2\chi_{\perp}e^{ip_{\perp}\cdot (X_{\perp}-\frac{\chi_{\perp}}{2})}e^{-ip^{\prime}_{\perp}\cdot(X_{\perp}+\frac{\chi_{\perp}}{2})}
\int dzdz^{\prime}e^{-i(z^{\prime}-z)\cdot p_{z}}\widetilde{\phi}(X_\perp-\frac{\chi_\perp}{2},z)\widetilde{\phi}^*(X_\perp+\frac{\chi_\perp}{2},z^{\prime})] \notag \\
=&\int d^2X_{\perp}e^{iq_{\perp}\cdot X_{\perp}}\int d^2\chi_{\perp}e^{-i\mathcal{P}_{\perp}\cdot \chi_{\perp}}\int dz \widetilde{\phi}(X_{\perp}-\frac{\chi_{\perp}}{2},z)\widetilde{\phi}^*(X_{\perp}+\frac{\chi_{\perp}}{2},z)\notag\\
=&\int d^2 X_{\perp}e^{iq_{\perp}\cdot X_{\perp}}W(X_{\perp},\mathcal{P}_{\perp}).
\end{align}
By first integrating with respect to $p_z$, we can obtain a delta function $\delta(z^\prime-z)$. Utilizing this delta function, the integration with respect to $z^{\prime}$ can be performed. Subsequently, the result of the first equal sign can be obtained. $W(X_\perp,\mathcal{P}_\perp)$ represents the Wigner function, which depends on both spatial coordinates and momentum. Its detailed definition is provided as 
\begin{align}
&W(X_{\perp},\mathcal{P}_{\perp})=\int d^2\chi_{\perp}e^{-i\mathcal{P}_{\perp}\cdot\chi_{\perp}}\int dz\widetilde{\phi}(X_{\perp}-\frac{\chi_{\perp}}{2},z)\widetilde{\phi}^*(X_{\perp}+\frac{\chi_{\perp}}{2},z).
\end{align}
Integrating with respect to the transverse momentum $\mathcal{P}_\perp$ for the Wigner function results in a delta function $\delta^{(2)}(\chi_\perp)$. Utilizing this delta function, the uncertainty $\chi_\perp$ in transverse spatial coordinates will be constrained to zero. 

By defining the transverse Wigner function, the wave packets corresponding to particles A and B can be represented as 
\begin{align}
&\int\frac{dp_{Az}}{2\pi}\phi_A(\vec{p}_A)\phi^*_A(\vec{p}^{\,\prime}_A)
=\int d^2X_{A\perp}e^{iq_{A\perp}\cdot X_{A\perp}}W(X_{A\perp},\mathcal{P}_{A\perp}),\\ 
&\int\frac{dp_{B_z}}{2\pi}\phi_B(\vec{p}_B)\phi^*_B(\vec{p}^{\prime}_B)=\int d^2X_{B_\perp}
e^{-iq_{B_\perp}\cdot X_{B_\perp}}W(X_{B_\perp},\mathcal{P}_{B_\perp}).
\end{align}
The aforementioned differential cross section can be represented as 
\begin{align}
&\frac{d\sigma_b}{dy_{l_1}d^2p_{l_1\perp}dy_{l_2}d^2p_{l_2\perp}d^2b_\perp}\\ \notag
=
&\int d^2X_{A\perp}\int\frac{d^2\mathcal{P}_{A\perp}}{(2\pi)^2}W(X_{A\perp},\mathcal{P}_{A\perp})\int d^2X_{B\perp}\int\frac{d^2\mathcal{P}_{B\perp}}{(2\pi)^2}W(X_{B\perp},\mathcal{P}_{B\perp})\\ \notag
&\times\int\frac{d^2q_{A\perp}}{(2\pi)^2}\frac{d^2q_{B\perp}}{(2\pi)^2}e^{-iq_{A\perp}\cdot x_{A\perp}+iq_{B\perp}\cdot x_{B\perp}}
 e^{iq_{A_\perp}\cdot X_{A_\perp}-iq_{B_\perp}\cdot X_{B_\perp}}(2\pi)^2\delta^{(2)}(q_{A\perp}-q_{B\perp})\\ \notag
&\times\frac{1}{[2(2\pi)^3]^2}\frac{1}{2s}\int d^4x e^{-i(p_{l_1}+p_{l_2})\cdot x}L^{\dagger}_{\mu^{\prime}\nu^{\prime}}L_{\mu\nu}
\mathop{\ooalign{$\sum$\cr$\displaystyle{\int}$\cr}}\limits_{\{p_x\}}[\bra{p^{\prime}_Ap^{\prime}_B}\hat{M}^{\dagger}(\frac{x}{2})\ket{\{p_x\}}\bra{\{p_x\}}\hat{M}(-\frac{x}{2})\ket{p_Ap_B}].
\end{align}
We must now proceed to integrate the Wigner function out. In this context, we adopt one of the two approximate conditions outlined at the beginning of our passage. Initially, we define the transverse distance in the reaction plane as the distance from the heavy ion's center to the point where two photons scattering $r_{i_\perp}\equiv X_\perp-x_{i_\perp}$, $i=A,B$. Additionally, we establish the equality $r_{A_\perp}-r_{B_\perp}=b_\perp$, where $b_\perp$ denotes the impact parameter. $X_{i_\perp}$, where $i = A, B$, represents the average value of the position coordinates at which photons around heavy ion $i$ undergo scattering. This average refers to the mean of the position coordinates corresponding to the amplitude and its conjugate amplitude. $X_{A_\perp}-X_{B_\perp}=\Delta x_{\perp}$ characterizes the uncertainty associated with the transverse coordinates of two scattering photons. Under the initial approximation condition denoted as $|b_\perp|\gg \Delta x_\perp$, the expression for the integration of the Wigner function can be formulated as 
\begin{align}
    &\int d^2X_{A\perp}\int\frac{d^2\mathcal{P}_{A\perp}}{(2\pi)^2}\,W(X_{A\perp},\mathcal{P}_{A\perp})\int d^2X_{B\perp}\int\frac{d^2\mathcal{P}_{B\perp}}{(2\pi)^2}\,W(X_{B\perp},\mathcal{P}_{B\perp})e^{iq_{\perp}(X_{A\perp}-X_{B\perp})}\\ \notag
    \simeq&\int d^2X_{A\perp}\int\frac{d^2\mathcal{P}_{A\perp}}{(2\pi)^2}W(X_{A\perp},\mathcal{P}_{A\perp})\int d^2 X_{B\perp}\int\frac{d^2\mathcal{P}_{B\perp}}{(2\pi)^2}W(X_{B\perp},\mathcal{P}_{B\perp})\\
    &\int d^2X_{A\perp}\int\frac{d^2\mathcal{P}_{A\perp}}{(2\pi)^2} \,W(X_{A\perp},\mathcal{P}_{A\perp})\\ \notag
    =&\int d^2 X_{A\perp}\int\frac{d^2\mathcal{P}_{A\perp}}{(2\pi)^2}\int d^2\chi_{A\perp}\,e^{-i\mathcal{P}_{A\perp}\cdot\chi_{A\perp}}\int dz\,\tilde{\phi}_A(X_{A\perp}-\frac{\chi_{A\perp}}{2},z)\tilde{\phi}^*_A(X_{A\perp}+\frac{\chi_{A\perp}}{2},z)\\ \notag
    =&\int d^2X_{A\perp}\int dz\,\tilde{\phi}_A(X_{A\perp},z)\tilde{\phi}_A^*(X_{A\perp},z)\\ \notag
    =&1.
\end{align}       
Between the first and second equal signs in second function, our primary procedure entails the integration with respect to $d^2\mathcal{P}_{A_\perp}$, which results in the subsequent generation of $\delta^{(2)}(\chi_{A_\perp})$. Sequentially, the integration with respect to $d^2\chi_{A_\perp}$ is executed. The ultimate outcome 1 arises from the integration over total spatial space pertaining to the probability density. We must point out that more rigorous mathematical calculations entail performing the Wigner function integration after completing the power counting of the collinear matrix elements, as indicated in equation \eqref{equation(19)}. After performing power counting on the spacetime points of the photon field $\mathcal{A}$ in the collinear matrix elements, the transverse momenta $\mathcal{P}_{A\perp}$ and $\mathcal{P}_{B\perp}$ in the corresponding external states will also be neglected. Consequently, the external state transforms like
\begin{align*}
    \ket{p_Ap_B}\rightarrow \ket{\bar{n}_A\cdot \mathcal{P}_A,\frac{q_{A\perp}}{2}},~~\ket{\bar{n}_B\cdot\mathcal{P}_B,-\frac{q_{B\perp}}{2}}.
\end{align*}
At this point, the external states of the collinear matrix elements no longer depend on $\mathcal{P}_{A\perp}$ and $\mathcal{P}_{B\perp}$, allowing for the integration over $d^2\mathcal{P}_{A\perp}$ and $d^2\mathcal{P}_{B\perp}$ with confidence. Consequently, the Wigner function can be integrated out from the differential scattering cross-section. Integrating out the Wigner function at this stage elucidates the physical significance of subsequent calculations, ensuring that the reader remains focused on the underlying concepts rather than becoming entangled in complex mathematical expressions.

Using this method, the Wigner function is integrated out, leading to the expression of the differential cross section in terms of 
\begin{align}
&\frac{d\sigma_b}{dy_{l_1}d^2p_{l_1\perp}dy_{l_2}d^2p_{l_2\perp}d^2b_\perp}\\ \notag
=
&\int\frac{d^2q_{A\perp}}{(2\pi)^2}\frac{d^2q_{B\perp}}{(2\pi)^2}e^{-iq_{A\perp}\cdot x_{A\perp}+iq_{B\perp}\cdot x_{B\perp}}(2\pi)^2\delta^{(2)}(q_{A\perp}-q_{B\perp})\\ \notag
&\times \frac{1}{[2(2\pi)^3]^2}\frac{1}{2s}\int d^4xe^{-i(p_{l_1}+p_{l_2})\cdot x}L^{\dagger}_{\mu^{\prime}\nu^{\prime}}L_{\mu\nu}
\mathop{\ooalign{$\sum$\cr$\displaystyle{\int}$\cr}}\limits_{\{p_x\}}[\bra{p^{\prime}_Ap^{\prime}_B}\hat{M}^{\dagger}(\frac{x}{2})\ket{\{p_x\}}
\bra{\{p_x\}}\hat{M}(-\frac{x}{2})\ket{p_Ap_B}].\notag
\end{align}
In the subsequent step, it is essential to factorize the $\mathcal{M}$ matrix element employing the SCET method
\begin{align}
&L^{\dagger}_{\mu^{\prime}\nu^{\prime}}L_{\mu\nu}\mathop{\ooalign{$\sum$\cr$\displaystyle{\int}$\cr}}\limits_{\{p_x\}}[\bra{p^{\prime}_Ap^{\prime}_B}\hat{M}^{\dagger}(\frac{x}{2})\ket{\{p_x\}}
\bra{\{p_x\}}\hat{M}(-\frac{x}{2})\ket{p_Ap_B}]\\ \notag
=&L^{\dagger}_{\mu^{\prime}\nu^{\prime}}L_{\mu\nu}\mathop{\ooalign{$\sum$\cr$\displaystyle{\int}$\cr}}\limits_{\{p_x\}}
\bra{p^{\prime}_Ap^{\prime}_B}\mathcal{A}^{\dagger\nu^{\prime}}_{B}\mathcal{A}^{\dagger\mu^{\prime}}_A(\frac{x}{2})\ket{\{p_x\}}\bra{\{p_x\}}\mathcal{A}^{\mu}_A\mathcal{A}^{\nu}_B(-\frac{x}{2})\ket{p_Ap_B}\\ \notag
=&L^{\dagger}_{\mu^{\prime}\nu^{\prime}}L_{\mu\nu}\bra{p^{\prime}_Ap^{\prime}_B}\mathcal{A}^{\dagger\nu^{\prime}}_B\mathcal{A}^{\dagger\mu^{\prime}}_A(\frac{x}{2})\mathcal{A}^{\mu}_A\mathcal{A}^{\nu}_B(-\frac{x}{2})\ket{p_Ap_B}.
\end{align}
Employing the equivalent photon approximation, the excited electromagnetic field surrounding heavy ions can be treated as the photon fields, denoted as $\mathcal{A}^{\mu}_A$ and $\mathcal{A}^{\nu}_B$ in the matrix elements.
Due to the heavy ion's non-negligible mass and its inability to move at the speed of light, defining the spacetime coordinates of the photon fields on light cone, which are distributed around the heavy ion and move in tandem with it, is not feasible. 
Within the framework of SCET, the coordinates of the photon field can be expanded along the direction of the anti-light cone and perpendicular to the light cone using the method of power counting. Meanwhile, it is possible to separate coordinates of $\mathcal{A}^{\mu}_A$ and $\mathcal{A}^{\nu}_B$, both of which are defined at the same position on the light cone, by a distance in the anti-collinear direction,
\begin{align}
&L^{\dagger}_{\mu^{\prime}\nu^{\prime}}L_{\mu\nu}\bra{p^{\prime}_Ap^{\prime}_B}\mathcal{A}^{\dagger\nu^\prime}_B\mathcal{A}^{\dagger\mu^{\prime}}_A(\frac{x}{2})\mathcal{A}^{\mu}_A\mathcal{A}^{\nu}_B(-\frac{x}{2})\ket{p_Ap_B}\\ \notag
=&L^{\dagger}_{\mu^{\prime}\nu^{\prime}}L_{\mu\nu}\bra{p^{\prime}_Ap^{\prime}_B}\int dt^{\prime}_Adt^{\prime}_BC^*(t^{\prime}_A,t^{\prime}_B)\int dt_Adt_BC(t_A,t_B)\\ \notag
&\times\mathcal{A}^{\dagger\nu^{\prime}}_B(\frac{x}{2}+\frac{t^{\prime}_B}{2}\bar{n}_B)\mathcal{A}^{\dagger\mu^{\prime}}_A(\frac{x}{2}+\frac{t^{\prime}_A}{2}\bar{n}_A)\mathcal{A}^{\mu}_A(-\frac{x}{2}-\frac{t_A}{2}\bar{n}_A)
\mathcal{A}^{\nu}_B(-\frac{x}{2}-\frac{t_B}{2}\bar{n}_B)]\ket{p_Ap_B}.
\end{align}
The divergence that occurs when we pull apart operators is absorbed in the Wilson coefficient, $C(t_A,t_B)$ and $C^*(t^{\prime}_A,t^{\prime}_B)$. Here, $t_A$ represents how far we pull operator $\mathcal{A}^{\mu}_A$ along the anti-collinear direction $\bar{n}_A$, and $t_B$ represents the pulling distance of operator $\mathcal{A}^{\nu}_B$ along the anti-collinear direction $\bar{n}_B$.

The next step involves performing power counting. In UPC, a relation exists where $|k_\perp|\approx\frac{1}{\gamma}|k_{\parallel}|$, with $k_\perp$ representing the transverse momentum of the initial photon and $\gamma=\frac{1}{\sqrt{1-\frac{v^2}{c^2}}}$ denoting the Lorentz contraction factor. Subsequently, $\lambda=\frac{1}{\gamma}$ leads to $k_\perp\sim\mathcal{O}(\lambda)$. Based on the aforementioned rationale, we obtain the relation, where $l$ denotes the total momentum of final-state soft photons,
 \begin{align}
 k_{1\perp}+k_{2\perp}=p_{l_{1\perp}}+p_{l_{2\perp}}+l_{\perp}\sim\mathcal{O}(\lambda).
 \end{align}
Based on the oscillatory nature of the Fourier factors, it can be known that when $x\sim\mathcal{O}(1,~1,~\lambda^{-1})$, the phase factor $e^{-i(p_{l_1}+p_{l_2}+l)\cdot x}$ contributes maximally. As a consequence, for the initial-state photon 1 associated with ion $A$, the maximum momentum component is observed along the $n_A$ direction,
\begin{align}
    &k_1\sim \mathcal{O}(\lambda^2,~1,~\lambda),\\ \notag
    &k^-_1\cdot x^+\sim\mathcal{O}(1),~~~k^+_1\cdot x^-\sim\mathcal{O}(\lambda^2),~~~
    k^{\perp}_1\cdot x^{\perp}\sim\mathcal{O}(1).
\end{align}
After employing power counting, the coordinates of $\mathcal{A}_A$ translates to $\mathcal{A}_A(-\frac{n_A\cdot x+t_A}{2}\bar{n}_A,-\frac{x_\perp}{2},0)$. For the same reason, in the case of the initial photon 2, which surrounds ion $B$ and has the maximum momentum component along the direction of $n_B$, the coordinates of the field are translated to $\mathcal{A}_B(-\frac{n_B\cdot x+t_B}{2}\bar{n}_B,-\frac{x_\perp}{2},0)$. Given the aforementioned translation, the matrix element can be represented as 
\begin{align}
\label{equation(19)}
&L^{\dagger}_{\mu^{\prime}\nu^{\prime}}L_{\mu\nu}
\bra{p^{\prime}_Ap^{\prime}_B}\int dt^{\prime}_Adt^{\prime}_BC^*(t^{\prime}_A,t^{\prime}_B)\int dt_Adt_BC(t_A,t_B)\mathcal{A}^{\dagger\nu^{\prime}}_B(\frac{x}{2}+\frac{t^\prime_B}{2}\bar{n}_B)
\mathcal{A}^{\dagger\mu^\prime}_A(\frac{x}{2}+\frac{t^{\prime}_A}{2}\bar{n}_A)\notag\\ \notag
&\times\mathcal{A}^{\mu}_A(-\frac{x}{2}-\frac{t_A}{2}\bar{n}_A)
\mathcal{A}^{\nu}_B(-\frac{x}{2}-\frac{t_B}{2}\bar{n}_B)\ket{p_Ap_B}
\\&=L^{\dagger}_{\mu^{\prime}\nu^{\prime}}L_{\mu\nu}
\int dt^{\prime}_Adt^{\prime}_BC^*(t^{\prime}_A,t^{\prime}_B)\int dt_Adt_BC(t_A,t_B)\hat{P}^{\mu^{\prime}\mu}\hat{P}^{\nu^{\prime}\nu}\\ \notag
&\times\bra{\bar{n}_A\cdot\mathcal{P}_A,-\frac{q_{A\perp}}{2}}\mathcal{A}^{\dagger\mu^{\prime}}_A(\frac{n_A\cdot x+t^{\prime}_A}{2},\frac{x_\perp}{2},0)\hat{P}_{\mu^{\prime}\mu}\mathcal{A}^{\mu}_A(-\frac{n_A\cdot x+t_A}{2},-\frac{x_\perp}{2},0)
\ket{\bar{n}_A\cdot\mathcal{P}_A,\frac{q_{A\perp}}{2}}\\ \notag
&\times\bra{\bar{n}_B\cdot \mathcal{P}_B,\frac{q_{B\perp}}{2}}\mathcal{A}^{\dagger\nu^{\prime}}_B(\frac{n_B\cdot x+t^{\prime}_B}{2},\frac{x_\perp}{2},0)\hat{P}_{\nu^{\prime}\nu}\mathcal{A}^{\nu}_B(-\frac{n_B\cdot x+t_B}{2},-\frac{x_\perp}{2},0)
\ket{\bar{n}_B\cdot \mathcal{P}_B,-\frac{q_{B\perp}}{2}}.
\end{align}
Due to photon field's transverse polarization, we introduce Lorentz projectors $\hat{P}_{\mu^{\prime}\mu}$ and $\hat{P}_{\nu^{\prime}\nu}$ for projecting the polarization of field onto the transverse plane:
\begin{align}
&\hat{P}_{\mu^{\prime}\mu}=\frac{1}{d-2}(-g_T)_{\mu^{\prime}\mu},\\ \notag
&(g_T)_{\mu^{\prime}\mu}=g_{\mu^{\prime}\mu}-\frac{(n_A)_{\mu^{\prime}}(n_B)_{\mu}+(n_A)_{\mu}(n_B)_{\mu^{\prime}}}{2}.
\end{align}
We define the photon's distribution function
 \begin{align}
&\mathcal{T}_i(q_{i\perp},n_i\cdot x+t_i,\frac{x_\perp}{2})\\ \notag
&=\bra{\bar{n}_i\cdot \mathcal{P}_i,-\frac{q_{i\perp}}{2}}\mathcal{A}^{\dagger\beta}_i(\frac{n_i\cdot x+t_i^{\prime}}{2},\frac{x_\perp}{2},0)\hat{P}_{\beta\mu}\mathcal{A}_i^{\mu}(-\frac{n_i\cdot x+t_i}{2},-\frac{x_\perp}{2},0)\ket{\bar{n}_i\cdot \mathcal{P}_i,\frac{q_{i\perp}}{2}}.
\end{align} 
Subsequently, We apply a Fourier transformation to convert the spacetime coordinate component in the anti-collinear direction to momentum space, and another function indicates the reverse transformation
\begin{align}
&\mathcal{T}_i(q_{i\perp},z,\frac{x_\perp}{2})\\ \notag
=&\int\frac{dt}{2\pi}e^{-it\cdot (z\bar{n}_i\cdot\mathcal{P}_i)}
\bra{\bar{n}_i\cdot\mathcal{P}_i,-\frac{q_{i\perp}}{2}}\mathcal{A}^{\dagger\beta}_i(\frac{t}{2}\bar{n},\frac{x_\perp}{2},0)\hat{P}_{\beta\mu}\mathcal{A}_i^{\mu}(-\frac{t}{2}\bar{n},-\frac{x_\perp}{2},0)]
\ket{\bar{n}_i\cdot\mathcal{P}_i,\frac{q_{i\perp}}{2}},\\ 
&\mathcal{T}_i(q_{i\perp},n_i\cdot x+t_i,\frac{x_\perp}{2})\\ \notag
=&\int d(z\bar{n}_i\cdot \mathcal{P}_i)\mathcal{T}_i(q_{i\perp},z,\frac{x_\perp}{2})e^{i[\frac{n_i\cdot x+t_i}{2}+\frac{n_i\cdot x+t^{\prime}_i}{2}](z\bar{n}_i\cdot\mathcal{P}_i)},
\end{align} 
where $\bar{n}_i\cdot\mathcal{P}_i\simeq\bar{n}_A\cdot p_A=\bar{n}_B\cdot p_B$ and z represents the fraction of photon energy involved in scattering with respect to the energy of heavy ions. The photon's four-dimensional momentum is represented by $k_1$ and $k_2$, and we additionally have 
 \begin{align}
     z_A\equiv\frac{\bar{n}_A\cdot k_1}{\bar{n}_A\cdot p_A},~~~
     z_B\equiv\frac{\bar{n}_B\cdot k_2}{\bar{n}_B\cdot p_B}.
 \end{align}
By embedding the aforementioned distribution function within the factorized function while simultaneously considering the transverse plane projection operator, we arrive at the outcome 
\begin{align}
\label{eq-photonmatrix}
&\int dt^{\prime}_Adt^{\prime}_BC^*(t^{\prime}_A,t^{\prime}_B)\int dt_Adt_BC(t_A,t_B)
\hat{P}^{\mu^{\prime}\mu}\hat{P}^{\nu^{\prime}\nu}\\ \notag
&\times\bra{\bar{n}_A\cdot\mathcal{P}_A,-\frac{q_{A_\perp}}{2}}\mathcal{A}^{\dagger\mu^{\prime}}_A(\frac{n_A\cdot x+t^{\prime}_A}{2},\frac{x_\perp}{2},0)\hat{P}_{\mu^{\prime}\mu}
\mathcal{A}^{\mu}_A(-\frac{n_A\cdot x+t_A}{2},-\frac{x_\perp}{2},0)\ket{\bar{n}_A\cdot\mathcal{P}_A,\frac{q_{A_\perp}}{2}}\\ \notag
&\times\bra{\bar{n}_B\cdot\mathcal{P}_B,\frac{q_{B_\perp}}{2}}\mathcal{A}^{\dagger\nu^{\prime}}_B(\frac{n_B\cdot x+t^{\prime}_B}{2},\frac{x_\perp}{2},0)\hat{P}_{\nu^{\prime}\nu}
\mathcal{A}^{\nu}_B(-\frac{n_B\cdot x+t_B}{2},-\frac{x_\perp}{2},0)\ket{\bar{n}_B\cdot\mathcal{P}_B,-\frac{q_{B_\perp}}{2}}\\ \notag
=&\int dt^{\prime}_Adt^{\prime}_BC^*(t^{\prime}_A,t^{\prime}_B)\int dt_Adt_BC(t_A,t_B)\hat{P}^{\mu^{\prime}\mu}\hat{P}^{\nu^{\prime}\nu}\\ \notag
&\times\int d(z_A\bar{n}_A\cdot \mathcal{P}_A)\mathcal{T}_A(q_{A\perp},z_A,\frac{x_\perp}{2})e^{i[\frac{n_A\cdot x+t_A}{2}+\frac{n_A\cdot x+t^{\prime}_A}{2}]\times[z_A\bar{n}_A\cdot\mathcal{P}_A]}\\ \notag
&\times\int d(z_B\bar{n}_B\cdot\mathcal{P}_B)\mathcal{T}_B(-q_{B\perp},z_B,\frac{x_\perp}{2})e^{i[\frac{n_B\cdot x+t_B}{2}+\frac{n_B\cdot x+t^{\prime}_B}{2}]\times[z_B\bar{n}_B\cdot\mathcal{P}_B]}.
\end{align} 
Upon transforming the Wilson coefficient into momentum space
\begin{align}
&\tilde{C}^*(z_A\bar{n}_A\cdot\mathcal{P}_A,z_B\bar{n}_B\cdot\mathcal{P}_B)=\int dt^{\prime}_Adt^{\prime}_BC^*(t^{\prime}_A,t^{\prime}_B)e^{i\frac{t^{\prime}_A}{2}(z_A\bar{n}_A\cdot\mathcal{P}_A)}e^{i\frac{t^{\prime}_B}{2}(z_B\bar{n}_B\cdot\mathcal{P}_B)},\\
&\tilde{C}(z_A\bar{n}_A\cdot\mathcal{P}_A,z_B\bar{n}_B\cdot\mathcal{P}_B)=
\int dt_Adt_BC(t_A,t_B)e^{i\frac{t_A}{2}(z_A\bar{n}_A\cdot \mathcal{P}_A)}e^{i\frac{t_B}{2}(z_B\bar{n}_B\cdot\mathcal{P}_B)},
\end{align} 
the function \eqref{eq-photonmatrix} can be represented as 
\begin{align}
&(\bar{n}_A\cdot\mathcal{P}_A)(\bar{n}_B\cdot\mathcal{P}_B)\int dz_A\mathcal{T}_A(q_{A\perp},z_A,\frac{x_\perp}{2})\int dz_B\mathcal{T}_B(-q_{B\perp},z_B,\frac{x_\perp}{2})\\ \notag
&\times\hat{P}^{\mu^{\prime}\mu}\hat{P}^{\nu^{\prime}\nu}
\tilde{C}^*(z_A\bar{n}_A\cdot\mathcal{P}_A,z_B\bar{n}_B\cdot\mathcal{P}_B)
\tilde{C}(z_A\bar{n}_A\cdot\mathcal{P}_A,z_B\bar{n}_B\cdot\mathcal{P}_B)e^{i(n_A\cdot x)[z_A\bar{n}_A\cdot \mathcal{P}_A]}e^{i(n_B\cdot x)[z_B\bar{n}_B\cdot\mathcal{P}_B]}.
\end{align}
Organizing the aforementioned results, the total scattering cross-section can be expressed as 
\begin{align}
\label{diff}
&\frac{d\sigma_b}{dy_{l_1}d^2p_{l_1\perp}dy_{l_2}d^2p_{l_2\perp}d^2b_\perp}\\ \notag
=
&\frac{1}{[2(2\pi)^3]^2}\frac{1}{\left|v^{\prime}_{Az}-v^{\prime}_{Bz}\right|}
\int d^4xe^{-i(p_{l_1}+p_{l_2})\cdot x}e^{i(n_A\cdot x)(z_A\bar{n}_A\cdot\mathcal{P}_A)}e^{i(\bar{n}_A\cdot x)(z_B\bar{n}_B\cdot\mathcal{P}_B)}\\ \notag
&\times\hat{P}^{\mu^\prime\mu}\hat{P}^{\nu^\prime\nu}L^{\dagger}_{\mu^{\prime}\nu^{\prime}}L_{\mu\nu}\\ \notag
&\times\int dz_A\int\frac{d^2q_{A\perp}}{(2\pi)^2}\mathcal{T}_A(q_{A\perp},z_A,\frac{x_\perp}{2})e^{-iq_{A\perp}\cdot x_{A\perp}}
\int dz_B\int\frac{d^2q_{B\perp}}{(2\pi)^2}\mathcal{T}_B(-q_{B\perp},z_B,\frac{x_\perp}{2})e^{iq_{B\perp}\cdot x_{B\perp}}\\ \notag
&\times(2\pi)^2\delta^{(2)}(q_{A\perp}-q_{B\perp})\widetilde{C}^*(z_A\bar{n}_A\cdot\mathcal{P}_A,z_B\bar{n}_B\cdot\mathcal{P}_B)\widetilde{C}(z_A\bar{n}_A\cdot\mathcal{P}_A,z_B\bar{n}_B\cdot\mathcal{P}_B), 
\end{align}
Now we consider about the function
\begin{align}
\label{PDFs}
&\int dz_A\int\frac{d^2q_{A\perp}}{(2\pi)^2}\mathcal{T}_A(q_{A\perp},z_A,\frac{x_\perp}{2})e^{-iq_{A\perp}\cdot x_{A_\perp}}
\int dz_B\int\frac{d^2q_{B\perp}}{(2\pi)^2}\mathcal{T}_B(-q_{B\perp},z_B,\frac{x_\perp}{2})e^{iq_{B\perp}\cdot x_{B_\perp}} \notag \\
&\times(2\pi)^2\delta^{(2)}(q_{A\perp}-q_{B\perp}). 
\end{align} 
At the preceding equation \eqref{PDFs}, if we directly compute $\delta^{(2)}(q_{A_\perp}-q_{B_\perp})$, we will ultimately derive the result
\begin{align}
        \int dz_A\mathcal{T}_A(z_A,\frac{x_\perp}{2},-\frac{1}{2}b_\perp)\int dz_B\mathcal{T}_B(z_B,\frac{x_\perp}{2},\frac{b_\perp}{2}).
\end{align}
The result indicates that photons situated at transverse coordinates $\frac{b_\perp}{2}$ and $-\frac{b_\perp}{2}$ undergo scattering. However, the constraint does not exist, and there is no reason to assume that only photons situated at symmetric positions about the zero point can be scattered. To handle the delta function more effectively, we adopt the following approach instead of directly integrating it here,
\begin{align}
        &\bra{\bar{n}_A\cdot\mathcal{P}_A,-\frac{q_{A_\perp}}{2}}
        \mathcal{A}^{\dagger\mu^{\prime}}_A(\frac{t}{2}\bar{n}_A,\frac{x_\perp}{2},0)
        \mathcal{A}^{\mu}_A(-\frac{t}{2}\bar{n}_A,-\frac{x_\perp}{2},0)\ket{\bar{n}_A\cdot\mathcal{P}_A,\frac{q_{A_\perp}}{2}}\\
        &\times\bra{\bar{n}_B\cdot\mathcal{P}_B,\frac{q_{B_\perp}}{2}}
        \mathcal{A}^{\dagger\nu^{\prime}}_B(\frac{t}{2}\bar{n}_B,\frac{x_\perp}{2},0)
        \mathcal{A}^{\nu}_B(-\frac{t}{2}\bar{n}_B,-\frac{x_\perp}{2},0)\ket{\bar{n}_B\cdot\mathcal{P}_B,-\frac{q_{B_\perp}}{2}}\notag\\
        &\times(2\pi)^2\delta^{(2)}(q_{A_\perp}-q_{B_\perp})\notag\\
        =&\int d^2X_\perp\notag\\
        &\times\bra{\bar{n}_A\cdot\mathcal{P}_A,-\frac{q_{A_\perp}}{2}}
        e^{i\hat{P}\cdot X_\perp}
        \mathcal{A}^{\dagger\mu^{\prime}}_A(\frac{t}{2}\bar{n}_A,\frac{x_\perp}{2},0)
e^{-i\hat{P}\cdot X_\perp} e^{i\hat{P}\cdot X_\perp}
       \mathcal{A}^{\mu}_A(-\frac{t}{2}\bar{n}_A,-\frac{x_\perp}{2},0)
       e^{-i\hat{P}\cdot X_\perp}
       \ket{\bar{n}_A\cdot\mathcal{P}_A,\frac{q_{A_\perp}}{2}}\notag\\
       &\times \bra{\bar{n}_B\cdot\mathcal{P}_B,\frac{q_{B_\perp}}{2}}
 e^{i\hat{P}\cdot X_\perp}\mathcal{A}^{\dagger\nu^{\prime}}_B
 (\frac{t}{2}\bar{n}_B,\frac{x_\perp}{2},0)e^{-i\hat{P}\cdot X_\perp}
e^{i\hat{P}\cdot X_{\perp}}
 \mathcal{A}^{\nu}_B(-\frac{t}{2}\bar{n}_B,-\frac{x_\perp}{2},0)
 e^{-i\hat{P}\cdot X_\perp}\ket{\bar{n}_B\cdot\mathcal{P}_B,-\frac{q_{B_\perp}}{2}}\notag\\
 =&\int d^2X_\perp\bra{\bar{n}_A\cdot\mathcal{P}_A,-\frac{q_{A_\perp}}{2}}\mathcal{A}^{\dagger\mu^{\prime}}_A(\frac{t}{2}\bar{n}_A,\frac{x_\perp}{2}+X_\perp,0)\mathcal{A}^{\mu}_A(-\frac{t}{2}\bar{n}_A,-\frac{x_\perp}{2}+X_\perp,0)\ket{\bar{n}_A\cdot\mathcal{P}_A,\frac{q_{A_\perp}}{2}}\notag\\
 &\times \bra{\bar{n}_B\cdot\mathcal{P}_B,\frac{q_{B_\perp}}{2}}
 \mathcal{A}^{\dagger\nu^{\prime}}_B(\frac{t}{2}\bar{n}_B,\frac{x_\perp}{2}+X_\perp,0)
 \mathcal{A}^{\nu}_B(-\frac{t}{2}\bar{n}_B,-\frac{x_\perp}{2}+X_\perp,0)\ket{\bar{n}_B\cdot \mathcal{P}_B,-\frac{q_{B_\perp}}{2}}.\notag
\end{align}
One can verify the equation by reverse calculation, starting from the lower equal sign and working back to the upper equal sign. Through the aforementioned calculations, equation \eqref{PDFs} undergoes a transformation,
\begin{align}
\label{equation(30)}
        &\int dz_A\int \frac{d^2q_{A_\perp}}{(2\pi)^2}
        \mathcal{T}_A(q_{A_\perp},z_A,\frac{x_\perp}{2})
        e^{-iq_{A_\perp}\cdot x_{A_\perp}}
        \int dz_B\int\frac{d^2q_{B_\perp}}{(2\pi)^2}
\mathcal{T}_B(-q_{B_\perp},z_B,\frac{x_\perp}{2})\\ \notag 
&\times e^{iq_{B_\perp}\cdot x_{B_\perp}}\times (2\pi)^2\delta^{(2)}
(q_{A_\perp}-q_{B_\perp})\\ \notag
=&\int dz_A\int dz_B\int\frac{d^2q_{A_\perp}}{(2\pi)^2}e^{-iq_{A_\perp}\cdot x_{A_\perp}}\int \frac{d^2q_{B_\perp}}{(2\pi)^2}
e^{iq_{B_\perp}\cdot x_{B_\perp}}\\ \notag
&\times \int d^2X_\perp\mathcal{T}_A(X_\perp,q_{A_\perp},z_A,\frac{x_\perp}{2})
\mathcal{T}_B(X_\perp,-q_{B_\perp},z_B,\frac{x_\perp}{2})\\ \notag
=&\int d^2X_\perp\int dz_Adz_B\mathcal{T}_A(X_\perp-b_\perp,z_A,\frac{x_\perp}{2})\mathcal{T}_B(X_\perp,z_B,\frac{x_\perp}{2}).
\end{align}
Additionally, we employ the definition $b_\perp=x_{B_\perp}-x_{A_\perp}$ introduced at the outset.

\section{The generalized transverse momentum distributions of photons}
\label{sec:TPS}
We start our analysis by considering the distribution function 
 \begin{align}
 \bra{\bar{n}\cdot\mathcal{P},-\frac{q_\perp}{2}}\mathcal{A}^{\dagger\beta}(\frac{t}{2}\bar{n},\frac{x_\perp}{2},0)\hat{P}_{\beta\mu}
 \mathcal{A}^{\mu}(-\frac{t}{2}\bar{n},-\frac{x_\perp}{2},0)\ket{\bar{n}\cdot\mathcal{P},\frac{q_\perp}{2}}.
 \end{align}
When calculating the scattering cross section, we apply Fourier transformation to map $q_\perp$ to $r_\perp$. Concomitantly, via Fourier transformation, we map $t\bar{n}^{\mu}$ to $z\bar{n}\cdot\mathcal{P}$, the momentum of the initial state photon. Consequently, we introduce a beam function that undergoes double Fourier transformations, denominated as the thickness beam function \cite{wu2021factorization}:
\begin{align}
&\mathcal{T}(r_\perp,z,\frac{x_\perp}{2})\\
&=\int\frac{d^2q_\perp}{(2\pi)^2}e^{iq_\perp\cdot r_\perp}\int \frac{dt}{2\pi}e^{-izt(\bar{n}\cdot \mathcal{P})}
\bra{\bar{n}\cdot\mathcal{P},-\frac{q_\perp}{2}}\mathcal{A}^{\dagger\beta}(\frac{t}{2}\bar{n},\frac{x_\perp}{2},0)\hat{P}_{\beta\mu}\mathcal{A}^{\mu}(-\frac{t}{2}\bar{n},-\frac{x_\perp}{2},0)\ket{\bar{n}\cdot\mathcal{P},\frac{q_\perp}{2}}. \notag
\end{align} 
Our ultimate objective is to obtain a photon distribution function that depends on the impact parameter. For this reason, it is necessary for the coordinates of the photon field $\mathcal{A}$  to also depend on the impact parameter. Using the translation operator, we shift the coordinates of $\mathcal{A}$ and $\mathcal{A}^{\dagger}$  by the same distance in the reaction plane, denoted as $r_\perp$, without any loss of generality,
\begin{align}
 &\mathcal{T}(r_\perp,z,\frac{x_\perp}{2})\\ \notag
=
&\int\frac{d^2q_\perp}{(2\pi)^2}e^{iq_\perp\cdot r_\perp}\frac{1}{2\pi}\int dt e^{-izt(\bar{n}\cdot\mathcal{P})}\hat{P}_{\beta\mu}
\bra{\bar{n}\cdot\mathcal{P},-\frac{q_\perp}{2}}e^{-i\hat{P}r_\perp}\mathcal{A}^{\dagger\beta}(\frac{t}{2}\bar{n},\frac{1}{2}x_\perp+r_\perp,0) e^{i\hat{P}r_\perp}\\ \notag
&\times e^{-i\hat{P}r_\perp}
\mathcal{A}^\mu(-\frac{t}{2}\bar{n},-\frac{x_\perp}{2}+r_\perp,0) e^{i\hat{P}r_\perp}\ket{\bar{n}\cdot\mathcal{P},\frac{q_\perp}{2}}\\ \notag
=&
\int\frac{d^2q_\perp}{(2\pi)^2}e^{i2q_\perp\cdot r_\perp}\frac{1}{2\pi}\int dte^{-izt(\bar{n}\cdot\mathcal{P})}\hat{P}_{\beta\mu}\\ \notag
&\times\bra{\bar{n}\cdot\mathcal{P},-\frac{q_\perp}{2}}\mathcal{A}^{\dagger\beta}(\frac{t}{2}\bar{n},\frac{1}{2}x_\perp+r_\perp,0)\mathcal{A}^{\mu}(-\frac{t}{2}\bar{n},-\frac{x_\perp}{2}+r_\perp,0)\ket{\bar{n}\cdot\mathcal{P},\frac{q_\perp}{2}}.
\end{align}
The thickness beam function can be expressed as:
\begin{align}
  &\mathcal{T}(r_\perp,z,\frac{x_\perp}{2})\\ \notag
=
&\int\frac{d^2q_\perp}{(2\pi)^2}e^{i2q_\perp\cdot r_\perp}\frac{1}{2\pi}\int dte^{-izt(\bar{n}\cdot\mathcal{P})}\hat{P}_{\beta\mu}\\ \notag
&\times\bra{\bar{n}\cdot\mathcal{P},-\frac{q_\perp}{2}}\mathcal{A}^{\dagger\beta}(\frac{t}{2}\bar{n},\frac{1}{2}x_\perp+r_\perp,0)\mathcal{A}^{\mu}(-\frac{t}{2}\bar{n},-\frac{x_\perp}{2}+r_\perp,0)\ket{\bar{n}\cdot\mathcal{P},\frac{q_\perp}{2}},
\end{align} 
where $z$ denotes the fraction of the parton's energy or, equivalently, the energy fraction of the initial-state scattered photon.

Next, we will demonstrate that performing a Fourier transform of the function $\mathcal{T}(r_\perp,z,\frac{x_\perp}{2})$ with respect to $x_\perp$ results in a $\delta$ function, which implies the conservation of momentum between the initial state photon and the final state lepton. The other functions, after the transformation, under the classical field approximation, correspond exactly to the photon distribution function in the Wood-Saxon model divided by $z$. The Fourier transformation of $\mathcal{T}(r_\perp,z,\frac{x_\perp}{2})$ denoted as $f\big(r_\perp,z,-\frac{1}{2}(p_{l_{1\perp}}+p_{l_{2\perp}})\big)$ is referred to as generalized transverse momentum dependent distributions (GTMDs) or transverse phase-space PDFs (TPS PDFs) \cite{wu2021factorization}. 

Initially, we simplify the thickness beam function using the classical field approximation. To apply the classical field results \cite{wang2021lepton,Krauss:1997vr}, we use Fourier transformation convert the coordinates of photon field into momentum space. It should be emphasized that $k$ and $k^{\prime}$ correspond respectively to the momenta of the initial state photon and its complex conjugate,
\begin{align}
&\frac{1}{2\pi}\int dte^{-izt(\bar{n}\cdot\mathcal{P})}
\bra{\bar{n}\cdot\mathcal{P},-\frac{q_\perp}{2}}\mathcal{A}^{\dagger\beta}(\frac{t}{2}\bar{n},\frac{1}{2}x_\perp+r_\perp,0)\mathcal{A}^{\mu}(-\frac{t}{2}\bar{n},-\frac{x_\perp}{2}+r_\perp,0)\ket{\bar{n}\cdot\mathcal{P},\frac{q_\perp}{2}}\notag \\
=
&\int \frac{dt}{2\pi} e^{-izt(\bar{n}\cdot\mathcal{P})}\\ \notag
&\times\int\frac{d(\bar{n}\cdot k^{\prime})}{2\pi}e^{i(\bar{n}\cdot k^{\prime})t\cdot\frac{1}{2}}\int\frac{d(\bar{n}\cdot k)}{2\pi}e^{i(\bar{n}\cdot k)t\frac{1}{2}}
\int\frac{d^2k^{\prime}_\perp}{(2\pi)^2}e^{ik^{\prime}_{\perp}\cdot(r_\perp+\frac{x_\perp}{2})}\int\frac{d^2k_\perp}{(2\pi)^2}e^{-ik_\perp\cdot(r_\perp-\frac{x_\perp}{2})}\\ \notag
&\times\int\frac{d(n\cdot k)}{2\pi}\int\frac{d(n\cdot k^{\prime})}{2\pi}
\bra{\bar{n}\cdot \mathcal{P},-\frac{q_\perp}{2}}\widetilde{\mathcal{A}}^{\dagger\beta}(\bar{n}\cdot k^{\prime},k^{\prime}_\perp,n\cdot k^{\prime})\widetilde{\mathcal{A}}^{\mu}(\bar{n}\cdot k,k_\perp,n\cdot k)\ket{\bar{n}\cdot\mathcal{P},\frac{q_\perp}{2}}.
\end{align}
Next, we employ the classical field approximation. In UPC, we can replace the photon field with a classical field or, equivalently, solutions of Maxwell's equations \cite{Krauss:1997vr}, expressed as: 
\begin{align}
    \mathcal{A}^{\mu}(k)=2\pi ze\delta(k\cdot u)\frac{F(-k^2)}{-k^2}[u^{\mu}-\frac{k^{\mu}}{\bar{n}\cdot k}\bar{n}\cdot u].
\end{align}
$u^{\mu}=\gamma(1,~0,~0,~v)$ represents the velocity of the heavy ion. 
The expansion of the variable in the function $\mathcal{A}^{\mu}(k)$ can be facilitated by employing light cone coordinates,
\begin{align}
    &u^{\mu}=n\cdot u\frac{\bar{n}^{\mu}}{2}+\bar{n}\cdot u\frac{n^{\mu}}{2}+u^{\mu}_T
    =\sqrt{\frac{1-v}{1+v}}\cdot\frac{\bar{n}^\mu}{2}+\sqrt{\frac{1+v}{1-v}}\cdot\frac{n^\mu}{2},\\
    &u\cdot k=(n\cdot k)(\bar{n}\cdot u)\frac{1}{2}+(\bar{n}\cdot k)(n\cdot u)\cdot\frac{1}{2}+0_\perp
    =(n\cdot k)\sqrt{\frac{1+v}{1-v}}\cdot\frac{1}{2}+(\bar{n}\cdot k)\sqrt{\frac{1-v}{1+v}}\cdot\frac{1}{2},\notag\\
    &\bar{n}\cdot u=\gamma(1+v)=\sqrt{\frac{1+v}{1-v}}.\notag
\end{align}
By applying the relativistic limit $v\rightarrow c=1$, we deduce the expression 
\begin{align}
    &u^\mu\simeq\sqrt{\frac{1+v}{1-v}}\cdot\frac{n^\mu}{2},\\ 
    &k\cdot u\simeq(n\cdot k)\sqrt{\frac{1+v}{1-v}}\cdot\frac{1}{2}\rightarrow\delta(k\cdot u)\simeq2\sqrt{\frac{1-v}{1+v}}\delta(n\cdot k).
\end{align}
Within the framework of the light cone coordinate system, $\mathcal{A}^\mu(k)$ can be expanded as follows,
\begin{align}
\mathcal{A}^{\mu}(k)\simeq& 4\pi ze\sqrt{\frac{1-v}{1+v}}\delta(n\cdot k)\frac{F(-k^2)}{-k^2}[\sqrt{\frac{1+v}{1-v}}\cdot\frac{n^\mu}{2}-\frac{k^\mu}{\bar{n}\cdot k}\sqrt{\frac{1+v}{1-v}}]\\
=&4\pi z e\delta(n\cdot k)\frac{F(-k^2)}{-k^2}[\frac{n^\mu}{2}-\frac{k^\mu}{\bar{n}\cdot k}].\notag
\end{align}
Utilizing a projection operator, the field's polarization is projected onto the transverse plane,
\begin{align}
    \hat{P}_{\beta\mu}\mathcal{A}^{*\beta}\mathcal{A}^{\mu}=
    (4\pi)^2z^2e^2\delta(n\cdot k)\delta(n\cdot k^{\prime})\frac{F(-k^2)}{-k^2}\frac{F(-k^{\prime 2})}{-k^{\prime 2}}\times[\frac{k_\perp\cdot k^{\prime}_\perp}{(\bar{n}\cdot k)(\bar{n}\cdot k^{\prime})}].
\end{align}
Subsequently, by introducing the classical field approximation to the correlation function of photons, the results can be expressed as 
\begin{align}
&\mathcal{T}(r_\perp,z,\frac{x_\perp}{2})\\ \notag
=
&\int\frac{d^2q_\perp}{(2\pi)^2}e^{i(2q_\perp)\cdot r_\perp}\frac{1}{2\pi}\int dte^{-izt(\bar{n}\cdot \mathcal{P})}\hat{P}_{\beta\mu}
\int\frac{d(\bar{n}\cdot k^{\prime})}{2\pi}e^{i(\bar{n}\cdot k^{\prime})t\cdot\frac{1}{2}}\int\frac{d(\bar{n}\cdot k)}{2\pi}e^{i(\bar{n}\cdot k)\frac{t}{2}}\\ \notag
&\times\int\frac{d^2k^{\prime}_{\perp}}{(2\pi)^2}e^{ik^{\prime}_\perp\cdot(r_\perp+\frac{x_\perp}{2})}\int\frac{d^2k_{\perp}}{(2\pi)^2}e^{-ik_\perp\cdot(r_\perp-\frac{x_\perp}{2})}\\ \notag
&\times\int\frac{d(n\cdot k)}{2\pi}\int\frac{d(n\cdot k^{\prime})}{2\pi}
\widetilde{A}^{*\beta}(k^{\prime})\widetilde{A}^{\mu}(k)2\sqrt{\frac{(\bar{n}\cdot\mathcal{P})}{2}\frac{(\bar{n}\cdot\mathcal{P})}{2}}2\pi\delta(\bar{n}\cdot k-\bar{n}\cdot k^{\prime}-\bar{n}\cdot\mathcal{P}+\bar{n}\cdot\mathcal{P}^{\prime})\\ \notag
&\times(2\pi)^2\delta^{(2)}(k_\perp-k^{\prime}_\perp-\frac{q_\perp}{2}-\frac{q_\perp}{2}).
\end{align}
Upon performing the integration with respect to the variable $t$, we derive the delta function denoted as $\delta[\frac{1}{2}(\bar{n}\cdot k^{\prime})+\frac{1}{2}(\bar{n}\cdot k)-z(\bar{n}\cdot\mathcal{P})]$. The delta function encapsulates a fundamental physical constraint: the energy of the scattering photon is $z$ times the energy of the heavy ion.

Meanwhile the delta function $\delta(\bar{n}\cdot k-\bar{n}\cdot k^{\prime}-\bar{n}\cdot\mathcal{P}+\bar{n}\cdot\mathcal{P}^{\prime})$ implies the condition that the component of a photon’s momentum along the light-cone direction is equal to that of its conjugate photon. It's important to note that we are operating with the implicit condition that $\bar{n}\cdot\mathcal{P}=\bar{n}\cdot\mathcal{P}^{\prime}$. This condition was clarified in text under equation \eqref{notation}.

By integrating $\int\frac{d^2q_\perp}{(2\pi)^2}$ with the delta function $(2\pi)^2\delta^{(2)}(k_\perp-k^{\prime}_\perp-q_\perp)$ and simultaneously substituting $q_\perp$ with $k_\perp-k^{\prime}_\perp$ in $e^{i(2q_\perp)\cdot r_\perp}$, the interference term we require emerges, where $r_\perp$ represents the impact parameter (without loss of generality, we refrain from using $b_\perp$ to denote the impact parameter),
\begin{align}
&\mathcal{T}(r_\perp,z,\frac{x_\perp}{2})\\ \notag
=&\frac{1}{2\pi}\hat{P}_{\beta\mu}\int\frac{d^2k^{\prime}_\perp}{(2\pi)^2}\int\frac{d^2k_\perp}{(2\pi)^2}e^{i(k_\perp-k^{\prime}_{\perp})\cdot r_\perp}e^{i(k^{\prime}_\perp+k_\perp)\cdot\frac{x_\perp}{2}}\int\frac{d(\bar{n}\cdot k)}{2\pi}(2\pi)\delta[(\frac{1}{2}(\bar{n}\cdot k^{\prime})+\frac{1}{2}(\bar{n}\cdot k)-z(\bar{n}\cdot\mathcal{P})]\\ \notag
&\times\int \frac{d(n\cdot k)}{2\pi}\int\frac{d(n\cdot k^{\prime})}{2\pi}\tilde{\mathcal{A}}^{* \beta}(k^{\prime})\tilde{A}^{\mu}(k)\times2\mathcal{W}_{\mathcal{P}}
\\ \notag
=&\frac{1}{2\pi}\hat{P}_{\beta\mu}\int\frac{d^2k^{\prime}_\perp}{(2\pi)^2}\frac{d^2k_\perp}{(2\pi)^2}e^{i(k_\perp-k^{\prime}_\perp)\cdot r_\perp}e^{i(k^{\prime}_\perp+k_\perp)\cdot \frac{x_\perp}{2}}
\int\frac{d(n\cdot k)}{2\pi}\int\frac{d(n\cdot k^{\prime})}{2\pi}\\ \notag
&\times 2\mathcal{W}_{\mathcal{P}}
\left\{ 4\pi Ze\delta(n\cdot k^{\prime})\frac{F(-k^{\prime 2})}{-k^{\prime 2}}[\frac{n^{\beta}}{2}-\frac{k^{\prime \beta}}{\bar{n}\cdot k^{\prime}}]\right\}\\ \notag
&\times\left\{4\pi Ze\delta(n\cdot k)\frac{F(-k^2)}{-k^2}[\frac{n^{\mu}}{2}-\frac{k^\mu}{\bar{n}\cdot k}]\right\}\\ \notag
=&\frac{4}{\pi}\int\frac{d^2k^{\prime}_{\perp}}{(2\pi)^2}\int\frac{d^2k_\perp}{(2\pi)^2}e^{i(k_\perp-k_\perp^{\prime})\cdot r_\perp}e^{i(k_\perp^{\prime}+k_\perp)\cdot\frac{x_\perp}{2}}\mathcal{W}_{\mathcal{P}}
 Z^2e^2\frac{F(-k^{\prime 2})}{-k^{\prime 2}}\frac{F(-k^2)}{-k^2}[\frac{k^{\prime}_\perp\cdot k_\perp}{(\bar{n}\cdot k^{\prime})(\bar{n}\cdot k)}].
\end{align}
Substituting $\bar{n}\cdot k^{\prime}$ and $\bar{n}\cdot k$ with $2\mathcal{W}$, concurrently exchanging $\mathcal{W}_{\mathcal{P}}$ with $\frac{\mathcal{W}}{z_i}$, wherein $\mathcal{W}$ represents the frequency of the photon, $\mathcal{W}_{\mathcal{P}}$ denotes the frequency associated with the wave packet, and $z_i$ signifies the energy fraction corresponding to the photon labeled as $i$,
\begin{align}
    \mathcal{T}(r_\perp,z,\frac{x_\perp}{2})
 =&\frac{Z^2e^2}{\pi}\int\frac{d^2k_\perp^{\prime}}{(2\pi)^2}\int\frac{d^2k_\perp}{(2\pi)^2}e^{i(k_\perp-k^{\prime}_\perp)\cdot r_\perp}e^{i(k^{\prime}_\perp+k_\perp)\cdot \frac{x_\perp}{2}}\\
 &\times\frac{\mathcal{W}}{z_i}
\frac{F(-k^{\prime 2})}{-k^{\prime 2}}\frac{F(-k^2)}{-k^2}[\frac{k^{\prime}_\perp\cdot k_\perp}{2\mathcal{W}\cdot 2\mathcal{W}}]\times 4. \notag
\end{align}
Now, it is imperative to consider the thickness beam function corresponding to ions A and B simultaneously. By incorporating the factor $\int d^2x_\perp e^{-i(p_{l_{1\perp}}+p_{l_{2\perp}})\cdot x_\perp}$ from the scattering cross-section, we will employ Fourier transformation to convert $x_\perp$ in $\mathcal{T}_i(r_\perp,z,\frac{x_\perp}{2})$. The product of the transformed functions $f_A\big(r_{A_\perp},z_A,-\frac{1}{2}(p_{l_{1\perp}}+p_{l_{2\perp}})\big)$ and $f_B\big(r_{B_\perp},z_B,-\frac{1}{2}(p_{l_{1\perp}}+p_{l_{2\perp}})\big)$, within the framework of UPC and utilizing the classical field approximation, can be expressed as the product of photon distribution functions divided by $z_Az_B$ and accompanied by a delta function. This delta function implies the conservation of transverse momentum between initial state photons and final state leptons. The function $f_i\big(r_{i_\perp},z_i,-\frac{1}{2}(p_{l_{1\perp}}+p_{l_{2\perp}})\big)$ is denoted as generalized transverse momentum dependent distributions (GTMDs) or transverse phase-space PDFs(TPS PDFs)\cite{wu2021factorization}. Now write down the complete function we need to deal with
\begin{align}
    &\int d^2 x_\perp e^{-i(p_{l_{1\perp}}+p_{l_{2\perp}})\cdot x_\perp}\mathcal{T}_A(r_{A_\perp},z_A,\frac{x_\perp}{2})\mathcal{T}_B(r_{B_\perp},z_B,\frac{x_\perp}{2})\\ \notag
    \equiv& f_A\big(r_{A_\perp},z_A,-\frac{1}{2}(p_{l_{1\perp}}+p_{l_{2\perp}})\big)
    f_B\big(r_{B_\perp},z_B,-\frac{1}{2}(p_{l_{1\perp}}+p_{l_{2\perp}})\big).
\end{align}
We neglect the contribution of final-state soft radiation in this analysis. Incorporating soft radiation will not alter the process of verification but will render the function more intricate in appearance. %To prevent readers from becoming lost in intricate mathematical calculations, I will explicitly present the necessary components and employ shorthand notation to simplify functions that do not require immediate attention. We utilize shorthand notation to represent $\mathcal{T}_i(r_{i_\perp},z_i,\frac{x_\perp}{2})$ as 
%\begin{align}
%    \mathcal{T}_i(r_{i_\perp},z_i,\frac{x_\perp}{2})
%    =\int \frac{d^2k^{\prime}_\perp}{(2\pi)^2}
%    \int\frac{d^2k_\perp}{(2\pi)^2}
%    e^{i(k^{\prime}_\perp+k_\perp)\cdot\frac{x_\perp}{2}}
%    \mathcal{F}_i(k_\perp,k^{\prime}_\perp,r_\perp,z_i),
%\end{align}
%where
%\begin{align}
%    \mathcal{F}_i(k_\perp,k^{\prime}_\perp,r_\perp,z_i)\equiv
%    \frac{Z^2e^2}{\pi}
%    e^{i(k_\perp-k^{\prime}_\perp)\cdot r_\perp}
%    \frac{\mathcal{W}}{z_i}
%    \frac{F(-k^{\prime 2})}{-k^{\prime 2}}\frac{F(-k^2)}{-k^2}\times\big[ \frac{k^{\prime}_\perp\cdot k_\perp}{2\mathcal{W}\cdot2\mathcal{W}} \big]\times 4.
%\end{align}
%And then we have
%\begin{align}
%    &f_A\big(r_{A_\perp},z_A,-\frac{1}{2}(p_{l_{1\perp}}+p_{l_{2\perp}})\big)
%    f_B(r_{B_\perp},z_B,-\frac{1}{2}(p_{l_{1\perp}}+p_{l_{2\perp}})\big)\\ \notag
%    =&
%    \int d^2x_\perp e^{-i(p_{l_{1\perp}}+p_{l_{2\perp}})\cdot x_\perp}
%    \int \frac{d^2k^{\prime}_{1\perp}}{(2\pi)^2}
%    \frac{d^2k_{1\perp}}{(2\pi)^2}
%    e^{i(k^{\prime}_{1\perp}+k_{1\perp})\cdot\frac{x_\perp}{2}}\\ \notag
%    &\times \int\frac{d^2k^{\prime}_{2\perp}}{(2\pi)^2}
%    \frac{d^2k_{2\perp}}{(2\pi)^2}
%    e^{i(k^{\prime}_{2\perp}+k_{2\perp})\cdot\frac{x_\perp}{2}}\mathcal{F}_A(k_{1\perp},k^{\prime}_{1\perp},r_{A_\perp},z_A)\mathcal{F}_B(k_{2\perp},k^{\prime}_{2\perp},r_{B_\perp},z_B).
%\end{align}
With the relationship between the transverse momentum of the initial-state photon and that of its conjugate photon, we have
\begin{align}
[k^{\prime}_{1\perp}+k_{1\perp}+k^{\prime}_{2\perp}+k_{2\perp}]\times\frac{1}{2}=
    k_{1\perp}+k_{2\perp}.
\end{align}
Then, the following results are obtained
\begin{align}
\label{(3)}
&f_A\big(r_{A_\perp},z_A,-\frac{1}{2}(p_{l_{1\perp}}+p_{l_{2\perp}})\big)f_B\big(r_{B_\perp},z_B,-\frac{1}{2}(p_{l_{1\perp}}+p_{l_{2\perp}})\big)\\ \notag
    =&\int d^2x_\perp e^{-i(p_{l_{1\perp}}+p_{l_{2\perp}})\cdot x_\perp}
    \mathcal{T}_A(z_A,\frac{x_\perp}{2},r_{A_\perp})\mathcal{T}_B
    (z_B,\frac{x_\perp}{2}, r_{B_\perp})\\ \notag
    =& \frac{Z^2e^2}{\pi}\int
    \frac{d^2k^{\prime}_{1\perp}}{(2\pi)^2}\int\frac{d^2k_{1\perp}}{(2\pi)^2}
     e^{i(k_{1\perp}-k^{\prime}_{1\perp})\cdot r_{A_\perp}}
    \frac{\mathcal{W}_1}{z_A}\\ \notag
    &\times\frac{F(-k^{\prime 2}_1)}{-k^{\prime 2}_1}\frac{F(-k^2_1)}{-k^2_1}[\frac{k^{\prime}_{1\perp}\cdot k_{1\perp}}{2\mathcal{W}_1\cdot 2\mathcal{W}_1}]\times 4\\ \notag
    &\times\frac{Z^2e^2}{\pi}\int\frac{d^2k^{\prime}_{2\perp}}{(2\pi)^2}
    \int\frac{d^2k_{2\perp}}{(2\pi)^2}e^{i(k^{\prime}_{2\perp}-k_{2\perp})\cdot r_{B_\perp}}\frac{\mathcal{W}_2}{z_B}\frac{F(-k^{\prime 2}_2)}{-k^{\prime 2}_2}\frac{F(-k^2_2)}{-k^2_2}\\ \notag
    &\times[\frac{k^{\prime}_{2\perp}\cdot k_{2\perp}}{2\mathcal{W}_2\cdot2\mathcal{W}_2}]\times4
    \int d^2x_\perp e^{i(k_{1\perp}+k_{2\perp}-p_{l_{1\perp}}-p_{l_{2\perp}})\cdot x_\perp}\\ \notag
    =&\frac{Z^2e^2}{\pi\mathcal{W}_1}
    |\int\frac{d^2k_{1\perp}}{(2\pi)^2}e^{ik_{1\perp}\cdot r_{A_\perp}}\frac{F(-k^2_1)}{-k^2_1}k_{1\perp}|^2\frac{1}{z_A}
    \times\frac{Z^2e^2}{\pi\mathcal{W}_2}|
    \int\frac{d^2k_{2\perp}}{(2\pi)^2}e^{ik_{2\perp}\cdot r_{B_\perp}}
    \frac{F(-k^2_2)}{-k^2_2}k_{2\perp}|^2\frac{1}{z_B}\\ \notag
    &\times (2\pi)^2\delta^{(2)}(k_{1\perp}+k_{2\perp}-p_{l_{1\perp}}-p_{l_{2\perp}})
    \\ \notag
    \equiv&\frac{n(\mathcal{W}_1,r_{A_\perp})}{z_A}\frac{n(\mathcal{W}_2,r_{B_\perp})}{z_B}
    \times (2\pi)^2\delta^{(2)}(k_{1\perp}+k_{2\perp}-p_{l_{1\perp}}-p_{l_2\perp}),
\end{align}
where $n(\mathcal{W},r_{i\perp})$ corresponds to the classical photon distribution function in the Wood-Saxon model. Thus far, introduce \eqref{(3)} and \eqref{equation(30)} into equation \eqref{diff} the complete representation of the differential cross section is formulated as
\begin{align}
&\frac{d\sigma_b}{dy_{l_1}d^2p_{l_1\perp}dy_{l_2}d^2p_{l_2\perp}d^2b_\perp}\\ \notag
=
&\int d^2X_\perp\frac{1}{[2(2\pi)^3]^2}\frac{1}{\left|v^{\prime}_{Az}-v^{\prime}_{Bz}\right|}
\int d^4x e^{i(k_{1\perp}+k_{2\perp}-p_{l_{1\perp}}-p_{l_{2\perp}})\cdot x_\perp}\\ \notag
&\times e^{i(n_A\cdot x)[z_A\bar{n}_A\cdot\mathcal{P}_A-\bar{n}_A\cdot(p_{l_1}+p_{l_2})]}e^{i(\bar{n}_A\cdot x)[n_A\cdot(z_B\mathcal{P}_B)-n_A\cdot(p_{l_1}+p_{l_2})]}
\hat{P}^{\mu^{\prime}\mu}\hat{P}^{\nu^{\prime}\nu}L^{\dagger}_{\mu^{\prime}\nu^{\prime}}L_{\mu\nu}
\\ \notag
&\times\int\frac{dz_A}{z_A}n(\mathcal{W}_1,X_\perp-b_\perp)\int\frac{dz_B}{z_B}n(\mathcal{W}_2,X_\perp)
\widetilde{C}^*(z_A\bar{n}_A\cdot\mathcal{P}_A,z_B\bar{n}_B\cdot\mathcal{P}_B)\widetilde{C}(z_A\bar{n}_A\cdot \mathcal{P}_A,z_B\bar{n}_B\cdot \mathcal{P}_B)\\ \notag
=
&\int d^2X_\perp\frac{1}{[2(2\pi)^3]^2}\frac{1}{\left|v^{\prime}_{Az}-v^{\prime}_{Bz}\right|}
\int\frac{dz_A}{z_A}n(\mathcal{W}_1,X_\perp-b_\perp)\int\frac{dz_B}{z_B}n(\mathcal{W}_2,X_\perp)\\
\notag
&\times(2\pi)^4\delta(z_A\bar{n}_A\cdot\mathcal{P}_A-\bar{n}_A\cdot p_{l_1}-\bar{n}_A\cdot p_{l_2})\delta(z_B\bar{n}_B\cdot \mathcal{P}_B-n_A\cdot p_{l_1}-n_A\cdot p_{l_2})\\ \notag
&\times\delta^{(2)}[(k_{1\perp}+k_{2\perp})-(p_{l_1\perp}+p_{l_2\perp})]\\ \notag
&\times\widetilde{C}^*(z_A\bar{n}_A\cdot\mathcal{P}_A,z_B\bar{n}_B\cdot\mathcal{P}_B)\widetilde{C}(z_A\bar{n}_A\cdot\mathcal{P}_A,z_B\bar{n}_B\cdot\mathcal{P}_B)
\hat{P}^{\mu^{\prime}\mu}\hat{P}^{\nu^{\prime}\nu}L^{\dagger}_{\mu^{\prime}\nu^{\prime}}L_{\mu\nu},
\end{align}  
where we use the relation $\bar{n}_B=n_A$. The total scattering cross section can be represented as 
\begin{align}
\label{cross}
\sigma_b=&
\int d^2b_\perp\int d^2X_\perp\times\int d\mathcal{W}_1d\mathcal{W}_2 n(\mathcal{W}_1,X_\perp-b_\perp)n(\mathcal{W}_2,X_\perp)\\ \notag
&\times\int dy_{l_1}d^2p_{l_1\perp}dy_{l_2}d^2p_{l_2\perp}\frac{1}{[2(2\pi)^3]^2\cdot\left|v^{\prime}_{Az}-v^{\prime}_{Bz}\right|\cdot (E_AE_B)}\\ \notag
&\times(2\pi)^4\delta(z_A\bar{n}_A\cdot\mathcal{P}_A-\bar{n}_A\cdot p_{l_1}-\bar{n}_A\cdot p_{l_2})\delta(z_B\bar{n}_B\cdot \mathcal{P}_B-n_A\cdot p_{l_1}-n_A\cdot p_{l_2})\\ \notag
&\times\delta^{(2)}[(k_{1\perp}+k_{2\perp})-(p_{l_1\perp}+p_{l_2\perp})]\\ \notag
&\times\widetilde{C}^*(z_A\bar{n}_A\cdot\mathcal{P}_A,z_B\bar{n}_B\cdot \mathcal{P}_B)\widetilde{C}(z_A\bar{n}_A\cdot\mathcal{P}_A,z_B\bar{n}_B\cdot\mathcal{P}_B)
\hat{P}^{\mu^{\prime}\mu}\hat{P}^{\nu^{\prime}\nu}L^{\dagger}_{\mu^{\prime}\nu^{\prime}}L_{\mu\nu}
\\ \notag
=&\int d^2b_\perp\int d^2X_\perp\int d(z_A\bar{n}_A\cdot\mathcal{P}_A)d(z_B\bar{n}_B\cdot\mathcal{P}_B) n(\frac{z_A\bar{n}_A\cdot\mathcal{P}_A}{2},X_\perp-b_\perp)n(\frac{z_B\bar{n}_B\cdot\mathcal{P}_B}{2},X_\perp)\\ \notag
&\times\sigma_{\gamma\gamma\rightarrow l\bar{l}},
\end{align}
where 
\begin{align}
&\sigma_{\gamma\gamma\rightarrow l\bar{l}}\\ \notag
=
&\int\frac{dy_{l_1}d^2p_{l_1\perp}}{2(2\pi)^3}\int\frac{dy_{l_2}d^2p_{l_2\perp}}{2(2\pi)^3}\frac{1}{2s}
(2\pi)^4\delta(z_A\bar{n}_A\cdot\mathcal{P}_A-\bar{n}_A\cdot p_{l_1}-\bar{n}_A\cdot p_{l_2})\\ \notag
&\times\delta(z_Bn_A\cdot\mathcal{P}_B-n_A\cdot p_{l_1}-n_A\cdot p_{l_2})
\delta^{(2)}[(k_{1\perp}+k_{2\perp})-(p_{l_1\perp}+p_{l_2\perp})]\\ \notag
&\times\widetilde{C}^*(z_A\bar{n}_A\cdot\mathcal{P}_A,z_B\bar{n}_B\cdot\mathcal{P}_B)\widetilde{C}(z_A\bar{n}_A\cdot\mathcal{P}_A,z_B\bar{n}_B\cdot\mathcal{P}_B)
\hat{P}^{\mu^{\prime}\mu}\hat{P}^{\nu^{\prime}\nu}L^{\dagger}_{\mu^{\prime}\nu^{\prime}}L_{\mu\nu}.
\end{align}  
The $\sigma_{\gamma\gamma\rightarrow l\bar{l}}$ pertains to the high-energy cross section, which is amenable to perturbation theory calculations and corresponds to the process in which two collinear photons scatter, leading to the production of lepton and antilepton in the final state.

\section{Decoupling transformation of the soft function}
\label{sec:dec}
In the forthcoming section, we will advance our cross-section refinement by incorporating the soft function and decoupling it from the high-energy cross section and photon PDFs. Consequently, It is essential to modify the constraints on four-dimensional momentum to accommodate the contributions from soft photons radiated in the final state and institute a mathematical convolution procedure.

We commence our analysis with Equation \eqref{cross}. Initially, we incorporate soft radiation into Equation \eqref{cross}, followed by the introduction of a convolution product. Subsequently, we employ the standard SCET approach to express the differential cross-section.

In chapter 3, We utilize translation operators to displace the coordinates of the photon field and that of its conjugate field within the non-perturbative matrix element by distances 
$-\frac{x}{2}$ and $\frac{x}{2}$, respectively. During that step, we did not explicitly represent the soft photon field. If we take into account the soft photon field, at this stage, the space-time coordinates of the soft photon field and its conjugate field are also individually translated to $-\frac{x}{2}$ and $\frac{x}{2}$. With the aforementioned analysis, the factorized matrix elements can be explicitly expressed, where $l$ represents the total momentum of soft photons radiated in the final state,
    \begin{align}
        \bra{0}\bar{S}^{\dagger}_2S_1(\frac{x}{2})\ket{l}\bra{l}\bar{S}^{\dagger}_1S_2(-\frac{x}{2})\ket{0}.
    \end{align}
By employing the power counting method, we can expand the space-time coordinates of soft photons,
    \begin{align}
        p_{l_1}+p_{l_2}&\sim \mathcal{O}(1,~1,~\lambda),\\ \notag
        x&\sim\mathcal{O}(1,~1,~\frac{1}{\lambda}).
    \end{align}
Should the transverse momentum of final-state lepton pair be influenced by the momentum of soft photons, the latter's magnitude must be comparable to that of the transverse momentum of lepton pair,
    \begin{align}
        l&\sim\mathcal{O}(\lambda,~\lambda,~\lambda),\\ \notag
        l\cdot x&\sim \mathcal{O}(\lambda,~\lambda,~1),
    \end{align}
So that the contributions of $\bar{n}_A\cdot x$ and $n_A\cdot x$ within soft photons can be neglected as small-order quantities. Subsequently, the soft photon can be expressed as 
    \begin{align}
        \mathcal{S}(x_\perp)=
\mathop{\ooalign{$\sum$\cr$\displaystyle{\int}$\cr}}\limits_{\{l\}}
\bra{0}\bar{S}^{\dagger}_2S_1(\frac{x_\perp}{2})\ket{l}
\bra{l}\bar{S}^{\dagger}_1S_2(-\frac{x_\perp}{2})\ket{0}.
    \end{align}
Let us recall the derivation of the four-dimensional delta function in equation \eqref{cross}, where longitudinal momentum conservation can be obtained by integrating the Fourier factor with respect to $\bar{n}_A\cdot x$ and $n_A\cdot x$. However, incorporating the soft function, which is dependent on $x_\perp$, the integration with respect to $x_\perp$ would not yield transverse momentum conservation.

Building upon the preceding analysis, and incorporating the soft function, the differential scattering cross-section associated with Equation \eqref{cross} is represented as: 
    \begin{align}
        &\frac{d\sigma_b}{d^2b_\perp dy_{l_1}d^2p_{l_{1\perp}}dy_{l_2}d^2p_{l_{2\perp}}}\\ \notag
        =&\frac{1}{[2(2\pi)^3]^2}\frac{1}{2s}\int d(z_A\bar{n}_A\cdot\mathcal{P}_A)
        d(z_B\bar{n}_B\cdot\mathcal{P}_B)\int d^2X_\perp\\ \notag
        &\times n(\frac{1}{2}z_A\bar{n}_A\cdot\mathcal{P}_A,X_\perp-b_\perp)
        n(\frac{1}{2}z_B\bar{n}_B\cdot\mathcal{P}_B,X_\perp)\\ \notag
        &\times(2\pi)^2\delta(z_A\bar{n}_A\cdot\mathcal{P}_A-\bar{n}_A\cdot p_{l_1}-\bar{n}_A\cdot p_{l_2})\delta(z_B\bar{n}_B\cdot\mathcal{P}_B-n_A\cdot p_{l_1}-n_A\cdot p_{l_2})\\ \notag
        &\times\tilde{C}^*(z_A\bar{n}_A\cdot\mathcal{P}_A,z_B\bar{n}_B\cdot\mathcal{P}_B)
        \tilde{C}(z_A\bar{n}_A\cdot \mathcal{P}_A,z_B\bar{n}_B\cdot\mathcal{P}_B)
\hat{P}^{\mu^{\prime}\mu}\hat{P}^{\nu^{\prime}\nu}L^{\dagger}_{\mu^{\prime}\nu^{\prime}}L_{\mu\nu}
\\ \notag
        &\times\int d^2x_\perp e^{i[(k_{1\perp}+k_{2\perp})-(p_{l_{1\perp}}+p_{l_{2\perp}})]\cdot x_\perp}\mathcal{S}(x_\perp),
    \end{align}
Substituting $(k_{1\perp}+k_{2\perp})-(p_{l_{1\perp}}+p_{l_{2\perp}})$ with $l_\perp$ and integrating with respect to $x_\perp$, the Fourier transformation of $\mathcal{S}(x_\perp)$ is expressed as $\tilde{\mathcal{S}}(l_\perp)$.

Next, we define the hard function,
    \begin{align}
        \mathcal{H}(z_A,z_B;p_{l_1},p_{l_2})\equiv
        &\tilde{C}^*(z_A\bar{n}_A\cdot\mathcal{P}_A,z_B\bar{n}_B\cdot\mathcal{P}_B)\tilde{C}(z_A\bar{n}_A\cdot\mathcal{P}_A,z_B\bar{n}_B\cdot\mathcal{P}_B)\\ \notag
&\times\hat{P}^{\mu^{\prime}\mu}\hat{P}^{\nu^{\prime}\nu}L^{\dagger}_{\mu^{\prime}\nu^{\prime}}L_{\mu\nu}.
    \end{align}
Finally, we perform convolution using the following definition
    \begin{align}
        g_\perp=p_{l_{1\perp}}+p_{l_{2\perp}}+l_{\perp},
    \end{align}
By inserting the identity,
    \begin{align}
        \int\frac{d^2g_\perp}{(2\pi)^2}(2\pi)^2\delta^{(2)}(g_\perp-p_{l_{1\perp}}-p_{l_{2\perp}}-l_{\perp})=1,
    \end{align}
the differential cross section can be expressed as 
    \begin{align}
        &\frac{d\sigma_b}{d^2b_\perp d\Gamma_{l_1} d\Gamma_{l_2}}\\ \notag
        =&\frac{1}{2s}\int d(z_A\bar{n}_A\cdot\mathcal{P}_A)d(z_B\bar{n}_B\cdot\mathcal{P}_B)
        (2\pi)^2\delta(z_A\bar{n}_A\cdot\mathcal{P}_A-\bar{n}_A\cdot p_{l_1}-\bar{n}_A\cdot p_{l_2})\\ \notag
        &\times\delta(z_B\bar{n}_B\cdot\mathcal{P}_B-n_A\cdot p_{l_1}-n_A\cdot p_{l_2})\int d^2X_\perp n(\frac{1}{2}z_A\bar{n}_A\cdot\mathcal{P}_A,X_\perp-b_\perp)n(\frac{1}{2}z_B\bar{n}_B\cdot\mathcal{P}_B,X_\perp)\\ \notag
        &\times\int \frac{d^2g_\perp}{(2\pi)^2}(2\pi)^2\delta^{(2)}(g_\perp-p_{l_{1\perp}}-p_{l_{2\perp}}-l_{\perp})
        \mathcal{H}(z_A,z_B;p_{l_1},p_{l_2})\tilde{\mathcal{S}}(l_\perp)\\ \notag
        =&\frac{1}{2s}\int d^2X_\perp\mathcal{T}(z_A,X_\perp-b_\perp)\mathcal{T}(z_B,X_\perp)\otimes\mathcal{H}(z_A,z_B;g_\perp-l_\perp)\otimes\tilde{\mathcal{S}}(l_\perp).
    \end{align}
In accordance with the SCET convention, I substitute $\mathcal{T}$ for $n$ to denote the GTMD function. The integration with respect to the transverse coordinate $X_\perp$ will result in the constraint $q_{A_\perp}=q_{B_\perp}$, and we will demonstrate this operation in the subsequent calculations. For the purposes of facilitating numerical calculations and subsequently comparing theoretical results with experimental outcomes, further variable substitutions are required for the transverse momentum and rapidity of the final-state lepton pairs. We redefine the variables in the phase space,
\begin{align}
    p_{l_{1\perp}}+p_{l_{2\perp}}=q_{l_\perp},~~~\frac{(y_1+y_2)}{2}=y,\\ \notag
    \frac{(p_{l_{1\perp}}-p_{l_{2\perp}})}{2}=\mathcal{P}_{l\perp},~~~y_1-y_2=\Delta y.
\end{align}
It is important to note that in this context, we are dealing with the difference and summation of momenta for the final-state leptons $l_1$ and 
$l_2$. These must be differentiated from the previously defined momenta $q_{A_\perp}=p_{A_\perp}-p^{\prime}_{A_\perp}$ and $\mathcal{P}_{A_\perp}=\frac{1}{2}(p_{A_\perp}+p^{\prime}_{A_\perp})$, which represent the difference and summation of momentum between the wave packet and its complex conjugate wave packet, respectively. Through the redefinition of variables, we obtain 
\begin{align}
    d^2p_{l_{1\perp}}d^2p_{l_{2\perp}}&=d^2q_{l_\perp}d^2\mathcal{P}_{l_\perp},\\ \notag
    dy_1dy_2&=2d\Delta ydy.
\end{align}
Since the final-state leptons are produced in a approximately back-to-back configuration, we establish the relationship $q_{l_\perp}\ll\mathcal{P}_{l_\perp}$. Furthermore, given the small transverse momentum of soft photons, only $q_{l_\perp}$ will experience explicit modifications, while $\mathcal{P}_{l_\perp}$ remains largely unaffected.

To explicitly express the results, I will denote the differential cross section, incorporating soft radiation modification, as $d\sigma^{(1)}_b$, and express the Born-level differential cross section as $d\sigma^{(0)}_b$. And then we have
    \begin{align}
        \frac{d\sigma^{(1)}_b}{d^2b_\perp d\Gamma_{l_1}d\Gamma_{l_2}}
        =\frac{d\sigma^{(0)}_b}{d^2b_{\perp}d\Gamma_{l_1}d\Gamma_{l_2}}\otimes\tilde{\mathcal{S}}(l_\perp),
    \end{align}
where
    \begin{align}
        &\frac{d\sigma^{(0)}_b}{d^2b_\perp d\Gamma_{l_1}d\Gamma_{l_2}}\\ \notag
        \equiv& \frac{1}{2s}\int d(z_A\bar{n}_A\cdot\mathcal{P}_A)d(z_B\bar{n}_B\cdot\mathcal{P}_B)
        (2\pi)^2\delta(z_A\bar{n}_A\cdot\mathcal{P}_A-\bar{n}_A\cdot p_{l_1}-\bar{n}_A\cdot p_{l_2})\\ \notag
        &\times\delta(z_B\bar{n}_B\cdot\mathcal{P}_B-\bar{n}_B\cdot p_{l_1}-\bar{n}_B\cdot p_{l_2})\int d^2X_\perp n(\frac{1}{2}z_A\bar{n}_A\cdot\mathcal{P}_A,X_\perp-b_\perp)n(\frac{1}{2}z_B\bar{n}_B\cdot\mathcal{P}_B,X_\perp)\\ \notag
        &\times\mathcal{H}(z_A,z_B;g_\perp-l_\perp).
    \end{align}
We substitute $d^2\mathcal{P}_{l_\perp}d^2q_{l_\perp}$ for $\prod\limits^{2}\limits_{i=1}d^2p_{l_{i\perp}}$ within $\prod\limits^{2}\limits_{i=1}d\Gamma_{l_i}$, yielding the result expressed as 
    \begin{align}
        &\frac{d\sigma^{(1)}_b(g_\perp)}{d^2b_\perp\prod\limits^{2}\limits_{j=1}dy_{l_j}d^2\mathcal{P}_{l_\perp}d^2g_{\perp}}\\ \notag
       =&\int d^2q_{l_\perp}\delta^{(2)}(g_\perp-q_{l_\perp}-l_\perp)
\frac{d\sigma^{(0)}_b(q_{l_\perp})}{d^2b_\perp\prod\limits^{2}\limits_{j=1}dy_{l_j}d^2\mathcal{P}_{l_\perp}d^2q_{l_\perp}}\tilde{\mathcal{S}}(l_\perp)\\ \notag
=&\int \frac{d^2q_{l_\perp}}{(2\pi)^2}
\frac{d\sigma^{(0)}_b(q_{l_\perp})}{d^2b_\perp\prod\limits^{2}\limits_{j=1}dy_{l_j}d^2\mathcal{P}_{l_\perp}d^2q_{l_\perp}}
\mathop{\ooalign{$\sum$\cr$\displaystyle{\int}$\cr}}\limits_{\{l\}}
\int d^2x_\perp e^{il_\perp\cdot x_\perp}\bra{0}\bar{S}^{\dagger}_2S_1(\frac{x_\perp}{2})\ket{l}\bra{l}\bar{S}^{\dagger}_1S_2(-\frac{x_\perp}{2})\ket{0}\\ \notag
&\times(2\pi)^2\delta^{(2)}(g_\perp-q_{l_\perp}-l_{\perp}).
    \end{align}
Employing the Fourier representation of the delta function, the differential cross section incorporating with soft function can be obtained,
    \begin{align}
        &\frac{d\sigma^{(1)}_b(g_\perp)}{d^2b_\perp\prod\limits^{2}\limits_{j=1}dy_{l_j}d^2\mathcal{P}_{l_\perp}d^2g_{\perp}}\\ \notag
        =&\int d^2 r_\perp e^{ig_\perp\cdot r_\perp} \int \frac{d^2 q_{l_\perp}}{(2\pi)^2} e^{-iq_{l_\perp}\cdot r_\perp} \frac{d\sigma^{(0)}_b(q_{l_\perp})}{d^2b_\perp\prod\limits^2\limits_{j=1}dy_{l_j}d^2\mathcal{P}_{l_\perp}d^2q_{l_\perp}}\\ \notag
        &\times\mathop{\ooalign{$\sum$\cr$\displaystyle{\int}$\cr}}\limits_{l}
        e^{-il_\perp\cdot r_\perp} \times \bigg[\int d^2x_\perp e^{il_\perp\cdot x _\perp} \bra{0}\bar{S}^{\dagger}_2 S_1 (\frac{x_\perp}{2})\ket{l} \bra{l} \bar{S}^{\dagger}_1 S_2 (-\frac{x_\perp}{2}) \ket{0}\bigg]\\ \notag
        =&\int d^2r_\perp e^{ig_\perp\cdot r_\perp} \tilde{\mathcal{S}}(r_\perp)
        \frac{d\tilde{\sigma}^{(0)}_b(r_\perp)}{d^2b_\perp \prod\limits^2\limits_{j=1}dy_{l_j} d^2\mathcal{P}_{l_\perp} d^2 r_\perp}.
        \end{align}
The result precisely matches Equation (8) in paper \cite{Shao:2022stc}.

\section{Angular correlation of transverse momentum in final-state lepton pairs}
\label{sec:ang}
\begin{figure}[ht]
\begin{center}
 %set default line width to 0.75pt        

\begin{tikzpicture}
%uncomment if require: \path (0,515); %set diagram left start at 0, and has height of 515

\draw[postaction={decorate,decoration={markings,mark=at position 0.5 with {\arrow{>}}}}] (-2.0,-0.3) -- (-4.0,-0.3);
\draw [postaction={decorate,decoration={markings,mark=at position 0.5 with {\arrow{>}}}}](-4.0,1.0) -- (-2.0,1.0);
\draw[black, decorate, decoration=snake] (-6.0,-0.3) -- (-4.0,-0.3);
\draw[black, decorate, decoration=snake] (-6.0,1.0) -- (-4.0,1.0);
\draw[postaction={decorate,decoration={markings,mark=at position 0.5 with {\arrow{>}}}}] (-4.0,-0.3) -- (-4.0,1.0);
\node at (-2.5,1.3) {$p_{l_1}$};
\node at (-2.5,-0.6) {$p_{l_2}$};
\node at (-6,1.3) {$k_1$};
\node at (-6,-0.6) {$k_2$};
\node at (-4,-0.6) {$\nu$};
\node at (-4,1.3) {$\mu$};
\draw[->] (-5.4,1.4) -- (-5.0,1.4);
\draw[->] (-5.4,-0.8) -- (-5.0,-0.8);
\draw[->] (-3,-0.1) -- (-2.5,-0.1);

\draw[postaction={decorate,decoration={markings,mark=at position 0.7 with {\arrow{>}}}}](4.0,-0.3) -- (6.0,1.0);
\draw [postaction={decorate,decoration={markings,mark=at position 0.3 with {\arrow{>}}}}](6.0,-0.3) -- (4.0,1.0);
\draw[black, decorate, decoration=snake]
(4.0,-0.3) -- (2.0,-0.3);
\draw[black, decorate, decoration=snake]
(4.0,1.0) -- (2.0,1.0);
\draw[postaction={decorate,decoration={markings,mark=at position 0.5 with {\arrow{>}}}}] (4.0,1.0) -- (4.0,-0.3);
\node at(6,1.3) {$p_{l_1}$};
\node at(6,-0.6){$p_{l_2}$};
\node at(2.5,1.3){$k_1$};
\node at(2.5,-0.6) {$k_2$};
\node at (4,-0.6) {$\nu$};
\node at (4,1.3) {$\mu$};
\draw[->] (4.3,1.0) -- (4.6,0.8);
\draw[->] (2.5,-0.1) -- (3,-0.1);
\end{tikzpicture}
\end{center}
\captionsetup{justification=raggedright,singlelinecheck=false}
\caption{High energy scattering amplitude for lepton pair production in double photon fusion.}
\label{high-energyfeynman}
\end{figure}
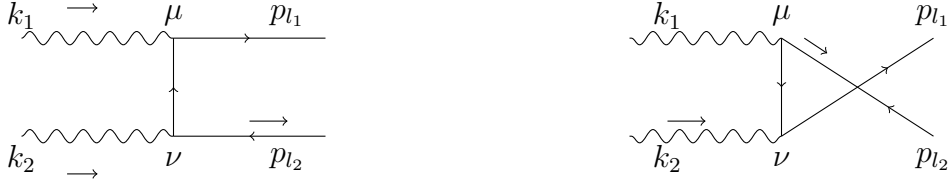

Subsequently, we proceed to calculate the angular correlation of the transverse momentum of the final-state lepton pair, commencing from
\begin{align}
\label{born-level equation}
    \frac{d\sigma^{(0)}_b(p_{l_{1\perp}}+p_{l_{2\perp}})}{\prod\limits^2_{j=1}dy_{l_j}d^2p_{l_{j\perp}}}=
    \int d^2b_{\perp}d^2X_{\perp}\int d\mathcal{W}_1d\mathcal{W}_2 n(\mathcal{W}_1,X_\perp-b_\perp)n(\mathcal{W}_2,X_\perp)
    \frac{d\sigma_{\gamma\gamma\rightarrow l\bar{l}}}{\prod\limits^2_{j=1}dy_{l_j}d^2p_{l_{j\perp}}}.
\end{align}
We represent $|\mathcal{P}_{l_\perp}|$ in a two-dimensional polarization coordinate system as
\begin{align}
d^2\mathcal{P}_{l_\perp}&=|\mathcal{P}_{l_\perp}|d|\mathcal{P}_{l_\perp}|d\phi_{\mathcal{P}_{l\perp}},\\ \notag
    d^2q_{l_\perp}&=|q_{l_\perp}|d|q_{l_\perp}|d\phi_{q_{l\perp}}. 
\end{align}
To establish the connection between $|\mathcal{P}_{l_\perp}|$ and the invariant mass $M$ of the lepton pair, and considering the soft photons momentum negligible under condition $|l|^2\ll M^2$, the relationship between the momentum of lepton pair and $M$ can be expressed by the equation
\begin{align}
    M^2&=(p_{l_1}+p_{l_2})^2=(k_1+k_2)^2=2k_1\cdot k_2\\ \notag
    &=2\times[\frac{1}{2}(\bar{n}\cdot k_1)(n\cdot k_2)+\frac{1}{2}(n\cdot k_1)(\bar{n}\cdot k_2)+k_{1\perp}\cdot k_{2\perp}],
\end{align}
where $p_{l_1}$ and $p_{l_2}$ denote the four-dimensional momenta of the final-state leptons. As previously discussed, both terms $(n\cdot k_1)(\bar{n}\cdot k_2)$ and $k_{1\perp}\cdot k_{2\perp}$ can be safely neglected up to $\mathcal{O}(\lambda^2)$ because they are power-suppressed. Consequently, we obtain the result: 
\begin{align}
    M^2\simeq 4\mathcal{W}_1\mathcal{W}_2.
\end{align}
In UPC, the four-momenta of the equivalent photons are denoted as $k_1=(\mathcal{W}_1,~k_{1\perp},~\frac{\mathcal{W}_1}{v})$ and $k_2=(\mathcal{W}_2,~k_{2\perp},~-\frac{\mathcal{W}_2}{v})$, while the four-momenta of the final-state leptons are represented as $p_{l_1}=(E_1,~p_{l_{1\perp}},~p_{l_{1z}})$ and $p_{l_2}=(E_2,~p_{l_{2\perp}},~p_{l_{2z}})$. Utilizing longitudinal momentum conservation, we can derive two equations
\begin{align}
\mathcal{W}_1+\mathcal{W}_2&=E_1+E_2,\\
\frac{\mathcal{W}_1}{v}-\frac{\mathcal{W}_2}{v}&=p_{l_{1z}}+p_{l_{2z}}.
\end{align}
Upon solving the equation, we have obtained the result 
\begin{align}
    \mathcal{W}_1=\frac{E_1+E_2+vp_{l_{1z}}+vp_{l_{2z}}}{2},~~~\mathcal{W}_2=\frac{E_1+E_2-vp_{l_{1z}}-vp_{l_{2z}}}{2}.
\end{align}
Through the application of the rapidity definition $y=\frac{1}{2}\ln{\frac{E+p_z}{E-p_z}}$, we have derived 
\begin{align}
    4\mathcal{W}_1\mathcal{W}_2\simeq(\mathcal{P}^2_{l_\perp}+m^2)(2+e^{y_1-y_2}+e^{y_2-y_1}),\\
    M^2\simeq(\mathcal{P}^2_{l_\perp}+m^2)(2+e^{y_1-y_2}+e^{y_2-y_1}),
\end{align}
where $m$ denotes the mass of final-state lepton. The variable $|\mathcal{P}_{l_\perp}|$ in the cross section can be replaced by the invariant mass $M$,
\begin{align}
    |\mathcal{P}_{l_\perp}|d|\mathcal{P}_{l_\perp}|=\frac{M}{2+e^{\Delta y}+e^{-\Delta y}}dM.
\end{align}
The measurement of the phase space can be redefined using the invariant mass $M$, and the angles $\phi_{q_{l\perp}}$ and $\phi_{\mathcal{P}_{l\perp}}$ between the transverse momentum of the lepton pair,
\begin{align}
    d^2q_{l_\perp} d^2\mathcal{P}_{l_\perp} dy_1dy_2=\frac{2M}{2+e^{\Delta y}+e^{-\Delta y}}dMd\phi_{\mathcal{P}_{l\perp}}|q_{l_\perp}|d|q_{l_\perp}|d\phi_{q_{l\perp}}d\Delta ydy.
\end{align}
Thus,
\begin{align}
    \frac{d\sigma^{(0)}_b}{\prod\limits^2_{j=1}dy_{l_j}d^2p_{l_{j\perp}}}
    ~~~\rightarrow~~~\frac{d\sigma^{(0)}_b}{dMd\phi_{\mathcal{P}_{l\perp}}|q_{l_\perp}|d|q_{l_\perp}|d\phi_{q_{l\perp}}d\Delta y dy}\times\frac{2+e^{\Delta y}+e^{-\Delta y}}{2M}.
\end{align}
Next, we perform calculations for the high-energy scattering process $\gamma+\gamma\rightarrow l+\bar{l}$, as illustrated in Fig\ref{high-energyfeynman}.
The results can be readily read using Feynman rules
\begin{align}
    \mathcal{M}_1+\mathcal{M}_2=\bar{u}(p_{l_1})[\gamma_{\mu}\frac{\slashed{p}_{l_1}-\slashed{k}_1+m}{(p_{l_1}-k_1)^2-m^2}\gamma_\nu+\gamma_\nu\frac{\slashed{p}_{l_1}-\slashed{k}_2+m}{(p_{l_1}-k_2)^2-m^2}\gamma_\mu]\nu(p_{l_2}),
\end{align}
where coefficient $e^2$ is absorbed in prefactors. The photon distribution function, which depends on the impact parameter, can be expressed as 
  \begin{align}
      n(\mathcal{W}_i,b_\perp)
      =\frac{4z^2\alpha}{\mathcal{W}_i}\left|\int \frac{d^2k_{i_\perp}}{(2\pi)^2}k_{i_\perp}\frac{F(k^2_{i_\perp}+\frac{\mathcal{W}^2_i}{\gamma^2})}{k^2_{i_\perp}+\frac{\mathcal{W}^2_i}{\gamma^2}}e^{-ib_\perp\cdot k_{i_\perp}}\right|^2,
  \end{align}
subsequently leading to the outcome 
\begin{align}
    &\int d^2X_\perp n(\mathcal{W}_1,X_\perp-b_\perp)n(\mathcal{W}_2,X_\perp)\\ \notag
    =&\int d^2X_\perp \frac{4z^2\alpha}{\mathcal{W}_1}\left|\int \frac{d^2k_{1\perp}}{(2\pi)^2}k_{1\perp}\frac{F(k^2_{1\perp}+\frac{\mathcal{W}^2_1}{\gamma^2})}{k_{1\perp}^2+\frac{\mathcal{W}^2_1}{\gamma^2}}e^{-i(X_\perp-b_\perp)\cdot k_{1\perp}}\right|^2\\ \notag
    &\times \frac{4z^2\alpha}{\mathcal{W}_2}\left|\int \frac{d^2k_{2\perp}}{(2\pi)^2}k_{2\perp}\frac{F(k_{2\perp}^2+\frac{\mathcal{W}^2_2}{\gamma^2})}{k^2_{2\perp}+\frac{\mathcal{W}^2_2}{\gamma^2}}e^{-iX_\perp\cdot k_{2\perp}}\right|^2\\ \notag
    =&\frac{[4z^2\alpha]^2}{\mathcal{W}_1\mathcal{W}_2}
    \int\frac{d^2k_{1\perp}}{(2\pi)^2}\frac{d^2k^{\prime}_{1\perp}}{(2\pi)^2}\frac{d^2k_{2\perp}}{(2\pi)^2}\frac{d^2k^{\prime}_{2\perp}}{(2\pi)^2}
    k_{1\perp}k_{1\perp}^{\prime}k_{2\perp}k_{2\perp}^{\prime}
    \frac{F(k^2_{1\perp}+\frac{\mathcal{W}^2_1}{\gamma^2})}{k^2_{1\perp}+\frac{\mathcal{W}_1^2}{\gamma^2}}
    \frac{F(k^{\prime 2}_{1\perp}+\frac{\mathcal{W}^2_1}{\gamma^2})}{k^{\prime 2}_{1\perp}+\frac{\mathcal{W}^2_1}{\gamma^2}}
    \frac{F(k^2_{2\perp}+\frac{\mathcal{W}^2_2}{\gamma^2})}{k^2_{2\perp}+\frac{\mathcal{W}^2_2}{\gamma^2}}\\ \notag
    &\times\frac{F(k^{\prime 2}_{2\perp}+\frac{\mathcal{W}^2_2}{\gamma^2})}{k^{\prime 2}_{2\perp}+\frac{\mathcal{W}^2_2}{\gamma^2}}
    \int d^2X_\perp e^{iX_\perp\cdot(k^{\prime}_{1\perp}-k_{1\perp}+k_{2\perp}^{\prime}-k_{2\perp})}e^{ib_\perp\cdot(k^{\prime}_{1\perp}-k_{1\perp})}\\ \notag
    =&\frac{[4z^2\alpha]^2}{\mathcal{W}_1\mathcal{W}_2}
    \int \frac{d^2k_{1\perp}}{(2\pi)^2}\frac{d^2k^{\prime}_{1\perp}}{(2\pi)^2}
    \frac{d^2k_{2\perp}}{(2\pi)^2}\frac{d^2k^{\prime}_{2\perp}}{(2\pi)^2}
    k_{1\perp}k^{\prime}_{1\perp}k_{2\perp}k^{\prime}_{2\perp}
    \frac{F(k^2_{1\perp}+\frac{\mathcal{W}^2_1}{\gamma^2})}{k^2_{1\perp}+\frac{\mathcal{W}^2_1}{\gamma^2}}
    \frac{F(k^{\prime 2}_{1\perp}+\frac{\mathcal{W}^2_1}{\gamma^2})}{k^{\prime 2}_{1\perp}+\frac{\mathcal{W}^2_1}{\gamma^2}}
    \frac{F(k^2_{2\perp}+\frac{\mathcal{W}^2_2}{\gamma^2})}{k^2_{2\perp}+\frac{\mathcal{W}^2_2}{\gamma^2}}\\ \notag
    &\times\frac{F(k^{\prime 2}_{2\perp}+\frac{\mathcal{W}^2_2}{\gamma^2})}{k^{\prime 2}_{2\perp}+\frac{\mathcal{W}^2_2}{\gamma^2}}
     e^{ib_\perp\cdot(k^{\prime}_{1\perp}-k_{1\perp})}(2\pi)^2\delta^{(2)}(k^{\prime}_{1\perp}-k_{1\perp}+k^{\prime}_{2\perp}-k_{2\perp}).
\end{align}
We introduce photon distribution and amplitude into the Born-level scattering cross-section \eqref{born-level equation},
\begin{align}
\label{(5)}
    \frac{d\sigma^{(0)}_b(p_{l_{1\perp}}+p_{l_{2\perp}})}{\prod\limits^2_{j=1}dy_{l_j}d^2p_{l_{j\perp}}d^2b_\perp}
    &=\frac{e^4}{[2(2\pi)^3]^2}\frac{1}{2s}
    \int d\mathcal{W}_1d\mathcal{W}_2(2\pi)^4\delta^{(4)}[k_1+k_2-(p_{l_1}+p_{l_2})]\\ \notag
    &\times\frac{[4z^2\alpha]^2}{\mathcal{W}_1\mathcal{W}_2}\int\frac{d^2k_{1\perp}}{(2\pi)^2}\frac{d^2k^{\prime}_{1\perp}}{(2\pi)^2}\frac{d^2k_{2\perp}}{(2\pi)^2}\frac{d^2k^{\prime}_{2\perp}}{(2\pi)^2}k_{1\perp}k^{\prime}_{1\perp}k_{2\perp}k^{\prime}_{2\perp}\\ \notag
    &\times\frac{F(k^2_{1\perp}+\frac{\mathcal{W}^2_1}{\gamma^2})}{k^2_{1\perp}+\frac{\mathcal{W}^2_1}{\gamma^2}}
 \frac{F(k^{\prime 2}_{1\perp}+\frac{\mathcal{W}^2_1}{\gamma^2})}{k^{\prime 2}_{1\perp}+\frac{\mathcal{W}^2_1}{\gamma^2}}
    \frac{F(k^2_{2\perp}+\frac{\mathcal{W}^2_2}{\gamma^2})}{k^2_{2\perp}+\frac{\mathcal{W}^2_2}{\gamma^2}}
    \frac{F(k^{\prime 2}_{2\perp}+\frac{\mathcal{W}^2_2}{\gamma^2})}{k_{2\perp}^{\prime 2}+\frac{\mathcal{W}^2_2}{\gamma^2}}\\ \notag
    &\times e^{ib_\perp\cdot(k^{\prime}_{1\perp}-k_{1\perp})}(2\pi)^2\delta^{(2)}(k^{\prime}_{1\perp}-k_{1\perp}+k^{\prime}_{2\perp}-k_{2\perp})
    \sum\limits_{Spin}\left|\mathcal{M}_1+\mathcal{M}_2\right|^2.
\end{align}
We shall compare the result with equation 2.40 in reference \cite{Krauss:1997vr}, where $q_\perp=k_{1\perp}-k^{\prime}_{1\perp}=k^{\prime}_{2\perp}-k_{2\perp}$. By inserting the identity 
\begin{align}
      \int d^2q_\perp\delta^{(2)}(k_{1\perp}-k^{\prime}_{1\perp}-q_\perp)=1,
\end{align}
we can derive the result denoted as 
\begin{align}
       &\int\frac{d^2k^{\prime}_{1\perp}}{(2\pi)^2}\frac{d^2k^{\prime}_{2\perp}}{(2\pi)^2}
       (2\pi)^2\delta^{(2)}[k_{1\perp}+k_{2\perp}-(p_{l_{1\perp}}+p_{l_{2\perp}})](2\pi)^2\delta^{(2)}(k^{\prime}_{1\perp}-k_{1\perp}+k^{\prime}_{2\perp}-k_{2\perp})\notag\\ \notag
       &\times\int\frac{d^2q_\perp}{(2\pi)^2}(2\pi)^2\delta^{(2)}(k_{1\perp}-k^{\prime}_{1\perp}-q_\perp)\\ 
       =&\int\frac{d^2q_\perp}{(2\pi)^2}(2\pi)^2\delta^{(2)}[k_{1\perp}+k_{2\perp}-(p_{l_{1\perp}}+p_{l_{2\perp}})].
\end{align} 
Reinsert the outcome into \eqref{(5)}. It is imperative to recognize the distinction between $q_{l_\perp}$ and $q_\perp$. Subsequently, the result can be derived
\begin{align}
      \frac{d\sigma^{(0)}_b(p_{l_{1\perp}}+p_{l_{2\perp}})}{\prod\limits^2_{j=1}dy_{l_j}d^2p_{l_{j\perp}}d^2b_\perp}
      =&\frac{e^4}{[2(2\pi)^3]^2}\frac{1}{2s}\frac{[4z^2\alpha]^2}{\mathcal{W}_1\mathcal{W}_2}
      \int d\mathcal{W}_1d\mathcal{W}_2\int\frac{d^2k_{1\perp}}{(2\pi)^2}\frac{d^2k_{2\perp}}{(2\pi)^2}\frac{d^2q_\perp}{(2\pi)^2}\\ \notag
      &\times(2\pi)^4\delta^{(4)}[k_1+k_2-(p_{l_1}+p_{l_2})]k_{1\perp}k^{\prime}_{1\perp}k_{2\perp}k^{\prime}_{2\perp}\\ \notag
      &\times\frac{F(k^2_{1\perp}+\frac{\mathcal{W}^2_1}{\gamma^2})}{k^2_{1\perp}+\frac{\mathcal{W}^2_1}{\gamma^2}}
 \frac{F(k^{\prime 2}_{1\perp}+\frac{\mathcal{W}^2_1}{\gamma^2})}{k^{\prime 2}_{1\perp}+\frac{\mathcal{W}^2_1}{\gamma^2}}
    \frac{F(k^2_{2\perp}+\frac{\mathcal{W}^2_2}{\gamma^2})}{k^2_{2\perp}+\frac{\mathcal{W}^2_2}{\gamma^2}}
    \frac{F(k^{\prime 2}_{2\perp}+\frac{\mathcal{W}^2_2}{\gamma^2})}{k_{2\perp}^{\prime 2}+\frac{\mathcal{W}^2_2}{\gamma^2}}\times e^{ib_\perp\cdot(k^{\prime}_{1\perp}-k_{1\perp})}\\ \notag
    &\times\sum\limits_{Spin}\left|\mathcal{M}_1+\mathcal{M}_2\right|^2,
 \end{align}
which precisely matches equation 2.40 in reference \cite{Krauss:1997vr}. 

Our investigation commences with the utilization of Equation \eqref{(5)} for the computation of angular correlations, followed by $|\mathcal{M}_1+\mathcal{M}_2|$ introduced into Equation. We analyze the angular correlation in the transverse momentum of the lepton pair by decomposing it into an angular-independent term, a term proportional to $\cos{2(\phi_{q_{l\perp}}-\phi_{\mathcal{P}_{l\perp}})}$, and a term proportional to $\cos{4(\phi_{q_{l\perp}}-\phi_{\mathcal{P}_{l\perp}})}$, leveraging the orthogonality of trigonometric functions,
\begin{align}
\label{(6)}
    &\frac{d\sigma^{(0)}_b}{d^2q_{l_\perp}d^2\mathcal{P}_{l_\perp}dy_1dy_2}\\ \notag
    =&\frac{2z^4\alpha^4_e}{(4\mathcal{W}_1\mathcal{W}_2)^2\pi^4}
    \int\frac{d^2b_\perp}{(2\pi)^2}\int d^2k_{1\perp}
    d^2k^{\prime}_{1\perp}d^2k_{2\perp}d^2k^{\prime}_{2\perp}\delta^{(2)}[k_{1\perp}+k_{2\perp}-(p_{l_{1\perp}}+p_{l_{2\perp}})]\\ \notag
    &\times\delta^{(2)}(k^{\prime}_{1\perp}+k^{\prime}_{2\perp}-k_{1\perp}-k_{2\perp})e^{ib_\perp\cdot(k^{\prime}_{1\perp}-k_{1\perp})}k_{1\perp}k^{\prime}_{1\perp}k_{2\perp}k^{\prime}_{2\perp}\\ \notag
    &\times\frac{F(k^2_{1\perp}+\frac{\mathcal{W}^2_1}{\gamma^2})}{k^2_{1\perp}+\frac{\mathcal{W}^2_1}{\gamma^2}}
 \frac{F(k^{\prime 2}_{1\perp}+\frac{\mathcal{W}^2_1}{\gamma^2})}{k^{\prime 2}_{1\perp}+\frac{\mathcal{W}^2_1}{\gamma^2}}
    \frac{F(k^2_{2\perp}+\frac{\mathcal{W}^2_2}{\gamma^2})}{k^2_{2\perp}+\frac{\mathcal{W}^2_2}{\gamma^2}}
    \frac{F(k^{\prime 2}_{2\perp}+\frac{\mathcal{W}^2_2}{\gamma^2})}{k_{2\perp}^{\prime 2}+\frac{\mathcal{W}^2_2}{\gamma^2}}\\ \notag
    &\times\sum\limits_{Spin}\left|\mathcal{M}_1+\mathcal{M}_2\right|^2\\ \notag
    \equiv&\frac{2z^4\alpha^4_e}{(4\mathcal{W}_1\mathcal{W}_2)^2\pi^4}\int
    \frac{d^2b_\perp}{(2\pi)^2}[A_0+A_2\,\cos{2(\phi_{q_{l_\perp}}-\phi_{\mathcal{P}_{l_\perp}})}
    +A_4\,\cos{4(\phi_{q_{l_\perp}}-\phi_{\mathcal{P}_{l_\perp}})}].
\end{align}
Given the relationships among the variables, the variables in equation \eqref{(6)} are replaced with the invariant mass denoted as $M$, along with the introduction of new rapidity variables $\Delta y$ and $y$,
\begin{align}
    &\frac{d\sigma^{(0)}_b}{dMd|q_{l_\perp}|^2d\Delta ydyd\phi_{\mathcal{P}_{l\perp}}d\phi_{q_{l\perp}}}\\ \notag
    =&\frac{M}{2+e^{\Delta y}+e^{-\Delta y}}\frac{d\sigma^{(0)}_b}{d^2q_{l_\perp}d^2\mathcal{P}_{l_\perp}dy_1dy_2}\\ \notag
    =&\frac{M}{2+e^{\Delta y}+e^{-\Delta y}}
    \frac{2z^4\alpha^4_e}{(4\mathcal{W}_1\mathcal{W}_2)^2\pi^4}
 \int\frac{d^2b_\perp}{(2\pi)^2}d^2k_{1\perp}
    d^2k^{\prime}_{1\perp}d^2k_{2\perp}d^2k^{\prime}_{2\perp}\\ \notag
    &\times\delta^{(2)}[k_{1\perp}+k_{2\perp}-(p_{l_{1\perp}}+p_{l_{2\perp}})]\\ \notag
    &\times\delta^{(2)}(k^{\prime}_{1\perp}+k^{\prime}_{2\perp}-k_{1\perp}-k_{2\perp})e^{ib_\perp\cdot(k^{\prime}_{1\perp}-k_{1\perp})}k_{1\perp}k^{\prime}_{1\perp}k_{2\perp}k^{\prime}_{2\perp}\\ \notag
    &\times\frac{F(k^2_{1\perp}+\frac{\mathcal{W}^2_1}{\gamma^2})}{k^2_{1\perp}+\frac{\mathcal{W}^2_1}{\gamma^2}}
 \frac{F(k^{\prime 2}_{1\perp}+\frac{\mathcal{W}^2_1}{\gamma^2})}{k^{\prime 2}_{1\perp}+\frac{\mathcal{W}^2_1}{\gamma^2}}
    \frac{F(k^2_{2\perp}+\frac{\mathcal{W}^2_2}{\gamma^2})}{k^2_{2\perp}+\frac{\mathcal{W}^2_2}{\gamma^2}}
    \frac{F(k^{\prime 2}_{2\perp}+\frac{\mathcal{W}^2_2}{\gamma^2})}{k^{\prime 2}_{2\perp}+\frac{\mathcal{W}^2_2}{\gamma^2}}\\ \notag
    &\times\sum\limits_{Spin}\left|\mathcal{M}_1+\mathcal{M}_2\right|^2. \notag
\end{align}
Integrate with respect to the azimuthal angles between the transverse momentum of lepton pairs to obtain the structure of $A_0$,
\begin{align}
          &\int \frac{d\sigma^{(0)}_b}{dMd|q_{l_\perp}|^2d\Delta ydyd\phi_{P_{l\perp}}d\phi_{q_{l\perp}}}d\phi_{\mathcal{P}_{l\perp}}d\phi_{q_{l\perp}}\\ \notag
          =&\frac{M}{2+e^{\Delta y}+e^{-\Delta y}}
          \frac{2z^4\alpha^4_e}{(4\mathcal{W}_1\mathcal{W}_2)^2\pi^4}\int \frac{d^2b_\perp}{(2\pi)^2}\int d\phi_{\mathcal{P}_{l\perp}}d\phi_{q_{l\perp}}A_0\\ \notag
          =&\frac{M}{2+e^{\Delta y}+e^{-\Delta y}}\frac{2z^4\alpha^4_e}{(4\mathcal{W}_1\mathcal{W}_2)^2\pi^4}\times 4\pi^2\int\frac{d^2b_\perp}{(2\pi)^2}A_0.
\end{align}
$A_0$ is structured in a manner \cite{li2020impact} that
\begin{align}
      A_0=&\int d^2k_{1\perp}d^2k_{2\perp}d^2k^{\prime}_{1\perp}d^2k^{\prime}_{2\perp}
      \delta^{(2)}(k_{1\perp}+k_{2\perp}-q_{l_\perp})\delta^{(2)}(k^{\prime}_{1\perp}+k^{\prime}_{2\perp}-q_{l_\perp})\\ \notag
      &\times e^{i(k_{1\perp}-k^{\prime}_{1\perp})\cdot b_\perp}k_{1\perp}k^{\prime}_{1\perp}k_{2\perp}k^{\prime}_{2\perp}\\ \notag
 &\times\frac{F(k^2_{1\perp}+\frac{\mathcal{W}^2_1}{\gamma^2})}{k^2_{1\perp}+\frac{\mathcal{W}^2_1}{\gamma^2}}
 \frac{F(k^{\prime 2}_{1\perp}+\frac{\mathcal{W}^2_1}{\gamma^2})}{k^{\prime 2}_{1\perp}+\frac{\mathcal{W}^2_1}{\gamma^2}}
    \frac{F(k^2_{2\perp}+\frac{\mathcal{W}^2_2}{\gamma^2})}{k^2_{2\perp}+\frac{\mathcal{W}^2_2}{\gamma^2}}
    \frac{F(k^{\prime 2}_{2\perp}+\frac{\mathcal{W}^2_2}{\gamma^2})}{k^{\prime 2}_{2\perp}+\frac{\mathcal{W}^2_2}{\gamma^2}}\\  \notag
    &\times\frac{1}{(\mathcal{P}^2_{l_\perp}+m^2)^2}[-2m^4\cos{(\phi_{k_{1\perp}}+\phi_{k^{\prime}_{1\perp}}-\phi_{k_{2\perp}}-\phi_{k^{\prime}_{2\perp}})}\\ \notag
    &+m^2(M^2-2m^2)\cos{(\phi_{k_{1\perp}}-\phi_{k^{\prime}_{1\perp}}-\phi_{k_{2\perp}}+\phi_{k^{\prime}_{2\perp}})}\\ \notag
    &+\mathcal{P}^2_{l_\perp}(M^2-2\mathcal{P}^2_{l_\perp})\cos{(\phi_{k_{1\perp}}-\phi_{k^{\prime}_{1\perp}}+\phi_{k_{2\perp}}-\phi_{k^{\prime}_{2\perp}})}].
\end{align}
After integrating with respect to $\phi_{k_{j\perp}}$ and $\phi_{k^{\prime}_{j\perp}}$, where $j=1,2$, $A_2$ is independent of the azimuthal angles $\phi_{q_{l\perp}}$ and $\phi_{\mathcal{P}_{l\perp}}$. By utilizing the orthogonality of trigonometric functions, the coefficient $A_2$ can be obtained,
\begin{align}
    &\int\frac{d\sigma^{(0)}_b}{dMd|q_{l_\perp}|^2d\Delta ydyd\phi_{\mathcal{P}_{l\perp}}d\phi_{q_{l\perp}}}[2\cos{2(\phi_{q_{l\perp}}-\phi_{\mathcal{P}_{l\perp}})}]d\phi_{\mathcal{P}_{l\perp}}d\phi_{q_{l\perp}}\\ \notag
    =&\frac{M}{2+e^{\Delta y}+e^{-\Delta y}}\frac{2z^4\alpha^4_e}{(4\mathcal{W}_1\mathcal{W}_2)^2\pi^4}\int\frac{d^2b_\perp}{(2\pi)^2}
    \int d\phi_{\mathcal{P}_{l\perp}}d\phi_{q_{l\perp}}2[\cos{2(\phi_{q_{l\perp}}-\phi_{\mathcal{P}_{l\perp}})}]^2A_2\\ \notag
    =&\frac{M}{2+e^{\Delta y}+e^{-\Delta y}}\frac{2z^4\alpha^4_e}{(4\mathcal{W}_1\mathcal{W}_2)^2\pi^4}\times 4\pi^2\int\frac{d^2b_\perp}{(2\pi)^2}A_2.
\end{align}
$A_2$ is structured in a manner \cite{li2020impact} that
\begin{align}
      A_2=&\int d^2k_{1\perp}d^2k_{2\perp}d^2k^{\prime}_{1\perp}d^2k^{\prime}_{2\perp}
      \delta^{(2)}(k_{1\perp}+k_{2\perp}-q_{l_\perp})\delta^{(2)}(k^{\prime}_{1\perp}+k^{\prime}_{2\perp}-q_{l_\perp})\\ \notag 
      &\times e^{i(k_{1\perp}-k^{\prime}_{1\perp})\cdot b_{\perp}}k_{1\perp}k^{\prime}_{1\perp}k_{2\perp}k^{\prime}_{2\perp}\\ \notag
    &\times\frac{F(k^2_{1\perp}+\frac{\mathcal{W}^2_1}{\gamma^2})}{k^2_{1\perp}+\frac{\mathcal{W}^2_1}{\gamma^2}}
 \frac{F(k^{\prime 2}_{1\perp}+\frac{\mathcal{W}^2_1}{\gamma^2})}{k^{\prime 2}_{1\perp}+\frac{\mathcal{W}^2_1}{\gamma^2}}
    \frac{F(k^2_{2\perp}+\frac{\mathcal{W}^2_2}{\gamma^2})}{k^2_{2\perp}+\frac{\mathcal{W}^2_2}{\gamma^2}}
    \frac{F(k^{\prime 2}_{2\perp}+\frac{\mathcal{W}^2_2}{\gamma^2})}{k^{\prime 2}_{2\perp}+\frac{\mathcal{W}^2_2}{\gamma^2}}\\ \notag
    &\times \frac{2\mathcal{P}^2_{l_\perp}}{M(\mathcal{P}^2_{l_\perp}+m^2)^2}4Mm^2\cos{(\phi_{k_{1\perp}}-\phi_{k_{2\perp}})}\cos{(\phi_{k^{\prime}_{1\perp}}+\phi_{k^{\prime}_{2\perp}}-2\phi_{q_{l\perp}})}.
\end{align}
Similar to $A_4$, we obtain 
\begin{align}
    &\int\frac{d\sigma^{(0)}_b}{dMd|q_{l_\perp}|^2d\Delta ydyd\phi_{\mathcal{P}_{l\perp}}d\phi_{q_{l\perp}}}[2\cos{4(\phi_{q_{l\perp}}-\phi_{\mathcal{P}_{l\perp}})}]d\phi_{\mathcal{P}_{l\perp}}d\phi_{q_{l\perp}}\\ \notag
    =&\frac{M}{2+e^{\Delta y}+e^{-\Delta y}}\frac{2z^4\alpha^4_e}{(4\mathcal{W}_1\mathcal{W}_2)^2\pi^4}\int\frac{d^2b_\perp}{(2\pi)^2}
    \int d\phi_{\mathcal{P}_{l\perp}}d\phi_{q_{l\perp}}2[\cos{4(\phi_{q_{l\perp}}-\phi_{\mathcal{P}_{l\perp}})}]^2A_4\\ \notag
    =&\frac{M}{2+e^{\Delta y}+e^{-\Delta y}}\frac{2z^4\alpha^4_e}{(4\mathcal{W}_1\mathcal{W}_2)^2\pi^4}\times 4\pi^2\int\frac{d^2b_\perp}{(2\pi)^2}A_4.
\end{align}  
$A_4$ is structured in a way \cite{li2020impact} that 
\begin{align}
      A_4=&\int d^2k_{1\perp}d^2k_{2\perp}d^2k^{\prime}_{1\perp}d^2k^{\prime}_{2\perp}
      \delta^{(2)}(k_{1\perp}+k_{2\perp}-q_{l_\perp})\delta^{(2)}(k^{\prime}_{1\perp}+k^{\prime}_{2\perp}-q_{l_\perp})\\ \notag
      &\times e^{i(k_{1\perp}-k^{\prime}_{1\perp})\cdot b_{\perp}}k_{1\perp}k^{\prime}_{1\perp}k_{2\perp}k^{\prime}_{2\perp}\\ \notag
    &\times\frac{F(k^2_{1\perp}+\frac{\mathcal{W}^2_1}{\gamma^2})}{k^2_{1\perp}+\frac{\mathcal{W}^2_1}{\gamma^2}}
 \frac{F(k^{\prime 2}_{1\perp}+\frac{\mathcal{W}^2_1}{\gamma^2})}{k^{\prime 2}_{1\perp}+\frac{\mathcal{W}^2_1}{\gamma^2}}
    \frac{F(k^2_{2\perp}+\frac{\mathcal{W}^2_2}{\gamma^2})}{k^2_{2\perp}+\frac{\mathcal{W}^2_2}{\gamma^2}}
    \frac{F(k^{\prime 2}_{2\perp}+\frac{\mathcal{W}^2_2}{\gamma^2})}{k^{\prime 2}_{2\perp}+\frac{\mathcal{W}^2_2}{\gamma^2}}\\ \notag
    &\times \frac{-2\mathcal{P}^4_{l_\perp}}{(\mathcal{P}^2_{l_\perp}+m^2)^2}\cos{(\phi_{k_{1\perp}}+\phi_{k^{\prime}_{1\perp}}+\phi_{k_{2\perp}}+\phi_{k^{\prime}_{2\perp}}-4\phi_{q_{l\perp}})}.
\end{align}
In order to align theoretical calculations with experimental results, it is necessary to represent the photon frequencies $\mathcal{W}_1$ and $\mathcal{W}_2$, which are variables within 
$A_0$, $A_2$, and $A_4$, in terms of the Mandelstam variable $s$, rapidity, lepton mass  $m$, and invariant mass $M$. By integrating the previously discussed constraints of energy conservation and the conservation of the z-component of momentum, we can derive the results 
\begin{align}
      &\mathcal{W}_1=\frac{p^0_{l_1}+p^0_{l_2}+p^3_{l_1}+p^3_{l_2}}{2},~~~\mathcal{W}_2=\frac{p^0_{l_1}+p^{0}_{l_2}-p^3_{l_1}-p^3_{l_2}}{2},\\
      &z_1=\frac{2\mathcal{W}_1}{\sqrt{s}},~~~z_2=\frac{2\mathcal{W}_2}{\sqrt{s}}.
\end{align}
The energy fraction of a photon is denoted as
\begin{align}
      &z_1=\frac{M(e^{y_1}+e^{y_2})}{\sqrt{s(2+e^{\Delta y}+e^{-\Delta y})}},~~~
      z_2=\frac{M(e^{-y_1}+e^{-y_2})}{\sqrt{s(2+e^{\Delta y}+e^{-\Delta y})}}.
\end{align}
Using the obtained results, we replace $\mathcal{W}_1$ and $\mathcal{W}_2$, within the nuclear charge form factor, with $z_1M_p$ and $z_2M_p$, where $M_p$ represents the mass of the proton. The ultimate result is expressed as 
\begin{align}
    \frac{F(k^2_{i\perp}+\frac{\mathcal{W}^2_i}{\gamma^2})}{k^2_{i\perp}+\frac{\mathcal{W}^2_i}{\gamma^2}}~~~\rightarrow~~~\frac{F(k^2_{i\perp}+z^2_iM^2_p)}{k^2_{i\perp}+z^2_iM^2_p}.
\end{align}
with $i=1,2$, and $F(|k|)$ is given by the STARlight generator.

Next, we calculate the angular average at the Born level, and represent the term that is independent of angle as
\begin{align}
    \frac{d\sigma^{(0)}_b}{d|q_{l_\perp}|^2}=&\frac{8z^4\alpha^4_e}{(4\mathcal{W}_1\mathcal{W}_2)^2\pi^2}\int_0^{y_{max}-y_{min}}d\Delta y
    \int^{M_{max}}_{M_{min}}dM\int^{y_{max}-\frac{\Delta y}{2}}_{y_{min}+\frac{\Delta y}{2}}dy
    (\int^{+\infty}_0\frac{d^2b_\perp}{(2\pi)^2}-\int^{2R}_0\frac{d^2b_\perp}{(2\pi)^2})\notag\\
    &\times\frac{M}{2+e^{\Delta y}+e^{-\Delta y}}A_0.
\end{align}
In UPC, it is necessary to integrate over the impact parameter from $2R$ to $+\infty$. Nonetheless, to simplify numerical calculations, we divide the integration interval into $(0,~+\infty)$ and $(0,~2R)$. When conducting integration over the interval $(0,~2R)$, we employ the following formula,
\begin{align}
    \int^{2R}_0e^{ib_\perp\cdot(k^{\prime}_{1\perp}-k_{1\perp})}d^2b_{\perp}=&\int^{2R}_0e^{i|b_\perp|\cdot|k^{\prime}_{1\perp}-k_{1\perp}|\cdot \cos{(\phi_{b_\perp}-\phi_{k^{\prime}_{1\perp}-k_{1\perp}})}}|b_\perp|d|b_\perp|d\phi_{b_\perp}\\ \notag
    =&\int^{2R}_0|b_\perp|d|b_\perp|2\pi J_0(|b_\perp|\cdot|k^{\prime}_{1\perp}-k_{1\perp}|)\\ \notag
    =&4\pi R\frac{J_1(2|k^{\prime}_{1\perp}-k_{1\perp}|\cdot R)}{|k^{\prime}_{1\perp}-k_{1\perp}|}.
\end{align}
Similarly, for the average of $\cos{2(\phi_{q_{l\perp}}-\phi_{\mathcal{P}_{l\perp}})}$, the result is 
\begin{align}
    \frac{d\sigma^{(0)}_b}{d|q_{l_{\perp}}|^2}\bigg|_{2\cos{2[\phi_{q_{l\perp}}-\phi_{\mathcal{P}_{l\perp}}]}}
    =&\frac{8z^4\alpha^4_e}{(4\mathcal{W}_1\mathcal{W}_2)^2\pi^2}
    \int^{y_{max}-y_{min}}_0d\Delta y\int^{M_{max}}_{M_{min}}dM\\ \notag
    &\times\int^{y_{max}-\frac{\Delta y}{2}}_{y_{min}+\frac{\Delta y}{2}}dy\left[\int^{+\infty}_0\frac{d^2b_\perp}{(2\pi)^2}-\int^{2R}_0\frac{d^2b_{\perp}}{(2\pi)^2}\right]
    \frac{M}{2+e^{\Delta y}+e^{-\Delta y}}A_2.
\end{align}
And for the average of $\cos{4(\phi_{q_{l\perp}}-\phi_{\mathcal{P}_{l\perp}})}$, the result is
\begin{align}
    \frac{d\sigma^{(0)}_b}{d|q_{l_\perp}|^2}\bigg|_{2\cos{4[\phi_{q_{l\perp}}-\phi_{\mathcal{P}_{l\perp}}]}}=&
    \frac{8z^4\alpha^4_e}{(4\mathcal{W}_1\mathcal{W}_2)^2\pi^2}
    \int^{y_{max}-y_{min}}_{0}d\Delta y\int^{M_{max}}_{M_{min}}dM\\ \notag
    &\times\int^{y_{max}-\frac{\Delta y}{2}}_{y_{min}+\frac{\Delta y}{2}}dy\left[\int^{+\infty}_{0}\frac{d^2b_{\perp}}{(2\pi)^2}-\int^{2R}_{0}\frac{d^2b_{\perp}}{(2\pi)^2}\right]\frac{M}{2+e^{\Delta y}+e^{-\Delta y}}A_4.
\end{align}
With normalization, the final result can be obtained
\begin{align}
      &\langle 2\cos{2[\phi_{q_{l\perp}}-\phi_{\mathcal{P}_{l\perp}}]}\rangle
      =\frac{\frac{d\sigma^{(0)}_b}{d|q_{l\perp}|^2}\bigg|_{2\cos{2[\phi_{q_{l\perp}}-\phi_{\mathcal{P}_{l\perp}}]}}}{\frac{d\sigma^{(0)}_b}{d|q_{l\perp}|^2}},\\
      &\langle 2\cos{4[\phi_{q_{l\perp}}-\phi_{\mathcal{P}_{l\perp}}]}\rangle
      =\frac{\frac{d\sigma^{(0)}_b}{d|q_{l\perp}|^2}\bigg|_{2\cos{4[\phi_{q_{l\perp}}-\phi_{\mathcal{P}_{l\perp}}]}}}{\frac{d\sigma^{(0)}_b}{d|q_{l\perp}|^2}}.
\end{align}

\section{Angular correlation with soft radiation corrections}
\label{sec:sof}

We will calculate the angular correlation with the inclusion of soft radiation corrections, starting from the following equation
\begin{align}
\label{(7)}
       \frac{d\sigma^{(1)}_b(g_\perp)}{d^2g_\perp d^2\mathcal{P}_{l\perp}\prod\limits^2_{j=1}dy_{l_j}}=\int d^2r_\perp e^{ig_\perp\cdot r_\perp}\tilde{\mathcal{S}}(r_\perp)\int \frac{d^2 q_{l_\perp}}{(2\pi)^2}e^{-iq_{l_\perp}\cdot r_\perp}
       \frac{d\sigma^{(0)}_b(q_{l_\perp})}{\prod\limits^2_{j=1}dy_{l_j}d^2\mathcal{P}_{l_\perp}d^2q_{l_\perp}}.
\end{align}
We initiate the calculation from \eqref{(6)}, and the soft function in coordinate space \cite{Shao:2022stc} is denoted by 
\begin{align}
      \tilde{\mathcal{S}}(r_\perp)=e^{-Sud(r_\perp)}
      [1+\frac{\alpha_e}{4\pi}(s_{11}+s_{22}+2s_{12})].
\end{align}
The one-loop Sudakov factor is expressed as 
  \begin{align}
      &Sud(r_\perp)=\frac{\alpha_e}{\pi}\ln{(\frac{\mathcal{P}^2_{l_\perp}}{\mu^2_r})}(1-\frac{1+\beta^2}{2\beta}\ln\frac{1-\beta}{1+\beta}),\\
      &\mu_r=\frac{2e^{-\gamma_E}}{|r_\perp|},\notag
  \end{align}
where $\gamma_E$ is Euler constant and the variable $s_{ij}$ is the product of various eikonal factors, given by
\begin{align}
      s_{11}=&s_{22}=\frac{-4c_r}{\sqrt{c^2_r+1}}\ln{(\sqrt{c^2_r+1}+c_r)},\\
      s_{12}=&-\frac{1+\beta^2}{2\beta}sign(c_r)
      [L_{\xi}[\xi(c_r,\alpha_r),\alpha_r]-L_\xi[\xi(-c_r,\alpha_r),\alpha_r]],
\end{align}
where 
\begin{align}
      &c_r=\frac{\mathcal{P}_{l_\perp}cos\phi_r}{m},~~~\beta=\sqrt{1-\frac{4m^2}{M^2}},\notag\\ \notag
      &\alpha_r=\frac{2\mathcal{P}^2_{l_\perp}cos^2 \phi_r}{-m^2+\mathcal{P}^2_{l_\perp}+(m^2+\mathcal{P}_{l_\perp}^2)cosh(y_1-y_2)},\\ \notag
      &\xi(a,b)=(a+\sqrt{1+a^2})(a+\sqrt{a^2+b}),\\ \notag
      &L_{\xi}(a,b)=2[-Li_2\left(\frac{a+b}{b-1}\right)+Li_2(-a)+\ln{(a+b)}\ln{(1-b)}]-\ln^2{\left(\frac{a}{a+b}\right)}+\frac{1}{2}\ln^2\left(\frac{a(a+1)}{a+b}\right).
\end{align}
Furthermore, $Sud(r_\perp)$ only depends on $|\mathcal{P}_{l_\perp}|^2$, $|r_\perp|$, and $\beta$, exhibiting no angular dependence. However, $s_{11}$, $s_{12}$, and $s_{22}$ are all angular-dependent. For this reason, we need to decompose the angular dependence of $s_{11}+s_{22}+2s_{12}$ into orthogonal trigonometric terms: an angle-independent term, and terms dependent on $\cos{2\phi_r}$ and $\cos{4\phi_r}$. Furthermore, $s_{12}$ exhibits a singularity at $\alpha_r=1$, necessitating the extraction of this singular point prior to numerical calculation. By decomposing the soft function using a set of orthogonal trigonometric functions and convoluting the result with the Born-level cross section, coupled with trigonometric reduction, we can derive the observables: the angle-independent term, $\langle \cos{2\phi_r}\rangle$, and $\langle\cos{4\phi_r}\rangle$. Next, we will provide a detailed calculation.
\begin{align}
    &\mathcal{S}\equiv\frac{\alpha_e}{4\pi}(s_{11}+s_{22}+2s_{12})
    =S_0+S_2 \cos{2(\phi_{r_\perp}-\phi_{\mathcal{P}_{l\perp}})}+
    S_4 \cos{4(\phi_{r_\perp}-\phi_{\mathcal{P}_{l\perp}})},\\
    &\int^{2\pi}_0S_0d\phi_{r_\perp}=\int^{2\pi}_0\frac{\alpha_e}{4\pi}(s_{11}+s_{22}+2s_{12})d\phi_{r_\perp},\\
   &\int^{2\pi}_0 S_2\cos^2{2(\phi_{r_\perp}-\phi_{\mathcal{P}_{l\perp}})}d\phi_{r_\perp}=
   \int^{2\pi}_0\frac{\alpha_e}{4\pi}(s_{11}+s_{22}+2s_{12})
   \cos{2(\phi_{r_\perp}-\phi_{\mathcal{P}_{l\perp}})}d\phi_{r_\perp},\\
   &\int^{2\pi}_0S_4\cos^2{4(\phi_{r_\perp}-\phi_{\mathcal{P}_{l\perp}})}d\phi_{r_\perp}=\int^{2\pi}_0\frac{\alpha_e}{4\pi}(s_{11}+s_{22}+2s_{12})\cos{4(\phi_{r_\perp}-\phi_{\mathcal{P}_{l\perp}})}d\phi_{r_\perp}.
\end{align}
From the above calculation, we obtain the results 
\begin{align}
      S_0=&\frac{1}{2\pi}
      \int^{2\pi}_0\frac{\alpha_e}{4\pi}(s_{11}+s_{22}+2s_{12})d\phi_{r_\perp},\\
      S_2=&\frac{1}{\pi}\int^{2\pi}_0\frac{\alpha_e}{4\pi}(s_{11}+s_{22}+2s_{12})\cos{2(\phi_{r_\perp}-\phi_{\mathcal{P}_{l\perp}})}d\phi_{r_\perp},\\
      S_4=&\frac{1}{\pi}\int^{2\pi}_0\frac{\alpha_e}{4\pi}
      (s_{11}+s_{22}+2s_{12})\cos{4(\phi_{r_\perp}-\phi_{\mathcal{P}_{l\perp}})}d\phi_{r_\perp},
\end{align}
followed by the derivation
\begin{align}
      \tilde{\mathcal{S}}(r_\perp)=&
      e^{-Sud(r_\perp)}[1+\frac{\alpha_e}{4\pi}(s_{11}+s_{22}+2s_{12})]\\ \notag
      =&e^{-Sud(r_\perp)}[1+S_0+S_2\cos{2(\phi_{r_\perp}-\phi_{\mathcal{P}_{l\perp}})}+S_4\cos{4(\phi_{r_\perp}-\phi_{\mathcal{P}_{l\perp}})}].
\end{align}
In the calculation of equation \eqref{(6)}, we decompose the Born-level differential cross section into orthogonal trigonometric functions, 
\begin{align}
&\frac{d\sigma^{(0)}_b}{d^2q_{l_\perp}d^2\mathcal{P}_{l_\perp}dy_1dy_2}\\ \notag
=&\frac{2z^4\alpha^4_e}{(4\mathcal{W}_1\mathcal{W}_2)^2\pi^4}
\int\frac{d^2b_\perp}{(2\pi)^2}[A_0+A_2\cos{2(\phi_{q_{l\perp}}-\phi_{\mathcal{P}_{l\perp}})}+A_4\cos{4(\phi_{q_{l\perp}}-\phi_{\mathcal{P}_{l\perp}})}].
\end{align}
Define a set of new coefficients,
\begin{align}
    &\frac{d\sigma^{(0)}_b}{d|q_{l_\perp}|^2d\Delta y dM d\phi_{\mathcal{P}_{l\perp}} d\phi_{q_{l\perp}}}\\ \notag
    =&
    \frac{M}{2+e^{\Delta y}+e^{-\Delta y}}\frac{2z^4\alpha_e^4}{(4\mathcal{W}_1\mathcal{W}_2)^2\pi^4}
    \int^{y_{max}-\frac{\Delta y}{2}}_{y_{min}+\frac{\Delta y}{2}}dy\int^{+\infty}_{2R}\frac{d^2b_\perp}{(2\pi)^2}[A_0
    +A_2\cos{2(\phi_{q_{l\perp}}-\phi_{\mathcal{P}_{l\perp}})}\\ \notag
    &+A_4\cos{4(\phi_{q_{l\perp}}-\phi_{\mathcal{P}_{l\perp}})}]\\ \notag
    =&B_0+B_2\cos{2(\phi_{q_{l\perp}}-\phi_{\mathcal{P}_{l\perp}})}+B_4\cos{4(\phi_{q_{l\perp}}-\phi_{\mathcal{P}_{l\perp}})},
\end{align}
where
\begin{align}
    B_0\equiv&\frac{M}{2+e^{\Delta y}+e^{-\Delta y}}\frac{2z^4\alpha_e^4}{(4\mathcal{W}_1\mathcal{W}_2)^2\pi^4}\int^{y_{max}-\frac{\Delta y}{2}}_{y_{min}+\frac{\Delta y}{2}}dy\int^{+\infty}_{2R}\frac{d^2b_\perp}{(2\pi)^2}A_0,\\
    B_2\equiv&\frac{M}{2+e^{\Delta y}+e^{-\Delta y}}
    \frac{2z^4\alpha_e^4}{(4\mathcal{W}_1\mathcal{W}_2)^2\pi^4}
    \int^{y_{max}-\frac{\Delta y}{2}}_{y_{min}+\frac{\Delta y}{2}} dy\int^{+\infty}_{2R}\frac{d^2b_\perp}{(2\pi)^2}A_2,\\
    B_4\equiv&\frac{M}{2+e^{\Delta y}+e^{-\Delta y}}\frac{2z^4\alpha_e^4}{(4\mathcal{W}_1\mathcal{W}_2)^2\pi^4}
    \int^{y_{max}-\frac{\Delta y}{2}}_{y_{min}+\frac{\Delta y}{2}}dy 
    \int^{+\infty}_{2R}\frac{d^2b_\perp}{(2\pi)^2}A_4.
\end{align}
Considering the substantial modifications to $q_{l_\perp}$ by the transverse momentum of soft radiation photons, we integrate the soft function into the differential cross section via convolution with $q_{l_\perp}$. By introducing the Born-level cross section and the angular-decomposed soft function into equation \eqref{(7)}, we obtain
\begin{align}
\label{born-soft}
    &\frac{d\sigma^{(1)}_b(g_\perp)}{dMd|g_\perp|^2d\Delta yd\phi_{\mathcal{P}_{l\perp}}d\phi_{g_{\perp}}}\\ \notag
    =&
    \int d^2r_\perp e^{ig_\perp\cdot r_\perp}\tilde{\mathcal{S}}(r_\perp)
    \int \frac{d^2q_{l_\perp}}{(2\pi)^2}e^{-iq_{l_\perp}\cdot r_\perp}
    \frac{d\sigma^{(0)}_b(q_{l_\perp})}{dMd|q_{l_\perp}|^2d\Delta yd\phi_{\mathcal{P}_{l_\perp}}d\phi_{q_{l\perp}}}\\ \notag
    =&\int d^2r_\perp e^{ig_\perp\cdot r_\perp}
    {e^{-Sud(r_\perp)}[1+S_0+S_2 \cos{2(\phi_{r_\perp}-\phi_{\mathcal{P}_{l\perp}})}+S_4\cos{4(\phi_{r_\perp}-\phi_{\mathcal{P}_{l\perp}})}]}\\ \notag
    &\times\int \frac{d^2q_{l_\perp}}{(2\pi)^2} e^{-iq_{l_\perp}\cdot r_\perp}
    [B_0+B_2\cos{2(\phi_{q_{l\perp}}-\phi_{\mathcal{P}_{l\perp}})}
    +B_4\cos{4(\phi_{q_{l\perp}}-\phi_{\mathcal{P}_{l\perp}})}].
\end{align}
Using the formula
\begin{align}
    &\int|q_{l_\perp}|d|q_{l_\perp}|\int d\phi_{q_{l\perp}}
    e^{-i|q_{l_\perp}|\cdot|r_\perp|\cos{(\phi_{q_{l\perp}}-\phi_{r_\perp})}}B_0
    =\int|q_{l_\perp}|d|q_{l_\perp}|
    2\pi J_0(|q_{l_\perp}|\cdot|r_\perp|)B_0,\\
    &\int |q_{l_\perp}|d|q_{l_\perp}|\int d\phi_{q_{l_\perp}} e^{-i|q_{l_\perp}|\cdot|r_\perp|\cos{(\phi_{q_{l\perp}}-\phi_{r_\perp})}}B_2 \cos{2(\phi_{q_{l\perp}}-\phi_{\mathcal{P}_{l\perp}})}\\ \notag
    =&\int|q_{l_\perp}|d|q_{l_\perp}|
    (-2\pi)J_2(|q_{l_\perp}|\cdot|r_\perp|)
    \cos{2(\phi_{\mathcal{P}_{l\perp}}-\phi_{r_\perp})}B_2,\\ 
    &\int|q_{l_\perp}|d|q_{l_\perp}|\int d\phi_{q_{l\perp}}
    e^{-i|q_{l_\perp}\cdot|r_\perp|\cos{(\phi_{q_{l\perp}}-\phi_{r_\perp})}}B_4 \cos{4(\phi_{q_{l\perp}}-\phi_{\mathcal{P}_{l\perp}})}\\ \notag
    =&\int|q_{l_\perp}|d|q_{l_\perp}|B_4\times2\pi J_4(|q_{l_\perp}|\cdot|r_\perp|)\cos{4(\phi_{r_\perp}-\phi_{\mathcal{P}_{l\perp}})},
\end{align}
equation \eqref{born-soft} can be rearranged into
\begin{align}
    &\frac{d\sigma^{(1)}_b(g_\perp)}{dMd|g_\perp|^2d\Delta yd\phi_{\mathcal{P}_{l\perp}}d\phi_{g_{\perp}}}\\ \notag
    =&\int \frac{d^2r_\perp}{(2\pi)^2} e^{ig_\perp\cdot r_\perp}e^{-Sud(r_\perp)}[1+S_0+S_2\cos{2(\phi_{r_\perp}-\phi_{\mathcal{P}_{l\perp}})}+S_4\cos{4(\phi_{r_\perp}-\phi_{\mathcal{P}_{l\perp}})}]\\ \notag
    &\times\int 2\pi|q_{l_\perp}|d|q_{l_\perp}|
    [J_0(|q_{l_\perp}|\cdot|r_\perp|)\cdot B_0
    -J_2(|q_{l_\perp}|\cdot|r_\perp|)\cos{2(\phi_{\mathcal{P}_{l\perp}}-\phi_{r_\perp})}B_2\\ \notag
    &+J_4(|q_{l_\perp}|\cdot|r_\perp|)
    \cos{4(\phi_{r_\perp}-\phi_{\mathcal{P}_{l\perp}})}B_4]\\ \notag
    =&\int \frac{d^2r_\perp}{(2\pi)^2} e^{ig_\perp\cdot r_\perp}e^{-Sud(r_\perp)}
    \int2\pi|q_{l_\perp}|d|q_{l_\perp}|\bigg\{(1+S_0)B_0J_0(|q_{l_\perp}|\cdot|r_\perp|)\\ \notag
    &-\big[(1+S_0)B_2J_2(|q_{l_\perp}|\cdot|r_\perp|)-S_2B_0J_0(|q_{l_\perp}|\cdot|r_\perp|)\big]\cos{2(\phi_{r_\perp}-\phi_{\mathcal{P}_{l\perp}})}\\ \notag
 &-S_2B_2J_2(|q_{l_\perp}|\cdot|r_\perp|)\cos^2{2(\phi_{\mathcal{P}_{l\perp}}-\phi_{r_\perp})}\\ \notag
 &-\big[S_4B_2J_2(|q_{l_\perp}|\cdot|r_\perp|)-S_2B_4J_4(|q_{l_\perp}|\cdot|r_\perp|)\big]\cos{2(\phi_{r_\perp}-\phi_{\mathcal{P}_{l\perp}})}\cos{4(\phi_{r_\perp}-\phi_{\mathcal{P}_{l\perp}})}\\ \notag
 &+\big[(1+S_0)J_4(|q_{l_\perp}|\cdot|r_\perp|)B_4+S_4B_0J_0(|q_{l_\perp}|\cdot|r_\perp|)\big]\cos{4(\phi_{r_\perp}-\phi_{\mathcal{P}_{l\perp}})}\\ \notag
 &+S_4B_4J_4(|q_{l_\perp}|\cdot|r_\perp|)\cos^2{4(\phi_{r_\perp}-\phi_{\mathcal{P}_{l\perp}})}
    \bigg\}.
\end{align}
By employing trigonometric reduction 
\begin{align}
    &\cos^2{2(\phi_{\mathcal{P}_{l\perp}}-\phi_{r_\perp})}
    =\frac{1}{2}[1+\cos{4(\phi_{\mathcal{P}_{l\perp}}-\phi_{r_\perp})}],\\
    &\cos{2(\phi_{r_\perp}-\phi_{\mathcal{P}_{l\perp}})}
    \cos{4(\phi_{r_\perp}-\phi_{\mathcal{P}_{l\perp}})}
    =\frac{1}{2}[\cos{2(\phi_{r_\perp}-\phi_{\mathcal{P}_{l\perp}})}+\cos{6(\phi_{r_\perp}-\phi_{\mathcal{P}_{l\perp}})}],\\
    &\cos^2{4(\phi_{\mathcal{P}_{l\perp}}-\phi_{r_\perp})}
    =\frac{1}{2}[1+\cos{8(\phi_{\mathcal{P}_{l\perp}}-\phi_{r_\perp})}],
\end{align}
and neglecting higher-order terms, the result can be expressed as
\begin{align}
   &\frac{d\sigma^{(1)}_b(g_\perp)}{dMd|g_\perp|^2d\Delta yd\phi_{\mathcal{P}_{l\perp}}d\phi_{g_{\perp}}}\\ \notag
   =&\int \frac{d^2r_\perp}{(2\pi)^2} e^{ig_\perp\cdot r_\perp}e^{-Sud(|r_\perp|)}\int2\pi|q_{l_\perp}|d|q_{l_\perp}|\bigg\{
   \big[(1+S_0)B_0J_0(|q_{l_\perp}|\cdot|r_\perp|)-\frac{1}{2}S_2B_2J_2(|q_{l_\perp}|\cdot|r_\perp|)\\ \notag
   &+\frac{1}{2}S_4B_4J_4(|q_{l_\perp}|\cdot|r_\perp|)\big]
   +\big[S_2B_0J_0(|q_{l_\perp}|\cdot|r_\perp|)-(1+S_0)B_2J_2(|q_{l_\perp}|\cdot|r_\perp|)+\frac{1}{2}S_2B_4J_4(|q_{l_\perp}|\cdot|r_\perp|)\\ \notag
   &-\frac{1}{2}S_4B_2J_2(|q_{l_\perp}|\cdot|r_\perp|)\big]\cos{2(\phi_{\mathcal{P}_{l\perp}}-\phi_{r_\perp})}\\ \notag
&+\big[(1+S_0)B_4J_4(|q_{l\perp}|\cdot|r_\perp|)+S_4B_0J_0(|q_{l_\perp}|\cdot|r_\perp|)-\frac{1}{2}S_2B_2J_2(|q_{l_\perp}|\cdot|r_\perp|)\big]\cos{4(\phi_{\mathcal{P}_{l\perp}}-\phi_{r_\perp})}\bigg\}.
\end{align}
Upon integrating with respect to angle $\phi_{r_\perp}$, we can obtain the final result for the scattering cross-section,
\begin{align}
    &\frac{d\sigma^{(1)}_b(g_\perp)}{dMd|g_\perp|^2d\Delta y d\phi_{\mathcal{P}_{l\perp}}d\phi_{g_{\perp}}}\\ \notag
    =&\int\frac{ |r_\perp|d|r_\perp|}{(2\pi)^2}e^{-Sud(|r_\perp|)}\int d\phi_{r_\perp} e^{i|g_\perp|\cdot|r_\perp|\cos{(\phi_{g_\perp}-\phi_{r_\perp})}}
    \int 2\pi|q_{l_\perp}|d|q_{l_\perp}|\\ \notag
    &\times\bigg\{\big[(1+S_0)B_0J_0-\frac{1}{2}S_2B_2J_2+\frac{1}{2}S_4B_4J_4\big]\\ \notag
    &+\big[S_2B_0J_0-(1+S_0)B_2J_2+\frac{1}{2}S_2B_4J_4-\frac{1}{2}S_4B_2J_2\big]\cos{2(
    \phi_{\mathcal{P}_{l\perp}}-\phi_{r_\perp})}\\ \notag
    &+\big[(1+S_0)B_4J_4+S_4B_0J_0-\frac{1}{2}S_2B_2J_2\big]\cos{4(\phi_{\mathcal{P}_{l\perp}}-\phi_{r_\perp})}\bigg\}\\ \notag
    =&\int|r_\perp|d|r_\perp| e^{-Sud(|r_\perp|)}\int|q_{l_\perp}|d|q_{l_\perp}|\\ \notag
    &\times\bigg\{\big[(1+S_0)B_0J_0(|q_{l_\perp}|\cdot|r_\perp|)-\frac{1}{2}S_2B_2J_2(|q_{l_\perp}|\cdot|r_\perp|)+\frac{1}{2}S_4B_4J_4(|q_{l_\perp}|\cdot|r_\perp|)\big]J_0(|g_\perp|\cdot|r_\perp|)\\ \notag
    &-\big[S_2B_0J_0(|q_{l_\perp}|\cdot|r_\perp|)-(1+S_0)B_2J_2(|q_{l_\perp}|\cdot|r_\perp|)+\frac{1}{2}S_2B_4J_4(|q_{l_\perp}|\cdot|r_\perp|)-\frac{1}{2}S_4B_2J_2(|q_{l_\perp}|\cdot|r_\perp|)\big]\\ \notag
    &\times J_2(|g_\perp|\cdot|r_\perp|)\cos{2(\phi_{\mathcal{P}_{l\perp}}-\phi_{g_\perp})}\\ \notag
&+\big[(1+S_0)B_4J_4(|q_{l_\perp}|\cdot|r_\perp|)+S_4B_0J_0(|q_{l_\perp}|\cdot|r_\perp|)-\frac{1}{2}S_2B_2J_2(|q_{l_\perp}|\cdot|r_\perp|)\big]\\ \notag
&\times J_4(|g_\perp|\cdot|r_\perp|)
    \cos{4(\phi_{\mathcal{P}_{l\perp}}-\phi_{g_\perp})}\bigg\}\\ \notag
    \equiv& C_0+C_2\cos{2(\phi_{\mathcal{P}_{l\perp}}-\phi_{g_\perp})}+C_4\cos{4(\phi_{\mathcal{P}_{l\perp}}-\phi_{g_\perp})}.
\end{align}
$C_0$, $C_2$, and $C_4$ are coefficients of 1, $\cos{2(\phi_{\mathcal{P}_{l\perp}}-\phi_{g_\perp})}$, and $\cos{4(\phi_{\mathcal{P}_{l\perp}}-\phi_{g_\perp})}$. It is important to note that the variables associated with the outermost Bessel functions $J_0$, $J_2$, and $J_4$ are $|g_\perp|\cdot|r_\perp|$, whereas the variables for the Bessel functions inside square brackets are $|q_{l_\perp}|\cdot|r_\perp|$. The subsequent calculation process closely resembles the Born-level calculation,
\begin{align}
    \frac{d\sigma^{(1)}_b(g_\perp)}{dMd|g_\perp|^2d\Delta yd\phi_{\mathcal{P}_\perp}d\phi_{g_{\perp}}}
    =C_0+C_2\cos{2(\phi_{\mathcal{P}_{l\perp}}-\phi_{g_{\perp}})}+C_4\cos{4(\phi_{\mathcal{P}_{l\perp}}-\phi_{g_\perp})}.
\end{align}
For the angle-independent term, $\int \frac{d\sigma}{d\mathcal{P}.\mathcal{S}.}\cos{\left[2 \Delta\phi\right ]}\,d\mathcal{P}.\mathcal{S}.$ term, and $\int\frac{d\sigma}{d\mathcal{P}.\mathcal{S}.}\cos{\left[ 4\Delta\phi \right]}\,d\mathcal{P}.\mathcal{S}.$ term, the result can be expressed separately as 
\begin{align}
    &\frac{d\sigma^{(1)}_b(g_\perp)}{d|g_\perp|^2}=\int dMd\Delta y\int^{2\pi}_0d\phi_{\mathcal{P}_\perp}d\phi_{g_\perp}C_0=4\pi^2\int dMd\Delta yC_0,\\
    &\frac{d\sigma^{(1)}_b(g_\perp)}{d|g_\perp|^2}\bigg|_{2\cos{2(\phi_{\mathcal{P}_{l\perp}}-\phi_{g_\perp})}}=\int dMd\Delta y\int^{2\pi}_{0}d\phi_{\mathcal{P}_\perp}d\phi_{g_\perp} 2\cos^2{2(\phi_{\mathcal{P}_{l\perp}}-\phi_{g_\perp})}C_2=4\pi^2\int dMd\Delta y C_2,\\
    &\frac{d\sigma^{(1)}_b(g_\perp)}{d|g_\perp|^2}\bigg|_{2\cos{4(\phi_{\mathcal{P}_{l\perp}}-\phi_{g_\perp})}}=\int dMd\Delta y\int^{2\pi}_{0}d\phi_{\mathcal{P}_\perp}d\phi_{g_\perp}2\cos^2{4(\phi_{\mathcal{P}_\perp}-\phi_{g_\perp})}C_4=4\pi^4\int dMd\Delta yC_4.
\end{align}
Through normalization, the average value of angular correlations is expressible as
\begin{align}
    \langle 2\cos{2(\phi_{\mathcal{P}_{l\perp}}-\phi_{g_\perp})}\rangle=
    \frac{\int dMd\Delta y~C_2}{\int dMd\Delta y~C_0},~~~\langle 2\cos{4(\phi_{\mathcal{P}_{l\perp}}-\phi_{g_\perp})}\rangle=\frac{\int dMd\Delta y ~C_4}{\int dMd\Delta y ~C_0}.
\end{align}
In QED, the soft function exhibits exponential behavior due to the nature of QED field,
\begin{align}
    \tilde{\mathcal{S}}(r_\perp)=e^{-Sud(r_\perp)+\frac{\alpha_e}{4\pi}(s_{11}+s_{22}+2s_{12})}.
\end{align}
Using the same calculation, when defining 
\begin{align}
  \tilde{\mathcal{S}}(r_\perp)=S_0+S_2\cos{2(\phi_{r_\perp}-\phi_{\mathcal{P}_{l\perp}})}+S_4\cos{4(\phi_{r_\perp}-\phi_{\mathcal{P}_{l\perp}})},
\end{align}
and taking advantage of the orthogonal properties of trigonometric functions, one can derive values for $S_0$, $S_2$, and $S_4$. Subsequently, through further calculation, the final result $\frac{d\sigma^{(all)}_b(g_\perp)}{d|g_\perp|^2}$, $\frac{d\sigma^{(all)}_b(g_\perp)}{d|g_\perp|^2}\bigg|_{2\cos{2(\phi_{\mathcal{P}_{l\perp}}-\phi_{g_\perp})}}$, $\frac{d\sigma^{(all)}_b(g_\perp)}{d|g_\perp|^2}\bigg|_{2\cos{4(\phi_{\mathcal{P}_{l\perp}}-\phi_{g_\perp})}}$emerges, encompassing modifications of all orders. Due to the minimal transverse momentum of soft photons from final-state lepton radiation within the UPC, the effects of modifications at the next-to-next-leading order are negligible. However, the method is readily generalizable. In certain QED processes, the contributions from two-loop and even three-loop calculations become significant, and the exponential results become observable.

\section{Numerical results for muon-muon and tau-tau}
\label{sec:Num}
Below, we present the numerical results. Figure \ref{fig:total} shows the outcomes from  UPCs in 200 GeV Au-Au collisions at the RHIC. The figure consists of three subfigures illustrating dimuon production under various kinematic conditions.
\begin{figure}[ht]
    \centering
    \begin{minipage}{0.3\textwidth}
        \centering
        \includegraphics[width=\linewidth]{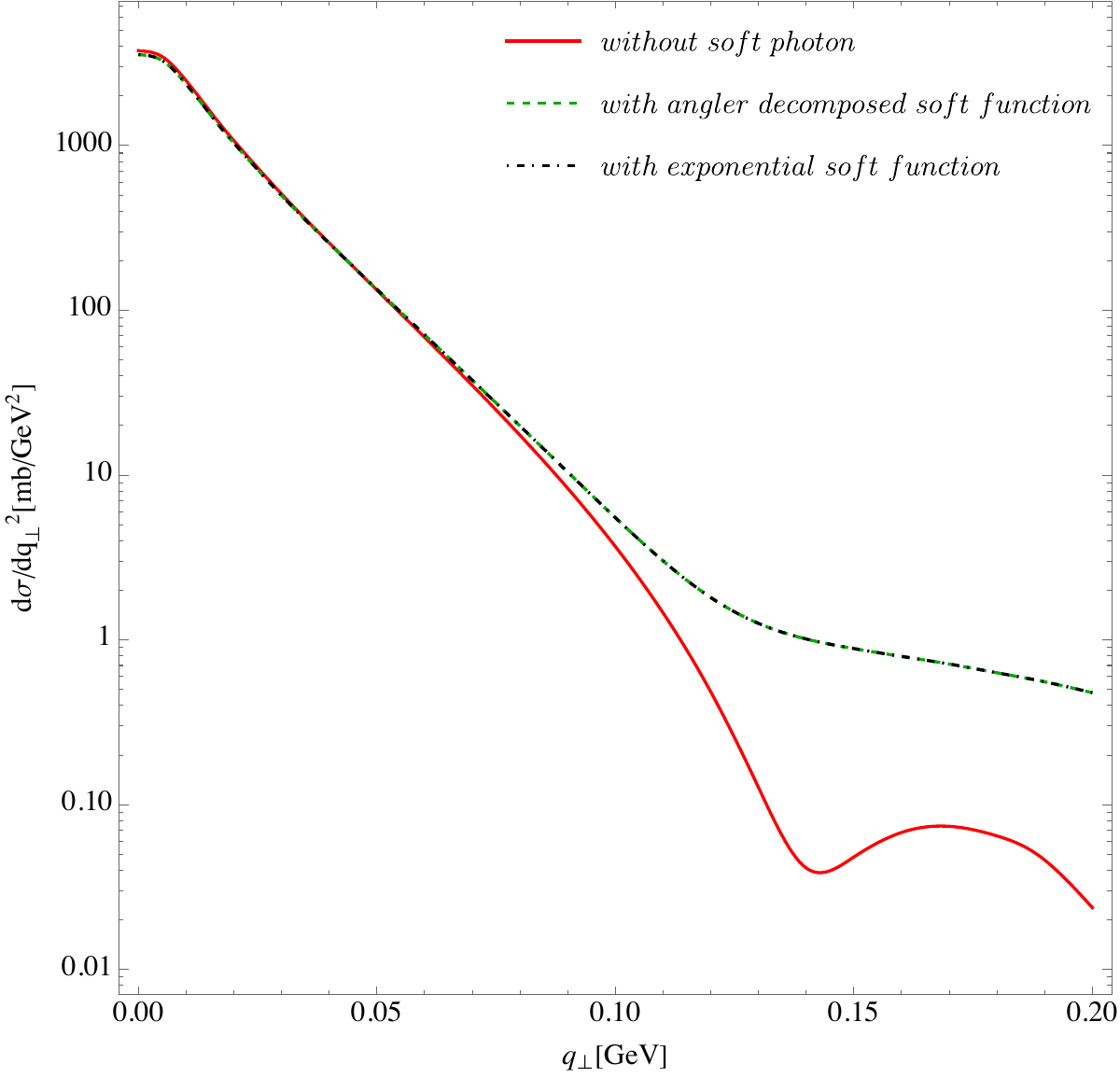}
        \label{fig:subfig1}
    \end{minipage}
    \hspace{0.0\textwidth} % 添加水平间距
    \begin{minipage}{0.3\textwidth}
        \centering
        \includegraphics[width=\linewidth]{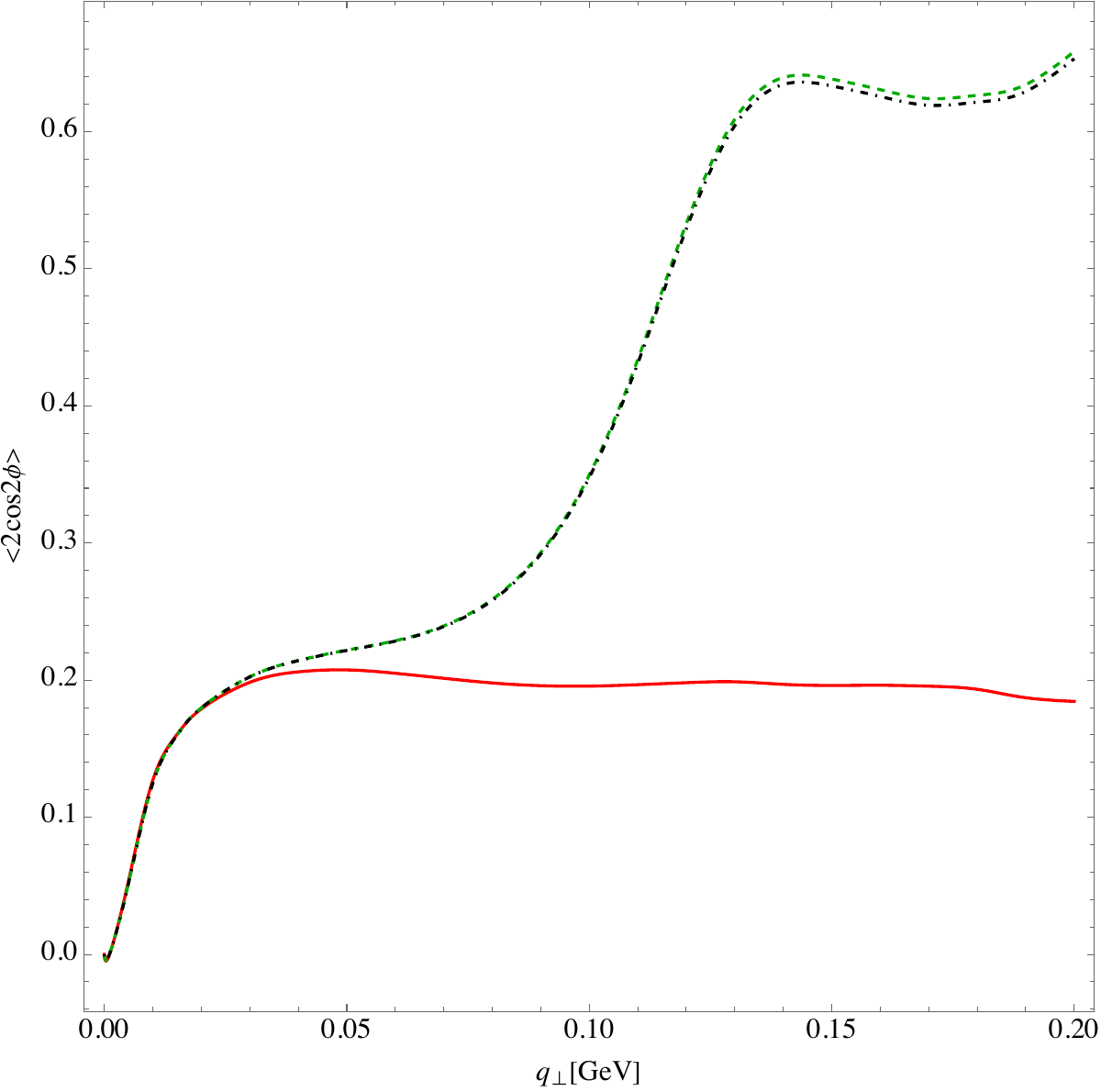}
        \label{fig:subfig2}
    \end{minipage}
    \vspace{0.0\textwidth} % 添加垂直间距
    \begin{minipage}{0.3\textwidth}
        \centering
        \includegraphics[width=\linewidth]{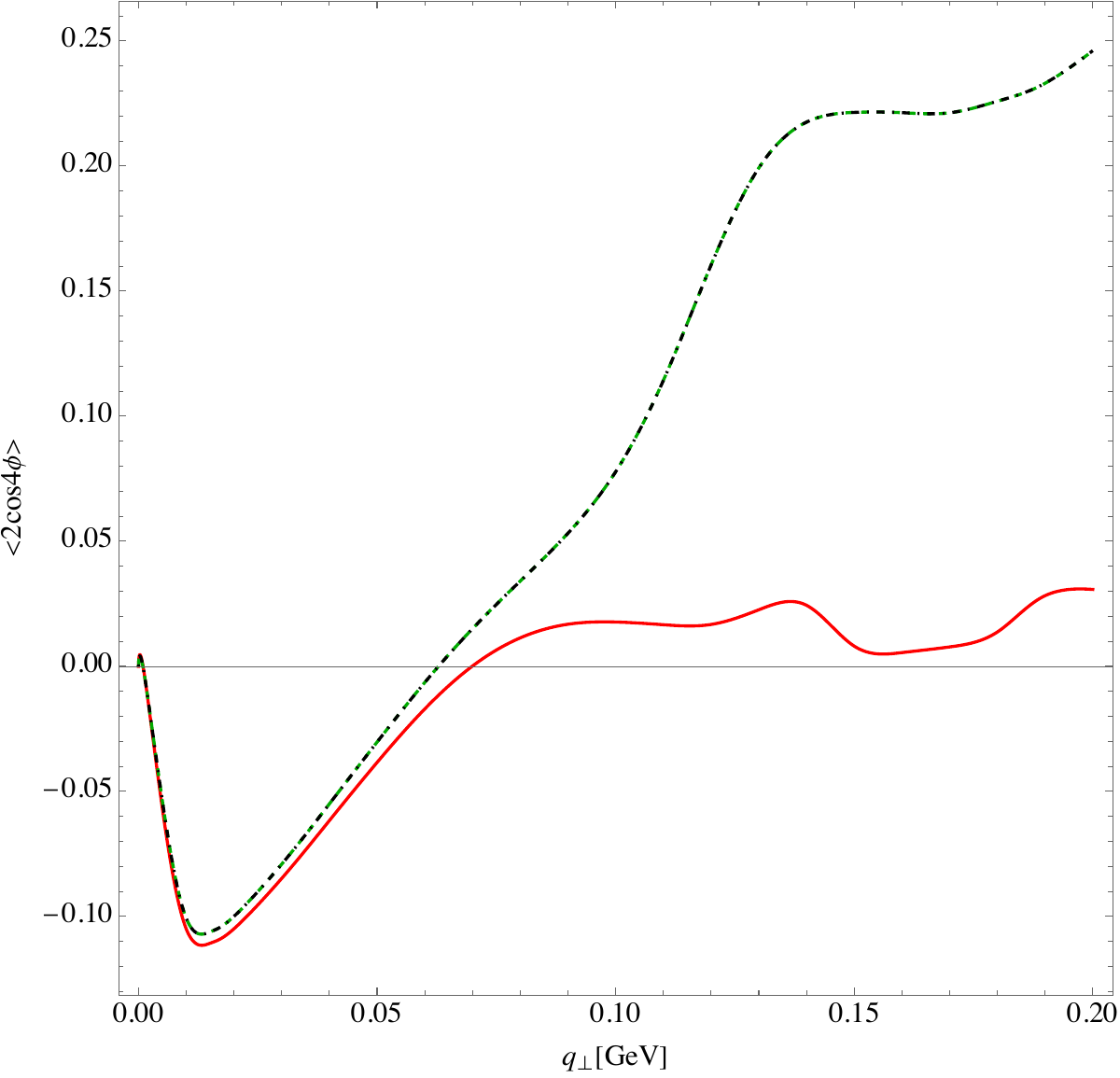}
        \label{fig:subfig3}
    \end{minipage}
    \caption{Dimuon production in unrestricted UPCs in Au-Au collisions. The following kinematic cuts are imposed: the $\mu$'s rapidities $|y_{1,2}|<0.8$, and the invariant mass of the dimuon $0.4\,{\rm GeV}<M_{\mu\mu}<0.64\,{\rm GeV}.$}
    \label{fig:total}
\end{figure}
The employed form factor, derived from the STARlight generator, is given by:
\begin{align}
F(|k|)=\frac{3[\sin(|k|R_A)-|k|R_A\cos(|k|R_A)]}{(|k|R_A)^3(a^2k^2+1)},
\end{align}
where $a=0.7$ fm, and the nuclear radius $R_A$ is defined as $R_A=1.1A^{\frac{1}{3}}$ fm for gold (Au) with $A=197$. The proton mass is $M_p=0.938$ GeV.

In the figures, we observe minimal modification by the soft function for the angular averages $\langle 2\cos 2\phi \rangle$ and $\langle 2\cos 4\phi \rangle$ in the region where $|q_\perp| < 0.01$ GeV. The primary attribution for the angular correlation asymmetry is the polarization of the initial photons. In this scenario, the polarization direction of the photons, induced by relativistic heavy ions, aligns parallel to their transverse momentum direction.

A clear physical explanation for the effectiveness is as follows \cite{pu2023coherent}: In the initial state, photons are fully linearly polarized. Assuming the fusion of two spin +1 photons results in the formation of a lepton pair, this pair in the final state possesses an angular momentum of +2. Neglecting the mass of the leptons, the back-to-back emission of the lepton pair in the final state results in a total spin of zero. The conservation of angular momentum dictates that the lepton pair possesses an orbital angular momentum of +2. In the amplitude, a lepton pair with an orbital angular momentum of +2 interferes with a conjugate amplitude where the lepton pair possesses an orbital angular momentum of -2. This interference results in angular modulation of the average angle correlations in the final state lepton pair, characterized by a $\cos 4\phi$ dependence. Taking into account the mass of the leptons, the mass term induces a helicity flip in the final-state leptons. Consequently, this results in back-to-back emitted lepton pairs carrying a total spin of either +1 or -1. Assuming that the lepton pair possesses a total spin of +1, it follows from the principle of angular momentum conservation that the pair exhibits an orbital angular momentum of +1. In the amplitude, the wave function of a lepton pair with an orbital angular momentum of +1 interferes with the conjugate amplitude's wave function, where the lepton pair has an orbital angular momentum of -1. This interference results in angular modulation of the average angle correlations in the final state lepton pair, characterized by a $\cos 2\phi$ dependence.

In regions of higher transverse momentum, for $\langle 2\cos 2\phi \rangle$ beyond $0.03$ GeV and for $\langle 2\cos 4\phi \rangle$ beyond $0.01$ GeV, the contributions from the soft function become significant, indicating that the azimuthal asymmetries in these regions are primarily due to the radiation of soft photons.
\begin{figure}[ht]
    \centering
    \begin{minipage}{0.3\textwidth}
        \centering
        \includegraphics[width=\linewidth]{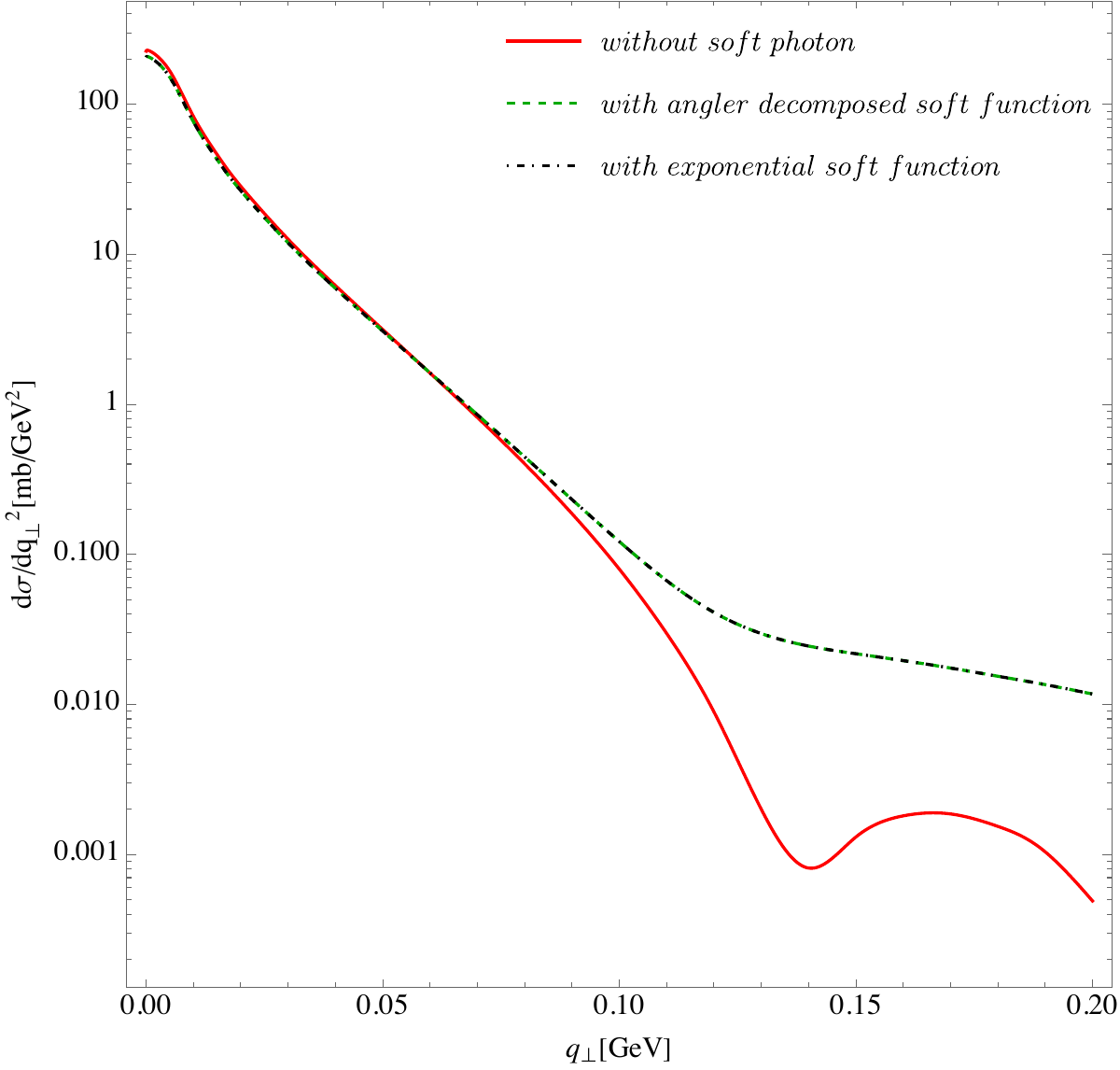}
    \end{minipage}
    \hspace{0.0\textwidth} 
    \begin{minipage}{0.3\textwidth}
        \centering
        \includegraphics[width=\linewidth]{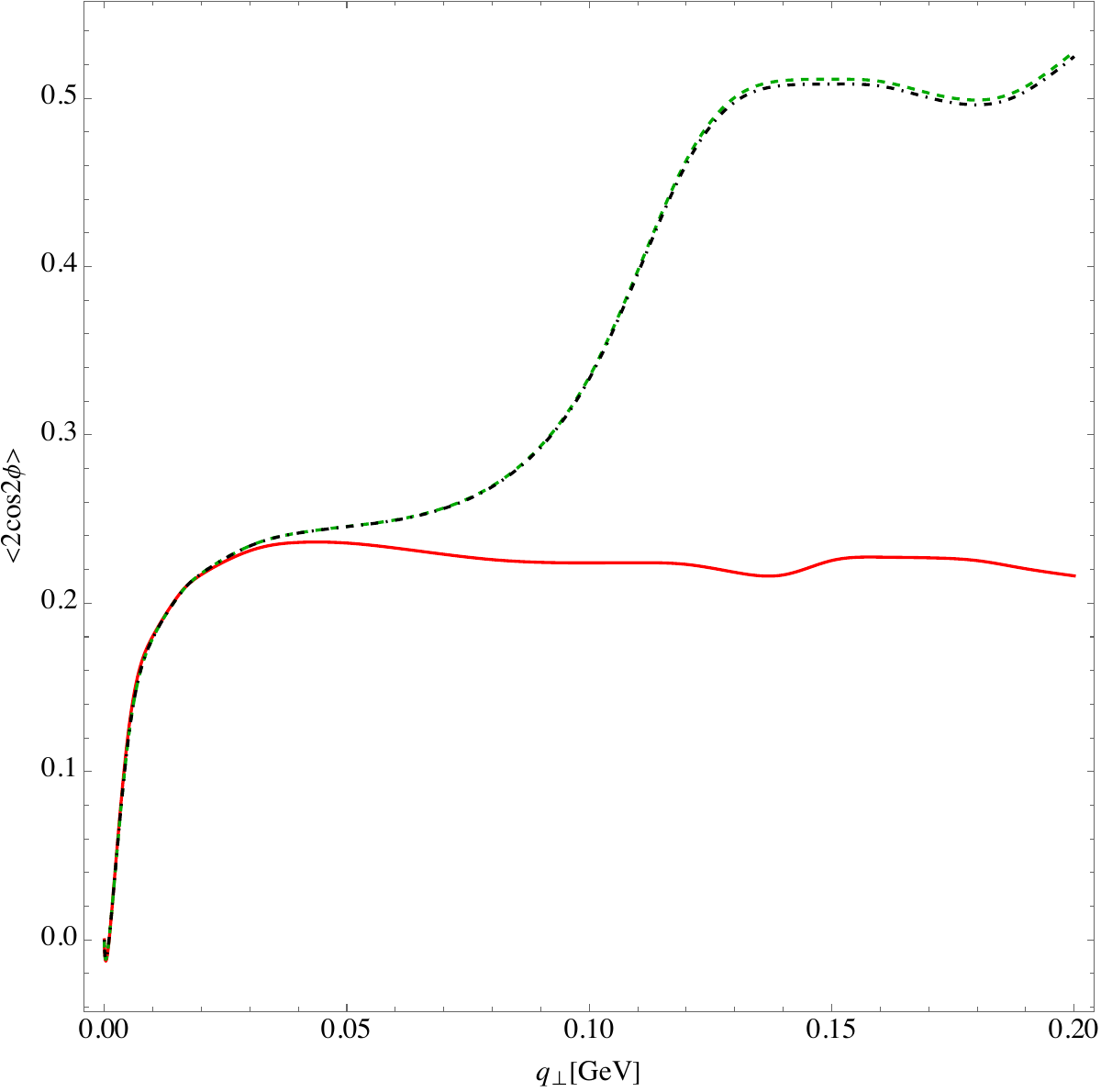}
    \end{minipage}
    \vspace{0.0\textwidth} 
    \begin{minipage}{0.3\textwidth}
        \centering
        \includegraphics[width=\linewidth]{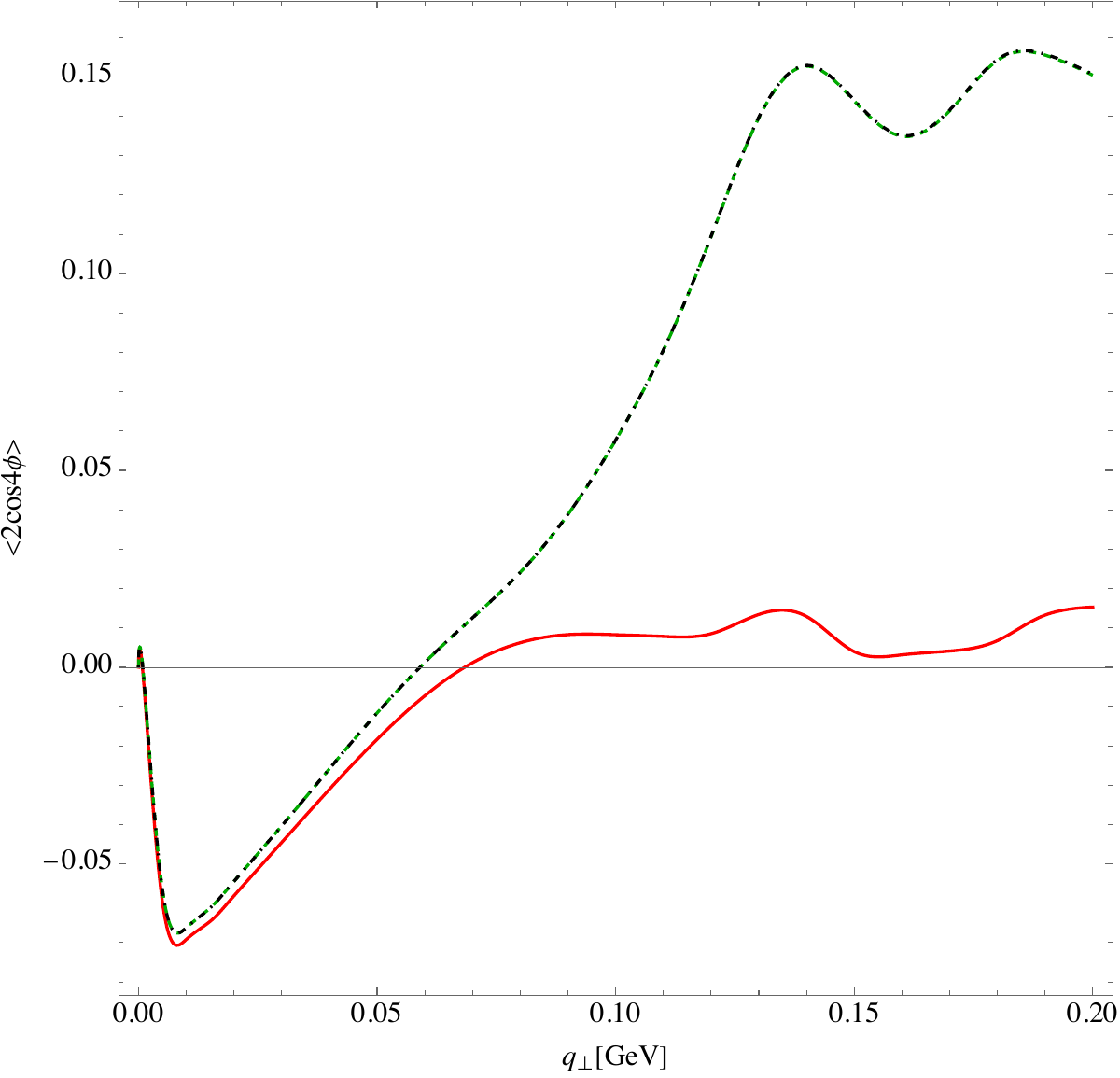}
    \end{minipage}
    \caption{Ditau production in unrestricted UPCs in Pb-Pb collisions. The following kinematic cuts are imposed: the $\tau$'s rapidities $|y_{1,2}|<1$, and the invariant mass of the ditau $6\,{\rm GeV}<M_{\tau\tau}<10\,{\rm GeV}.$}
    \label{fig:total2}
\end{figure}
Figure \ref{fig:total2} displays the outcomes from unrestricted ultra-peripheral collisions at $5020\,{\rm GeV}$ in Pb-Pb collisions for the process $\gamma+\gamma\rightarrow\tau^++\tau^-$, featuring $\tau$ leptons with a mass of $m_\tau=1.777\,{\rm GeV}$, the nucleon number of lead \( A=208 \), the nuclear radius \( R_A = 1.1A^{1/3} \), and rapidities \( |y_{1,2}| < 1 \). The figures illustrate the results for angular modulation of $\tau$ pairs, corresponding to angle-independent behavior (left), $\langle 2\cos{2\phi}\rangle$ (middle), and $\langle 2\cos{4\phi} \rangle$ (right), respectively. The results demonstrate that excluding the electromagnetic moment, the behavior of angular modulation in $\tau$ pairs would closely resemble that in $\mu$ pairs, differing only slightly at the peak value due to the larger mass of $\tau$.

\section{Conclusion}
In this paper, we begin by developing an effective Lagrangian and applying power counting alongside factorization techniques within SCET to factorize the differential cross section of UPC into a non-perturbative photon distribution function and a perturbative high-energy cross section. We then introduce and define the GTMD of initial photons, showing that the GTMD can be expressed as the ratio of the photon distribution function to the energy fraction, adjusted by a Fourier factor. Furthermore, we examine the influence of final state soft radiation and integrate a soft function into the differential cross section through convolution. In subsequent sections, we compute the Born-level angular correlation for the final state lepton pair, and analyze the angular correlation modifications due to soft photon emission in the final state. The culmination of our study is presented through numerical simulations. These results clearly show that in regions of higher transverse momentum, the impact of soft photon radiation on the azimuthal asymmetries, specifically $\langle\cos 2\phi\rangle$ and $\langle\cos 4\phi\rangle$, becomes significantly more pronounced.

%\onecolumngrid
%\newpage

%\section*{Supplemental material}

%\section{}\label{app:adim}
%\section{}

%\bibliography{refs.bib}{}
%\bibliographystyle{ieeetr}

\end{document}